\documentclass[%
reprint,
 amsmath,
 amssymb,
 aps,
 prx,
]{revtex4-2}

\usepackage{xcolor}
\usepackage[colorlinks=true,citecolor=blue]{hyperref}
\usepackage{graphicx}
\usepackage{subfigure}
\usepackage{bookmark}
\usepackage{hyperref}
\usepackage[normalem]{ulem}
\usepackage{dsfont} 
\usepackage{upgreek}
\usepackage{booktabs}
\usepackage{tabularx}

\newcommand{\newtext}[1]{{\color{black}{#1}}}

\renewcommand{\vec}[1]{\boldsymbol{#1}}

\newcommand{\comma}{~,}
\newcommand{\fullstop}{~.}
\newcommand{\hc}{\mathrm{H.c.}}
\newcommand{\abs}[1]{\left\vert #1 \right\vert}
\newcommand{\cabs}[1]{\vert #1 \vert}

\newcommand{\cerw}[1]{\langle #1 \rangle}
\newcommand{\komm}[2]{\left[ #1, #2 \right]}

\newcommand{\cakomm}[2]{\lbrace #1, #2 \rbrace}
\newcommand{\ket}[1]{\left\vert #1 \right\rangle}
\newcommand{\bra}[1]{\left\langle #1 \right\vert}

\renewcommand{\d}{\mathrm{d}}

\newcommand{\ncav}{n_\mathrm{cav}}

\newcommand{\1}{\mathds{1}}
\newcommand{\bfDelta}{\mathbf{\Delta}}
\newcommand{\Xidet}{\Xi_\mathrm{det}}
\newcommand{\Xidetsq}{\Xidet^2}
\newcommand{\Wineland}{\xi_\mathrm{R}^2}

\newcommand{\omc}{\omega_\mathrm{c}}
\newcommand{\oms}{\omega_\mathrm{s}}
\newcommand{\ome}{\omega_\mathrm{e}}
\newcommand{\chiopt}{\chi_\mathrm{ACF}}
\newcommand{\chimw}{\chi_\mathrm{SCF}}
\newcommand{\etaphi}{\eta_\phi}
\newcommand{\etarel}{\eta_\mathrm{rel}}

\newcommand{\gammalocmw}{\gamma_\mathrm{loc}^\mathrm{SCF}}
\newcommand{\gammalocopt}{\gamma_\mathrm{loc}^\mathrm{ACF}}
\newcommand{\Omegaopt}{\Omega_\mathrm{ACF}}
\newcommand{\Omegamw}{\Omega_\mathrm{SCF}}

\newcommand{\chitc}{\chi_\mathrm{TC}}
\newcommand{\Deltatc}{\Delta_\mathrm{TC}}
\newcommand{\gammarel}{\gamma_\mathrm{rel}}
\newcommand{\gammaphi}{\gamma_\phi}
\newcommand{\Gammaphi}{\Gamma_\phi}
\newcommand{\Gammarel}{\Gamma_\mathrm{rel}}
\newcommand{\gammalocz}{\gamma_z}
\newcommand{\gammalocp}{\gamma_+}
\newcommand{\gammalocm}{\gamma_-}
\newcommand{\tildeoms}{\tilde{\omega}_\mathrm{s}}
\newcommand{\tsqz}{t_\mathrm{sqz}}
\newcommand{\tunsqz}{t_\mathrm{unsqz}}
\newcommand{\tamp}{t_\mathrm{amp}}
\newcommand{\Gmet}{\mathcal{G}_\mathrm{met}}
\newcommand{\Gmetideal}{\Gmet^\mathrm{ideal}}
\newcommand{\Gmetest}{\Gmet^\mathrm{est}}

\def\FigOATCompareMetrologicalGainVsCollectiveCooperativityFormulaAsymptoteDashDot{0.1 N \eta}
\def\FigOATCompareMetrologicalGainVsCollectiveCooperativityValueXidetsq{1.0}
\def\FigOATCompareMetrologicalGainVsCollectiveCooperativityValueN{10^5}

\def\FigSCFSingleGainRescaledGainVsRescaledCooperativityFitParameterA{1.31}
\def\FigSCFSingleGainRescaledGainVsRescaledCooperativityFitParameterB{0.64}

\def\FigSCFdoubleGainRescaledGainVsRescaledCooperativityFitParameterA{1.20}
\def\FigSCFdoubleGainRescaledGainVsRescaledCooperativityFitParameterB{1.09}

\def\FigOATSCFDoubleMetrologicalGainVsCollectiveCooperativityValueXidetsq{1.0}

\def\FigOATACFDoubleMetrologicalGainVsCollectiveCooperativityFormulaAsymptoteDashDot{0.1 N \eta}

\def\FigOATSCFSqueezingWinelandVsCFormulaAsymptote{3/\sqrt{N \eta}}

\def\FigOATSCFOptimalDetuningVsCFormulaAsymptote{\sqrt{N \eta / 10}}

\def\FigSCFSingleGainOptimalDetuningVsCooperativityFormulaAsymptote{0.4 \sqrt{\eta}}

\def\FigSCFSingleGainOptimalTimeVsRescaledCooperativityFormulaAsymptote{0.5 \sqrt{N} \eta}

\def\FigACFSingleGainOptimalDetuningVsCooperativityFormulaAsymptote{0.4 \sqrt{\eta}}

\def\FigACFSingleGainOptimalTimeVsRescaledCooperativityFormulaAsymptote{0.5 \sqrt{N} \eta}

\begin{document}

\title{Revisiting the impact of dissipation on time-reversed one-axis-twist quantum-sensing protocols}
\author{Martin Koppenh\"ofer and A.\ A.\ Clerk}
\affiliation{Pritzker School of Molecular Engineering, University of Chicago, Chicago, IL 60637, USA}
\date{\today}

\begin{abstract}
Spin squeezing can increase the sensitivity of interferometric measurements of small signals in large spin ensembles beyond the standard quantum limit.  
In many practical settings, the ideal metrological gain is limited by imperfect readout of the sensor. 
To overcome this issue, protocols based on time reversal of unitary one-axis-twist (OAT) spin-squeezing dynamics have been proposed. 
Such protocols mitigate readout noise and, when implemented using cavity feedback, have been argued to also be robust against dissipation as long as the collective cooperativity of the system is sufficiently large [Davis \emph{et al.}, PRL \textbf{116}, 053601 (2016)]. 
Here, we perform a careful systematic study of dissipative effects on three different implementations of a OAT twist-untwist sensing scheme (based on symmetric as well as asymmetric cavity feedback and on a Tavis-Cummings interaction).
Our full treatment shows that the three approaches have markedly different properties and resilience when subject to dissipation. 
Moreover, the metrological gain for an implementation using symmetric cavity feedback is more sensitive to undesired dissipation than was previously appreciated. 
\end{abstract}

\maketitle


\section{Introduction}

Spin-based quantum sensing protocols use entangled states of large spin ensembles to improve interferometric measurements of small signals \cite{Giovannetti2006,Degen2017,Pezze2018}. 
Famous entangled states that provide a sensitivity improvement over any classical sensing strategy are GHZ and spin-squeezed states \cite{Kitagawa1993,Bollinger1996}. 
Both of these states can be generated by evolving a coherent-spin state, i.e., a product state of identically polarized spins, under a so-called one-axis-twist (OAT) Hamiltonian \cite{Kitagawa1993}.
Even though a two-axis-twist Hamiltonian provides faster and stronger spin squeezing in principle \cite{Kitagawa1993}, the OAT Hamiltonian has become the workhorse for numerous quantum sensing experiments thanks to its experimental simplicity \cite{Leroux2010,Gross2010,Riedel2010,Strobel2014,Muessel2015,Hosten2016,Hosten2016b,Bohnet2016,Braverman2019,PedrozoPenafiel2020,Colombo2022,Hines2023,Eckner2023,Franke2023}.  
It can be implemented in several different ways.  
One approach is to couple the two-level sensor spins directly to a detuned bosonic mode using a Tavis-Cummings (TC) interaction \cite{Bennett2013,Dooley2016,LewisSwan2018}; this is common in solid-state platforms \cite{Bennett2013,Dooley2016,Groszkowski2020}. 
An alternative approach is to implement a cavity-feedback protocol, where the sensor spins are coupled to the bosonic mode using auxiliary levels \cite{SchleierSmith2010,Leroux2010}; this is the preferred method in atomic platforms \cite{SchleierSmith2010,Leroux2010,Zhang2015,Davis2016,Hosten2016,Hosten2016b,%
Chu2021,Braverman2019,PedrozoPenafiel2020,Colombo2022,Li2022} but could also be useful in solid-state settings \cite{Zou2014}.
As explained below, there are two versions of the cavity-feedback approach, the symmetric cavity-feedback (SCF) scheme \cite{SchleierSmith2010,Leroux2010,Zhang2015,Davis2016,Hosten2016,Hosten2016b,Chu2021} and the asymmetric cavity-feedback (ACF) scheme \cite{Braverman2019,PedrozoPenafiel2020,Chu2021,Colombo2022,Li2022}.

\begin{figure}[t!]
	\centering
	\includegraphics[width=0.48\textwidth]{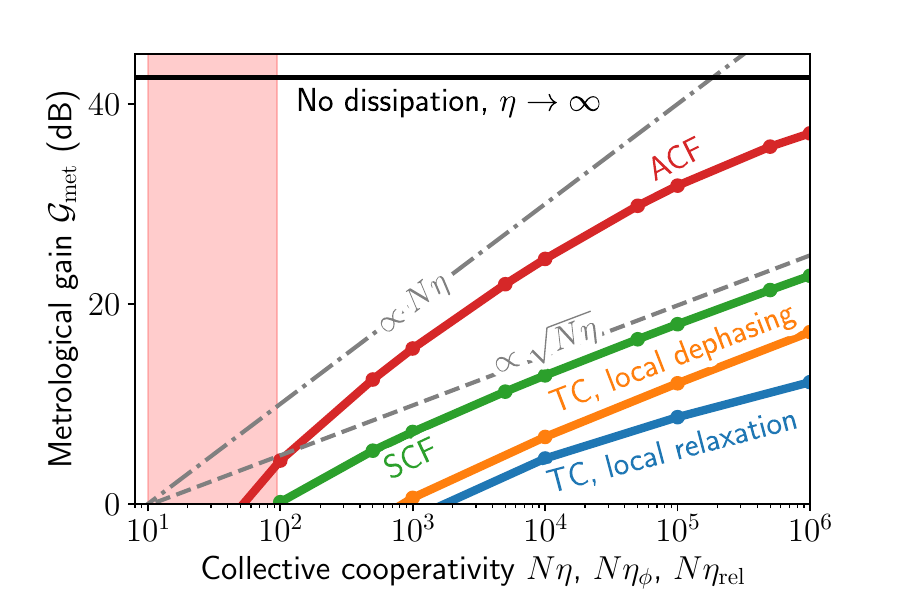}
	\caption{
		Metrological gain $\Gmet = 1/N (\mathbf{\Delta}\phi)^2$ as a function of the collective cooperativity for a sensing protocol using time-reversal of one-axis-twist (OAT) dynamics.  
        We use $N = \FigOATCompareMetrologicalGainVsCollectiveCooperativityValueN$ spins and include readout noise equal to the standard projection noise of a coherent-spin state ($\Xidetsq = \FigOATCompareMetrologicalGainVsCollectiveCooperativityValueXidetsq$). 
		Thick solid curves correspond to different implementations of the OAT Hamiltonian [symmetric cavity feedback (SCF), asymmetric cavity feedback (ACF), and a Tavis-Cummings (TC) interaction with single-spin dephasing or single-spin relaxation].
		Solid black line: ideal metrological gain $\Gmetideal$ that can be achieved at zero dissipation.
		The dashed gray curve shows the expression $\Gmetest$, defined in Eq.~\eqref{eqn:GmetEmily}, which has been derived in Ref.~\onlinecite{Davis2016} using a heuristic model of dissipative effects and applies to the SCF and TC schemes. 
		As shown by the thick SCF and TC curves, nonperturbative effects of dissipation can lead to strong deviations from this result even though the collective cooperativity $N \eta$ is large. 
        In particular, the collective cooperativity required to reach $\Gmet > 1$ is underestimated by almost an order of magnitude (shaded red area).
		The dash-dotted gray curve indicates $\Gmet = \FigOATCompareMetrologicalGainVsCollectiveCooperativityFormulaAsymptoteDashDot$. 
	}
	\label{fig:comparison}
\end{figure}

Even though all three of these schemes implement the same OAT Hamiltonian, they differ in the structure of undesired dissipative processes that arise as a byproduct of the OAT interaction. 
These differences may be crucial since dissipation competes with the unitary OAT dynamics and limits the attainable level of spin squeezing.
Nevertheless, it has been shown that all three schemes generate spin squeezing in the presence of dissipation if the \emph{collective} cooperativity is large \cite{SchleierSmith2010,LewisSwan2018,Li2022}.

\begin{figure*}
	\centering
    \includegraphics[width=0.3\textwidth]{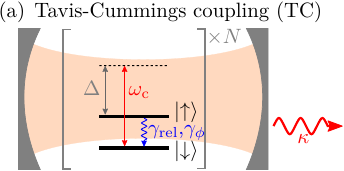}~~~~
    \includegraphics[width=0.3\textwidth]{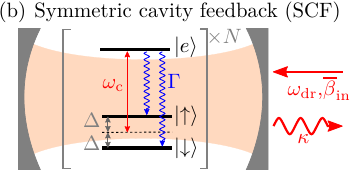}~~~~
    \includegraphics[width=0.3\textwidth]{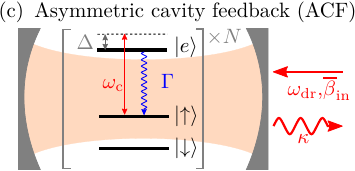}
	\caption{
        Sketch of the three considered schemes to experimentally implement the one-axis-twist (OAT) Hamiltonian $\hat{H}_\mathrm{OAT}$, defined in Eq.~\eqref{eqn:HOAT}, in an ensemble of $N$ spin-$1/2$ systems  with levels $\{ \ket{\uparrow}_i, \ket{\downarrow}_i \}_{i=1,\dots,N}$. 
        (a) $\hat{H}_\mathrm{OAT}$ can be implemented using a direct Tavis-Cummings (TC) coupling between the spin-$1/2$ systems and a detuned cavity mode with frequency $\omc$ and detuning $\Delta$.
        Undesired dissipative processes are single-spin relaxation (rate $\gammarel$), single-spin dephasing (rate $\gammaphi$), and cavity decay (rate $\kappa$). 
        Alternatively, the OAT Hamiltonian can be implemented using feedback from a driven cavity mode, with drive frequency $\omega_\mathrm{dr}$ and ampliude $\overline{\beta}_\mathrm{in}$, either by 
        (b) symmetric cavity feedback (SCF), where both $\ket{\uparrow_j}$ and $\ket{\downarrow_j}$ are coupled to the excited level $\ket{e_j}$ via the cavity mode with equal but opposite-in-sign detuning $\Delta$, or by 
        (c) asymmetric cavity feedback (ACF), where only $\ket{\uparrow_j}$ is coupled to $\ket{e_j}$ via the cavity mode with detuning $\Delta$. 
        Undesired dissipative processes in the cavity-feedback schemes are decay of the excited level (rate $\Gamma$) and cavity decay. 
	}
	\label{fig:system}
\end{figure*}

Another interesting feature of the OAT Hamiltonian is that it can be used for amplification of small signals in a spin system \cite{Davis2016,Colombo2022} (sometimes also called ``phase magnification'' \cite{Hosten2016}), 
which is an important technique to render quantum sensing protocols robust to imperfect readout \cite{Davis2016,Hosten2016,Koppenhoefer2022,Koppenhoefer2023}.
OAT-based spin amplification relies on two time-reversed applications of the OAT Hamiltonian (which is why it is sometimes called a ``twist-untwist protocol'') and, similar to the generation of spin squeezing, the question arises of how the amplification dynamics depends on the undesired dissipation and how this impacts the achievable metrological gain. 
In the seminal proposal of OAT amplification \cite{Davis2016}, an appealing intuitive argument suggested that, for a SCF-based implementation, a large collective cooperativity was enough to ensure metrological gain in the presence of dissipation.  
This heuristic argument assumed moderate readout noise and that the signal amplification was independent of dissipation. 
Surprisingly, to the best of our knowledge, this prediction has not been rigorously tested.  
Moreover, the robustness predicted in Ref.~\onlinecite{Davis2016} for the SCF-based implementation is not generic to all implementations of OAT amplification:
A recent analysis of a TC-based implementation of the OAT Hamiltonian \cite{Koppenhoefer2022} showed that undesired dissipative processes can strongly degrade the amplification, such that significant levels of gain are only possible in an experimentally very challenging regime of extremely large \emph{single-spin} cooperativities.

It is clearly crucial to applications to understand when one can achieve large metrological gain in the presence of readout noise and dissipation.  
In this work, we address this question by performing a careful analysis of the requirements on cooperativity that are needed to obtain significant levels of metrological gain and signal amplification in TC, SCF, and ACF-based OAT amplification schemes.
We use a combination of numerically exact solutions of the quantum master equation (QME) for small system sizes and a mean-field theory (MFT) analysis for larger system sizes.
We find that, depending on the level of readout noise, the requirement for achieving large metrological gain is very different in each scheme.  
This is in contrast to the simpler task of generating spin squeezing alone, where all implementations yield similar constraints.  
Moreover, we find that the metrological gain in the SCF and TC schemes is more sensitive to undesired dissipation than previously thought.

This paper is organized as follows. 
In Sec.~\ref{sec:System}, we introduce the three different schemes to implement a OAT Hamiltonian. 
For each of the schemes, we then derive an effective QME including only the spin degrees of freedom of interest in Sec.~\ref{sec:EffectiveQME}. 
In Sec.~\ref{sec:OATProtocol}, we review the twist-untwist sensing protocol proposed by Davis \emph{et al.} \cite{Davis2016} and their heuristic analysis for the scaling of the metrological in the presence of dissipation. 
We then analyze numerically which requirements need to be satisfied to achieve large amplification in the different schemes in Sec.~\ref{sec:Gain}. 
For the cavity-feedback based schemes, we find an interesting dependence of the gain on a scaled single-spin cooperativity, which we interpret using an intuitive argument based on MFT in Sec.~\ref{sec:MFT}. 
Finally, in Sec.~\ref{sec:MetrologicalGain}, we analyze the full metrological gain of the cavity-feedback-based schemes. 
We conclude in Sec.~\ref{sec:Conclusion}.

\section{The system}
\label{sec:System}

We consider an ensemble of $N$ atoms with two levels $\ket{\uparrow_j}$ and $\ket{\downarrow_j}$, $j \in \{1, \dots, N\}$, which form an effective two-level system that is used for sensing. 
The state $\ket{\uparrow_j}$ ($\ket{\downarrow_j}$) is the eigenstate of the Pauli operator $\hat{\sigma}_z^{(j)}$ with eigenvalue $+1$ ($-1$), and the lowering operator is $\hat{\sigma}_-^{(j)} = [\hat{\sigma}_+^{(j)}]^\dagger = \ket{\downarrow_j}\bra{\uparrow_j}$.
The ensemble of two-level systems can be described by a collective spin vector with components $\hat{S}_x = (\hat{S}_+ + \hat{S}_-)/2$, $\hat{S}_y = (\hat{S}_+ - \hat{S}_-)/2i$, and $\hat{S}_z = \sum_{j=1}^N \hat{\sigma}_z^{(j)}/2$, where $\hat{S}_\pm = \sum_{j=1}^N \hat{\sigma}_\pm^{(j)}$. 
Our goal is to generate a OAT Hamiltonian
\begin{align}
    \hat{H}_\mathrm{OAT} = \chi \hat{S}_z^2 \comma
    \label{eqn:HOAT}
\end{align}
where $\chi$ denotes the OAT strength and both its sign and magnitude can be controlled experimentally.
A Hamiltonian of the form of Eq.~\eqref{eqn:HOAT} can be engineered by coupling the spins to a bosonic mode with frequency $\omc$, damping rate $\kappa$, and annihilation operator $\hat{a}$. 
The system is described by a QME of the form ($\hbar = 1$)
\begin{align}
	\frac{\d}{\d t} \hat{\rho} &= - i \komm{\hat{H}_0 + \hat{H}_1}{\hat{\rho}} + \kappa \mathcal{D}[\hat{a}] \hat{\rho} + \mathcal{L}_\mathrm{spin} \hat{\rho} \comma 
	\label{eqn:Details:QMEStartingPoint}
\end{align}
where $\hat{H}_0 = \omc \hat{a}^\dagger \hat{a} + \oms \hat{S}_z$ and the specific form of the Hamiltonian $\hat{H}_1$ as well as the dissipative processes $\mathcal{L}_\mathrm{spin}$ of the spins depend on the details of the spin-cavity coupling scheme. 
\newtext{%
The superoperator $\mathcal{D}[\hat{O}]\hat{\rho} = \hat{O}\hat{\rho} \hat{O}^\dagger - \cakomm{\hat{O}^\dagger \hat{O}}{\hat{\rho}}/2$ denotes a standard Lindblad dissipator.
}%

For the TC-based implementation of OAT, shown in Fig.~\ref{fig:system}(a), they are given by
\begin{subequations}%
\begin{align}%
	\hat{H}_1^\mathrm{TC} &= g \left( \hat{S}_- \hat{a}^\dagger + \hat{S}_+ \hat{a} \right) \comma \displaybreak[1]\\
	\mathcal{L}_\mathrm{spin}^\mathrm{TC} \hat{\rho} &= \sum_{j=1}^N \left( \gammarel \mathcal{D}[\hat{\sigma}_-^{(j)}] + \frac{\gammaphi}{2} \mathcal{D}[\hat{\sigma}_z^{(j)}] \right) \hat{\rho} \comma
\end{align}%
\label{eqn:Details:TC}%
\end{subequations}%
where $g$ is the spin-cavity coupling strength and $\gammarel$ ($\gammaphi$) is the single-spin relaxation (dephasing) rate. 
To generate the OAT Hamiltonian~\eqref{eqn:HOAT} from the TC interaction, one considers the dispersive regime where the spins are highly detuned from the cavity mode, $\abs{\Deltatc} \equiv \abs{\omc - \oms} \gg g$.

The SCF-based implementation of OAT, shown in Fig.~\ref{fig:system}(b), uses a third, auxiliary atomic level $\ket{e_j}$ and is defined by
\begin{subequations}%
\begin{align}%
	\hat{H}_1^\mathrm{SCF} &= \sum_{j=1}^N \Big( \omc \ket{e_j}\bra{e_j} + g \left[ \hat{a} \left( \ket{e_j}\bra{\uparrow_j} + \ket{e_j} \bra{\downarrow_j} \right) + \hc \right] \Big) \nonumber \\
		&\phantom{=}\ + \sqrt{\kappa} \left[ \beta_\mathrm{in}(t) \hat{a}^\dagger + \beta_\mathrm{in}^*(t) \hat{a} \right] \comma \displaybreak[1]\\
	\mathcal{L}_\mathrm{spin}^\mathrm{SCF} \hat{\rho} &= \sum_{j=1}^N \Gamma \Big( \mathcal{D}\left[ \ket{\uparrow_j}\bra{e_j} \right] + \mathcal{D} \left[ \ket{\downarrow_j}\bra{e_j} \right] \Big) \hat{\rho} \fullstop
\end{align}%
\label{eqn:Details:SCF}%
\end{subequations}%
The levels $\ket{\uparrow_j}$ and $\ket{\downarrow_j}$ are equally detuned by an amount $\pm \Delta$ from a cavity-assisted transition to the auxiliary level $\ket{e_j}$, i.e., $\oms = 2 \Delta$, and the excited level can decay back into the levels $\ket{\uparrow_j}$ and $\ket{\downarrow_j}$ with equal rates $\Gamma$. 
This coupling scheme is used in microwave atomic clocks, where the levels $\ket{\uparrow_j}$ and $\ket{\downarrow_j}$ are typically long-lived states connected by a clock transition \cite{Davis2016,Hosten2016} such that relaxation and dephasing processes within the $\{\ket{\uparrow_j},\ket{\downarrow_j}\}$ manifold (which were present in $\mathcal{L}_\mathrm{spin}^\mathrm{TC}$) can now be neglected. 
The OAT strength $\chi$ and its sign are controlled by the amplitude $\overline{\beta}_\mathrm{in}$ and the detuning $\delta = \omega_\mathrm{dr} - \omc$ of a coherent drive of the optical cavity, 
\begin{align}
    \beta_\mathrm{in}(t) = \overline{\beta}_\mathrm{in} e^{-i \omega_\mathrm{dr} t} \comma
    \label{eqn:DefinitionDrive}
\end{align}
which leads to a steady-state intracavity photon number 
\begin{align}
    \ncav = \frac{\kappa \cabs{\overline{\beta}_\mathrm{in}}^2}{\delta^2 + \kappa^2/4} \fullstop
    \label{eqn:Definitionncav}
\end{align}

Finally, for the ACF-based implementation, shown in Fig.~\ref{fig:system}(c), only the level $\ket{\uparrow_j}$ is coupled to $\ket{e_j}$ and we have
\begin{subequations}%
\begin{align}%
	\hat{H}_1^\mathrm{ACF} &= \sum_{j=1}^N \Big( \omc \ket{e_j}\bra{e_j} + g \left[ \hat{a} \ket{e_j}\bra{\uparrow_j} + \hc \right] \Big) \nonumber \\
		&\phantom{=}\ + \sqrt{\kappa} \left[ \beta_\mathrm{in}(t) \hat{a}^\dagger + \beta_\mathrm{in}^*(t) \hat{a} \right] \comma \displaybreak[1]\\
	\mathcal{L}_\mathrm{spin}^\mathrm{ACF} \hat{\rho} &= \sum_{j=1}^N \Gamma \mathcal{D}\left[ \ket{\uparrow_j}\bra{e_j} \right] \hat{\rho} \comma
\end{align}%
\label{eqn:Details:ACF}%
\end{subequations}%
where $\omc = \Delta + \ome - \oms/2$. 
This coupling scheme is used in optical atomic clocks, where the levels $\ket{\uparrow_j}$ and $\ket{\downarrow_j}$ are again long-lived states connected by a clock transition. 
Since only $\ket{\uparrow_j}$ is coupled to the auxiliary $\ket{e_j}$ level, the level $\ket{e_j}$ can only decay back to $\ket{\uparrow_j}$ at a rate $\Gamma$.

\section{Effective quantum master equation}
\label{sec:EffectiveQME}

\begin{table}
\caption{%
	Parameters of the effective QME~\eqref{eqn:QME:EffectiveGeneral} describing the dynamics in the subspace spanned by the states $\{\ket{\uparrow_j},\ket{\downarrow_j}\}_{j=1,\dots,N}$ for the different implementations of the OAT Hamiltonian shown in Fig.~\ref{fig:system}.
	The single-spin cooperativity $\eta$, the number of photons $\ncav$ in the cavity due to a coherent drive, and the amplitude $\beta_\mathrm{in}(t)$ of the coherent drive are defined in Eqs.~\eqref{eqn:Definitioneta}, \eqref{eqn:Definitionncav}, and~\eqref{eqn:DefinitionDrive}, respectively. 
}
\label{tbl:Rates}
\newcolumntype{R}{>{\raggedleft\arraybackslash}p{.132\textwidth}}
\begin{tabular}{l|RRR}
	\toprule
	& \multicolumn {1}{c}{\textbf{TC}} & \multicolumn {1}{c}{\textbf{SCF}} & \multicolumn {1}{c}{\textbf{ACF}} \\
	\midrule
	$\chi$       & $\frac{g^2}{\Deltatc}$           & $4 \frac{g^4}{\Delta^2} \ncav \frac{\delta}{\delta^2 + \kappa^2/4}$   & $\frac{g^4}{\Delta^2} \ncav \frac{\delta}{\delta^2 + \kappa^2/4}$ \\
	$\tildeoms$  & $\oms$                           & $\oms$                                                                & $\oms + N \chiopt$ \\
	$\Gammaphi$  & $0$                              & $\chimw \frac{\kappa}{\delta}$                                        & $\chiopt \frac{\kappa}{\delta}$ \\
	$\Gammarel$  & $\chitc \frac{\kappa}{\Deltatc}$ & $0$                                                                   & $0$ \\
	$\gammalocz$ & $\gammaphi/2$                    & $\frac{1}{2} \chimw \frac{\delta^2 + \kappa^2/4}{\kappa \delta \eta}$ & $\chiopt \frac{\delta^2 + \kappa^2/4}{\delta \kappa \eta}$ \\
	$\gammalocp$ & $0$                              & $\chimw \frac{\delta^2 + \kappa^2/4}{\kappa \delta \eta}$             & $0$ \\
	$\gammalocm$ & $\gammarel$                      & $\chimw \frac{\delta^2 + \kappa^2/4}{\kappa \delta \eta}$             & $0$ \\
	\bottomrule
\end{tabular}
\end{table}

As discussed in the Introduction, all three coupling schemes shown in Fig.~\ref{fig:system} generate the desired OAT Hamiltonian~\eqref{eqn:HOAT}, but they differ in the specific form of additional dissipative terms. 
To derive an effective QME describing the dynamics in the subspace spanned by the states $\{\ket{\uparrow_j},\ket{\downarrow_j}\}_{j=1,\dots,N}$, we eliminate the auxiliary levels $\{\ket{e_j}\}_{j=1,\dots,N}$ and the cavity mode $\hat{a}$ for each coupling scheme.
As derived in Appendix~\ref{sec:App:DerivationEffectiveQME}, they all lead to an effective QME of the form
\begin{align}
	\frac{\d}{\d t} \hat{\rho} &= - i \komm{\tildeoms \hat{S}_z + \chi \hat{S}_z^2}{\hat{\rho}} + \Gammaphi \mathcal{D}[\hat{S}_z] \hat{\rho} + \Gammarel \mathcal{D}[\hat{S}_-] \hat{\rho} \nonumber \\
		&\phantom{=}\ + \sum_{j=1}^N \left( \gammalocz \mathcal{D}[\hat{\sigma}_z^{(j)}] + \gammalocm \mathcal{D}[\hat{\sigma}_-^{(j)}] + \gammalocp \mathcal{D}[\hat{\sigma}_+^{(j)}] \right) \hat{\rho} \comma
	\label{eqn:QME:EffectiveGeneral}
\end{align}
where the different rates are defined in Table~\ref{tbl:Rates}.
Since we are only interested in the effect of the OAT term $\chi \hat{S}_z^2$, we will work in a rotating frame at the frequency $\tildeoms$ such that the term $\tildeoms \hat{S}_z$ can be ignored.

As shown by Table~\ref{tbl:Rates}, the three implementations of the OAT Hamiltonian differ significantly in the form of their dissipative processes.
The TC scheme has a \emph{collective relaxation} process, whose strength can be tuned relative to the OAT strength $\chitc$ by changing the ratio $\kappa/\Deltatc$, but the single-spin relaxation and dephasing rates are \emph{independent} of $\kappa/\Deltatc$. 
Both the SCF and the ACF scheme have a \emph{collective dephasing} process, whose strength can be tuned relative to the OAT strength by changing the ratio $\kappa/\delta$, but the single-spin dissipation rates now \emph{depend} on the ratio $\kappa/\delta$, too.
Moreover, in the absence of the OAT Hamiltonian, the dissipative processes in the three schemes try to stabilize very different steady states: 
The dissipation in the TC scheme aims to relax each spin into its ground state, $\cerw{\hat{\sigma}_x^{(j)}} = \cerw{\hat{\sigma}_y^{(j)}} = 0$ and $\cerw{\hat{\sigma}_z^{(j)}} = -1$. 
The ACF scheme dephases the system in the $z$ basis, such that the steady state has $\cerw{\hat{\sigma}_x^{(j)}} = \cerw{\hat{\sigma}_y^{(j)}} = 0$ and $\cerw{\hat{\sigma}_z^{(j)}}$ is a constant of motion. 
The SCF scheme, finally, features a combination of single-spin dephasing as well as single-spin excitation and relaxation processes at equal rates, which evolve the system into a completely mixed state, $\cerw{\hat{\sigma}_x^{(j)}} = \cerw{\hat{\sigma}_y^{(j)}} = \cerw{\hat{\sigma}_z^{(j)}} = 0$.

These very different dissipative processes compete with the unitary OAT dynamics, which raises the question whether the schemes will have different requirements to achieve signal amplification and large metrological gain. 
An important figure of merit to compare the strength of the desired unitary OAT dynamics to the undesired dissipative processes is the single-spin cooperativity, which is defined as
\begin{align}
	\eta = \frac{4 g^2}{\kappa \Gamma} 
	\label{eqn:Definitioneta}
\end{align}
for the SCF and ACF schemes, and has the form
\begin{align}
	\eta_{\phi,\mathrm{rel}} = \frac{4 g^2}{\kappa \gamma_{\phi,\mathrm{rel}}}
\end{align}
for the TC scheme. 
As far as spin squeezing is concerned, it has already been shown that (despite the different dissipative processes) spin squeezing can be achieved in all three schemes if the collective  cooperativities $N \eta$ and $N \eta_{\phi,\mathrm{rel}}$ are large (see Refs.~\onlinecite{SchleierSmith2010,LewisSwan2018,Li2022} as well as Appendix~\ref{sec:App:SpinSqueezing}).

\section{OAT twist-untwist sensing protocol}
\label{sec:OATProtocol}
\subsection{Coherent dynamics}
\label{sec:Analysis:IdealProtocol}

We start by reviewing the ideal OAT twist-untwist protocol in the absence of any dissipation \cite{Davis2016,Hosten2016}, with a specific focus on the  underlying amplification process. 
The goal of the sensing protocol is to measure a small signal that changes the level-splitting frequency of all $N$ ensemble spins \cite{Degen2017,Pezze2018}. 
Such a signal causes a small excess precession angle $\phi$ in a Ramsey-type interferometric protocol. 
In the absence of any other noise sources, the estimation error $\bfDelta \phi$, with which the angle $\phi$ can be inferred from the final measurement, depends only on the projection noise of the sensing state in the direction of the signal acquisition (which we take to be the $\hat{S}_y$ direction without loss of generality):
\begin{align}
    (\bfDelta \phi)^2_\text{proj} = \frac{(\bfDelta S_y)^2}{\cabs{\partial_\phi \cerw{\hat{S}_y}}^2} \comma
\end{align}
where $(\bfDelta S_y)^2 = \cerw{\hat{S}_y^2} - \cerw{\hat{S}_y}^2$. 
Spin squeezing reduces this estimation error by reshaping the projection noise, and is quantified by the Wineland parameter \cite{Wineland1992,Pezze2018}
\begin{align}
	\Wineland = N \frac{(\mathbf{\Delta} S_\perp)^2}{\cabs{\cerw{\vec{\hat{S}}}}^2} \comma
	\label{eqn:Definition:Wineland}
\end{align}
where the variance $(\mathbf{\Delta}S_\perp)^2$ is minimized over all directions perpendicular to the polarization of the collective spin vector $\vec{\hat{S}} \equiv (\hat{S}_x, \hat{S}_y, \hat{S}_z)^\top$.

A key figure of merit is the metrological gain $\Gmet$, which quantifies how much the estimation error has been reduced compared to minimum error achievable using a product state:
\begin{align}
	\Gmet = \frac{(\bfDelta \phi)^2_\mathrm{SQL}}{(\bfDelta \phi)^2} = \frac{1}{N (\bfDelta \phi)^2} \fullstop
	\label{eqn:Definition:MetrologicalGain}
\end{align}
For a spin-squeezed state, and in the absence additional readout noise, one finds $\Gmet = 1/\Wineland$.  
Hence, the Wineland parameter quantifies the sensitivity improvement of a spin-squeezed state (which has $1/N \leq \Wineland < 1$) over a coherent-spin state (which has $\Wineland = 1$) in an ideal Ramsey-type measurement.

In practice, the final readout process of the  Ramsey interferometer will be imperfect, i.e., it will add (technical) detection noise.
This detection noise can be quantified in terms of an equivalent amount of projection noise and is typically given in multiples $\Xidetsq$ of the projection noise $N/4$ of a coherent-spin state \cite{Barry2020}.
Assuming that spin squeezing has not significantly reduced the polarization of the sensing state, $\cabs{\cerw{\vec{\hat{S}}}} \approx N/2$, the total estimation error in the presence of projection and detection noise is
\begin{align}
	(\bfDelta \phi)^2 
    = \frac{(\bfDelta S_y)^2 + \Xidetsq N/4}{\cabs{\partial_\phi \cerw{\hat{S}_y}}^2}
    \approx \frac{\Wineland + \Xidetsq}{N} \fullstop 
	\label{eqn:DeltaPhiPrinciple}
\end{align}
Thus,  in the presence of nonzero detection noise, spin squeezing reduces $\bfDelta \phi$ only if $\Xidetsq < \Wineland$, i.e., even a small level of detection noise will ultimately limit the estimation error if spin squeezing becomes strong enough.

\begin{figure}
    \centering
    \includegraphics[width=0.48\textwidth]{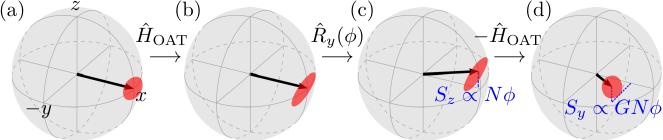}
    \caption{
        Sketch of spin amplification using unitary one-axis-twist (OAT) dynamics. 
        (a) The system is initialized in a coherent-spin state pointing along the $\hat{S}_x$ direction. 
        (b) The initial state is evolved under the OAT Hamiltonian~\eqref{eqn:HOAT} for a time $\tsqz$ to transform it into a spin-squeezed state. 
        (c) The signal of interest is encoded in the $\hat{S}_z$ component of the state using a rotation about the $\hat{S}_y$ axis.
        (d) Finally, the OAT Hamiltonian with opposite sign is applied for a time $\tunsqz$, which returns the fluctuations of the state to the level of a coherent-spin state and generates a nonzero $\hat{S}_y$ polarization that is an amplified version of the $\hat{S}_z$ polarization encoding the signal.
    }
    \label{fig:sketch_OAT_protocol}
\end{figure}

To benefit from spin squeezing even in cases where detection noise is appreciable, Davis \emph{et al.} \cite{Davis2016} proposed a twist-untwist sensing protocol that amplifies the signal encoded in the spin ensemble prior to readout.
It allows one to achieve a sensitivity below the standard quantum limit even if $\Xidetsq \approx 1$.
The basic principle of this amplification mechanism is sketched in Fig.~\ref{fig:sketch_OAT_protocol}:
The spin system is initialized in a coherent-spin state where all spins are polarized in the $x$ direction, $\hat{S}_x \ket{\psi_0} = N/2 \ket{\psi_0}$. 
In the first ``twist'' step, the OAT Hamiltonian $\hat{H}_\mathrm{OAT}$ defined in Eq.~\eqref{eqn:HOAT} is applied for a time $\tsqz$ to transform $\ket{\psi_0}$ into a spin-squeezed state $\ket{\psi_1} = e^{-i \hat{H}_\mathrm{OAT} \tsqz} \ket{\psi_0}$.
This state is used as the initial state of a sensing protocol that effectively encodes a small signal $\phi \ll 1$ of interest by rotating the state about the $y$ axis, 
\begin{align}
	\ket{\psi_2(\phi)} = e^{i \phi \hat{S}_y} \ket{\psi_1} \fullstop
	\label{eqn:rotatedpsi2}
\end{align}
The resulting nonzero $\hat{S}_z$ polarization $\bra{\psi_2(\phi)} \hat{S}_z \ket{\psi_2(\phi)} \propto \phi$ is amplified in the last ``untwist'' step, where $\ket{\psi_2(\phi)}$ is evolved using $-\hat{H}_\mathrm{OAT}$ for a time $\tunsqz$, $\ket{\psi_3(\phi)} = e^{+ i \hat{H}_\mathrm{OAT} \tunsqz} \ket{\psi_2(\phi)}$. 
Choosing $\tunsqz = \tsqz$ ensures that $\ket{\psi_3(\phi)}$ is close to a coherent-spin state but, due to the nonlinearity of the OAT Hamiltonian, the final $\hat{S}_y$ polarization is a gain factor $G \gg 1$ larger than if one simply rotated the initial state $\ket{\psi_0}$ by an angle $\phi$ about the $z$ axis, $\bra{\psi_3(\phi)} \hat{S}_y \ket{\psi_3(\phi)} = G N \phi/2$.

The amplification dynamics generated by the second application of $\hat{H}_\mathrm{OAT}$ can be understood in terms of Heisenberg equations of motion of the system \cite{Koppenhoefer2022}. 
The OAT Hamiltonian~\eqref{eqn:HOAT} generates a dynamics of the form
\begin{align}
	\frac{\d}{\d t} \hat{S}_y = 2 \chi \hat{S}_z \hat{S}_x \fullstop
	\label{eqn:HeisenbergOATSy}
\end{align}
Since $\hat{S}_z$ commutes with $\hat{H}_\mathrm{OAT}$, $\cerw{\hat{S}_z}$ is a constant of motion and the dynamics has a quantum-nondemolition structure: 
As long as $\cerw{\hat{S}_x} \approx N/2$ is positive, the $\cerw{\hat{S}_y}$ polarization will grow linearly in time at rate that is $\approx N \chi \cerw{\hat{S}_z}$. 
The curvature of the Bloch sphere leads to deviations of this linear growth because $\cerw{\hat{S}_x}$ decreases with increasing $\cerw{\hat{S}_y}$, such that $\cerw{\hat{S}_y}$ takes a maximum value after an evolution time $\tamp^\mathrm{ideal} \approx 1/\chi \sqrt{N}$ for $N \gg 1$. 
At this time, $\cerw{\hat{S}_y}$ is an amplified version of the original signal stored in $\bra{\psi_2(\phi)}\hat{S}_z\ket{\psi_2(\phi)} \propto \phi$ with a gain factor
\begin{align}
	G^\mathrm{ideal}_\mathrm{max} \equiv \lim_{\phi \to 0} \frac{\cerw{\hat{S}_y(\tamp^\mathrm{ideal})}}{N \phi/2} = \sqrt{\frac{N}{e}} \fullstop
	\label{eqn:GOAT}
\end{align}
The gain $G^\mathrm{ideal}_\mathrm{max}$ quantifies the amplification of the signal encoded in the spin ensemble and should not be confused with the metrological gain $\Gmet$ defined in Eq.~\eqref{eqn:Definition:MetrologicalGain}.

Since the detection noise is unchanged by the amplification process, the second OAT step improves the signal-to-noise ratio of the readout process such that the final estimation error is given by
\begin{align}
	(\bfDelta \phi)^2_\mathrm{amp} 
	= \frac{1 + \Xidetsq}{(G_\mathrm{max}^\mathrm{ideal})^2 N} \fullstop
	\label{eqn:AmplificationOATIdeal}
\end{align}
Comparing this result with Eq.~\eqref{eqn:DeltaPhiPrinciple}, one can interpret the impact of amplification in Eq.~\eqref{eqn:AmplificationOATIdeal} as a suppression of the detection noise, $\Xidetsq \to \Xidetsq / (G_\mathrm{max}^\mathrm{ideal})^2 = e \Xidetsq/N$, combined with an effective squeezing operation, $\Wineland \to 1/(G_\mathrm{max}^\mathrm{ideal})^2 = e/N$. 
The corresponding metrological gain $\Gmetideal = 1/N(\bfDelta \phi)^2_\mathrm{amp} = N / [ e (1 + \Xidetsq)]$ is shown by the solid black line in Fig.~\ref{fig:comparison}.

Note that, instead of using a twist and an untwist step with different signs of the OAT interaction strength, one can also rotate the squeezed state after the first twist step, encode the signal by a rotation about the $\hat{S}_y$ axis, and then use another twist step with the \emph{same} sign of the OAT interaction strength to amplify the signal and convert the fluctuations back to those of a coherent spin state.  
This strategy was implemented in Ref.~\cite{Hosten2016}.

\subsection{Impact of dissipation on the sensing protocol}
\label{sec:Analysis:Dissipation}

The dissipative terms in the effective QME~\eqref{eqn:QME:EffectiveGeneral} lead to important modifications of the ideal amplification protocol outlined in Sec.~\ref{sec:Analysis:IdealProtocol}: 
both collective and local dissipation cause dephasing of the sensor that reduces $\cerw{\hat{S}_x}$ and $\cerw{\hat{S}_y}$ during the amplification step and thus competes with the unitary OAT dynamics that attempts to increase $\cerw{\hat{S}_y}$. 
Moreover, local relaxation and excitation processes break the conservation of $\cerw{\hat{S}_z}$ and destroy the quantum-nondemolition structure of the amplification process shown in Eq.~\eqref{eqn:HeisenbergOATSy}, where a static signal encoded in the $\cerw{\hat{S}_z}$ polarization is transduced into the $\cerw{\hat{S}_y}$ polarization.
Both of these effects reduce the gain to 
\begin{align}
    G = \max_{t_\mathrm{amp}} \lim_{\phi \to 0} \frac{\cerw{\hat{S}_y(t_\mathrm{amp})}}{N \phi/2} \leq G_\mathrm{max}^\mathrm{ideal} \fullstop
    \label{eqn:GainWithDissipation}
\end{align}
Finally, the effects of decoherence during the first ``twist'' step will be amplified during the subsequent ``untwist'' step, such that the projection noise of the final state may be larger than that of a coherent-spin state: 
\begin{align}
    (\bfDelta S_y)^2(t_\mathrm{amp}) = (1 + \sigma_\mathrm{diss}^2) \frac{N}{4} \comma
    \label{eqn:DefinitionSigmadisssq}
\end{align}
where $\sigma_\mathrm{diss}^2 \geq 0$ captures the enhanced fluctuations of the final state, expressed in multiples of the fluctuations $N/4$ of a coherent-spin state. 
Note that Eq.~\eqref{eqn:DefinitionSigmadisssq} is an implicit definition of $\sigma_\mathrm{diss}^2$, i.e., no additional assumptions on the form of $\sigma_\mathrm{diss}^2$ are necessary.
One can extract $G$ and $\sigma_\mathrm{diss}^2$ from a (numerical) solution of Eq.~\eqref{eqn:QME:EffectiveGeneral} for small but nonzero $\phi$, which provides one with $\cerw{\hat{S}_y(\tamp)}$ and $(\bfDelta S_y)^2(t_\mathrm{amp})$.

Combining all of these effects, one obtains the following form of the estimation error in the presence of dissipation:
\begin{align}
	(\mathbf{\Delta}\phi)^2 = \frac{1 + \sigma^2_\mathrm{diss} + \Xidetsq}{G^2 N} \fullstop
	\label{eqn:CompleteModelDeltaphisq}
\end{align}
This result shows that, in the limit of large readout noise, $\Xidetsq \gg 1 + \sigma_\mathrm{diss}^2$, the metrological gain depends only on the ability to create large gain $G$ in the amplification process, $\Gmet \approx G^2 / \Xidetsq$. 
In the opposite regime of small readout noise, $\Xidetsq \lesssim 1 + \sigma_\mathrm{diss}^2$, the maximum metrological gain depends both on the enhanced fluctuations $\sigma_\mathrm{diss}^2$ and on the gain $G$, $\Gmet \approx G^2 / (1 + \sigma_\mathrm{diss}^2)$.

Davis \emph{et al.} \cite{Davis2016} analyzed the performance of the SCF scheme in the latter regime using a phenomenologically motivated expression for $\sigma_\mathrm{diss}^2$, 
\begin{align}
	\frac{\sigma_\mathrm{diss}^2}{G^2} = \frac{2 N \chi t \kappa }{\delta} \frac{1}{(G_\mathrm{max}^\mathrm{ideal})^2} + \frac{2 (\mathbf{\Delta} S_z^\mathrm{sc})^2}{N} \fullstop
	\label{eqn:EmilysModel}
\end{align}
Here, the first term describes growth of the projection noise of the squeezed state due to collective dephasing and the last term accounts for fluctuations of the $\hat{S}_z$ component due to spin-flip processes mediated by the excited state, where $(\mathbf{\Delta}S_z^\mathrm{sc})^2 = N \chi t (2 \delta/\kappa + \kappa / 2 \delta) /6 \eta$. 
\newtext{%
A similar analysis has been performed by Chu \emph{et al.} \cite{Chu2021}, which is based on analytical solutions for the special cases where only collective dephasing or only local dissipation is present and then combines these results to find an approximate solution of the full problem. 
A key assumption of these works 
}%
is that the gain as a function of time is not reduced by dissipation, $G \approx G_\mathrm{max}^\mathrm{ideal}$, such that the contributions to the estimation error stemming from projection noise and detection noise are negligible compared to $\sigma_\mathrm{diss}^2/G^2$.
The detuning $\delta$ can then be chosen to minimize Eq.~\eqref{eqn:EmilysModel}, leading to a scaling of the metrological gain with the collective cooperativity, 
\begin{align}
    \Gmetest = \sqrt{\frac{3 N \eta}{32}} \fullstop
    \label{eqn:GmetEmily}
\end{align}

This estimate suggests that achieving a large metrological gain simply requires  having a sufficiently large collective cooperativity $N \eta$. 
In Fig.~\ref{fig:comparison}, we compare this result (dashed grey line) with a numerical optimization of $\Gmet$ based on MFT (thick green curve), see Sec.~\ref{sec:MFT} for details. 
We see that the result of Davis \emph{et al.} underestimates the minimum collective cooperativity
$N \eta$ required to achieve $\Gmet > 1$ by almost an order of magnitude (highlighted by the red shaded area in Fig.~\ref{fig:comparison}), even though it does capture the correct scaling of $\Gmet$ with $ N \eta$ (up to a prefactor of $\approx 1.5$) at extremely large cooperativities.  
The discrepancy at modest $N \eta $ suggests that dissipation impacts $\Gmet$ more severly than what is captured in the perturbative analysis of Ref.~\onlinecite{Davis2016}. 
Indeed, as we show in Sec.~\ref{sec:Gain} below, dissipation can actually \emph{reduce} the gain $G$ even if $N \eta \gg 1$, such that the first term in Eq.~\eqref{eqn:EmilysModel} is underestimated, which in turn affects the optimization of the detuning $\delta$ and the optimal twisting time. 
These all lead to the discrepancy between the full results and the simple estimate seen in Fig.~\ref{fig:comparison}.

\section{Gain in the presence of dissipation}
\label{sec:Gain}

We now start our more rigorous analysis of how dissipation impacts the different implementations of the OAT twist-untwist amplification protocol.  
In a first step, we analyze how the dissipative terms in the QME~\eqref{eqn:QME:EffectiveGeneral} reduce the achievable gain $G$. 
As discussed in Sec.~\ref{sec:OATProtocol}, this gain $G$ is the relevant quantity to be maximized in the limit of large readout noise, $\Xidetsq \gg 1$. 
In a second step, the impact of dissipation on the metrological gain $\Gmet$ will be analyzed in Sec.~\ref{sec:MetrologicalGain}.

Since spin-squeezing is not improving the estimation error $(\bfDelta \phi)$ in the regime $\Xidetsq \gg 1$, we ignore the first squeezing step for now, i.e., we set $\tsqz = 0$ and use a coherent spin state rotated about the $\hat{S}_y$ axis as our initial state, 
\begin{align}
	\ket{\psi(\phi)} = e^{i \phi \hat{S}_y} \ket{\psi_0} \fullstop
	\label{eqn:InitialStatePhi}
\end{align}
The restriction $\tsqz = 0$ does not matter if there is only unitary dynamics generated by $\hat{H}_\mathrm{OAT}$. 
If there is additional dissipative dynamics [as described by Eq.~\eqref{eqn:QME:EffectiveGeneral}], setting $\tsqz$ to zero makes a difference since dissipation acting during the first twist step can depolarize the state $\ket{\psi_2(\phi)}$ and increase its fluctuations, which in turn affects the subsequent amplification dynamics during the untwist step.
However, we found that the results for $G$ do not change significantly if the initial squeezing step is omitted. 
More details can be found in Appendix~\ref{sec:App:AdditionalDetails:SCF}, where we analyze the gain $G$ of the full twist-untwist protocol.

In analogy with the definitions in Eqs.~\eqref{eqn:GOAT} and~\eqref{eqn:GainWithDissipation}, we define the maximum gain $G(\eta)$ in the presence of dissipation with a given single-spin cooperativity $\eta$ as
\begin{align}
	G(\eta) = \max_{\tamp} \max_{\lambda} \lim_{\phi \to 0} \frac{\cerw{\hat{S}_y(\tamp)}}{N \phi/2} \fullstop
	\label{eqn:DefinitionGain}
\end{align}
Here, $\tamp$ is the time over which $\ket{\psi(\phi)}$ has been evoved under $\hat{H}_\mathrm{OAT}$ and $\lambda$ is a detuning parameter that can be optimized experimentally: 
$\lambda = \Delta_\mathrm{TC}/\kappa$ for the TC scheme and $\lambda = \delta/\kappa$ for the SCF and ACF schemes.  
It effectively controls the relative strength of the collective and single-spin dissipation (cf.\ Table~\ref{tbl:Rates}).

\subsection{TC implementation of OAT dynamics}
\label{sec:Gain:TC}

The impact of dissipative processes on the gain in a TC implementation of OAT dynamics has been analyzed in detail in Ref.~\onlinecite{Koppenhoefer2022}: 
the collective and single-spin relaxation processes cause $\cerw{\hat{S}_z}$ to decay towards the joint ground state, which leads to a nonzero $\cerw{\hat{S}_y}$ polarization even if no signal has been applied, $\phi = 0$. 
Therefore, the definition~\eqref{eqn:DefinitionGain} of the gain needs to be modified by subtracting this background, 
\begin{align}
	G^\mathrm{sub}(\eta) = \max_{\tamp} \max_\lambda \lim_{\phi \to 0} \frac{\delta \cerw{\hat{S}_y(\tamp)}}{N \phi/2} \comma
	\label{eqn:Gsub}
\end{align}
where $\delta \cerw{\hat{S}_y(\tamp)} = \cerw{\hat{S}_y(\tamp,\phi)} - \cerw{\hat{S}_y(\tamp,0)}$.
The need for a background subtraction makes the TC implementation of OAT dynamics practically useless in many sensing applications. 
Even with this background subtraction, it was found that a significant fraction of the ideal gain, e.g., $G^\mathrm{sub}(\eta)/[\lim_{\eta \to \infty} G^\mathrm{sub}(\eta)] = 1/2$, can only be achieved for large single-spin cooperativities $\eta_\phi \gg \sqrt{N}$ and $\eta_\mathrm{rel} \gg N^{0.9}$ \cite{Koppenhoefer2022}.
This condition is much more stringent than a condition in terms of the collective cooperativity $N \eta_{\phi,\mathrm{rel}}$, since it becomes harder to satisfy with increasing size of the spin ensemble.

\subsection{SCF implementation of OAT dynamics}
\label{sec:Gain:SCF}

\begin{figure*}
	\centering
	\includegraphics[width=0.31\textwidth]{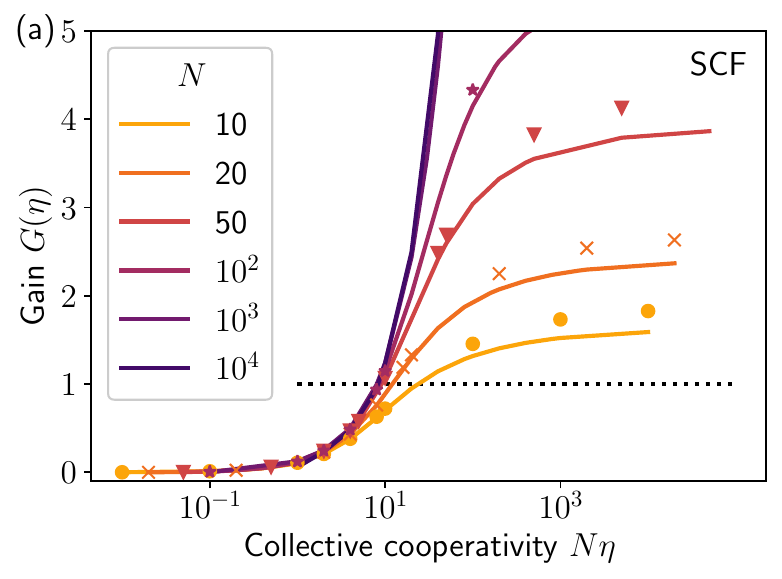}
	\includegraphics[width=0.32\textwidth]{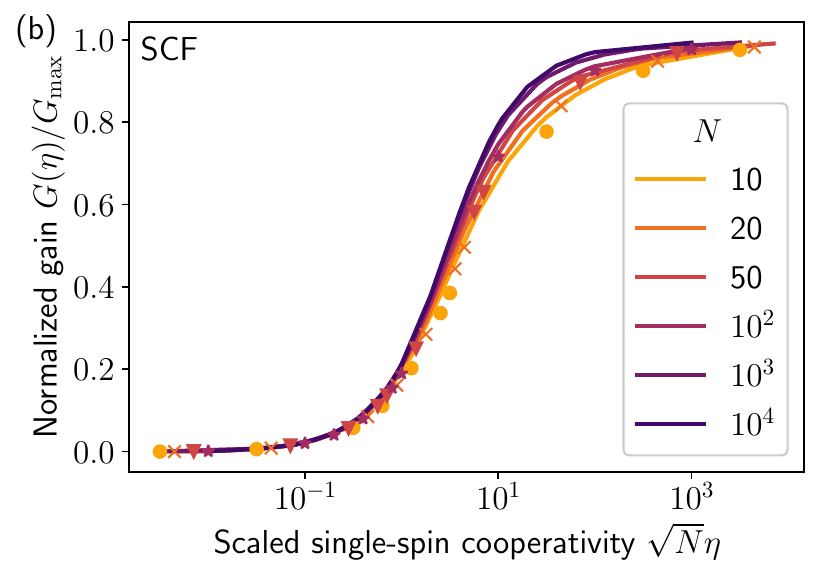}
	\includegraphics[width=0.32\textwidth]{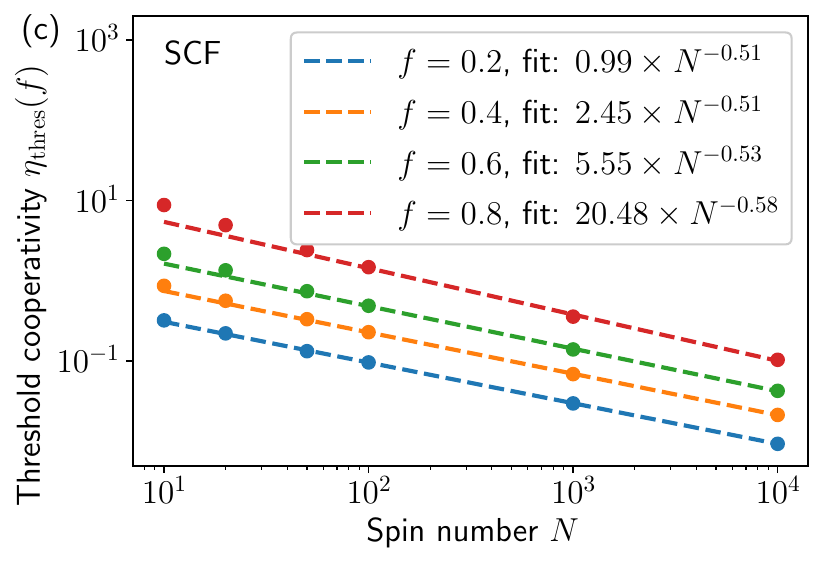}
	\caption{
		(a) Gain $G(\eta)$ of the SCF scheme in the presence of undesired collective and single-spin dissipation, defined in Eq.~\eqref{eqn:DefinitionGain}, as a function of the collective cooperativity $N \eta$. 
		The solid lines have been obtained using a numerical optimization of the gain using a second-order mean-field theory (MFT), see Sec.~\ref{sec:MFT}. 
		The data points of the corresponding color have been obtained from a numerically exact solution of the quantum master equation~\eqref{eqn:QME:EffectiveGeneral} for small spin numbers $N$, see Sec.~\ref{sec:Analysis:Dissipation} and Appendix~\ref{sec:App:AdditionalDetails:SCF}. 
		The collective cooperativity controls only the onset of gain at $N \eta \lesssim 1$ and $G(\eta) \ll 1$. A gain of unity, $G=1$ is indicated by the dotted black line.
		(b) Same data for $G(\eta)$, normalized to the ideal value $G_\mathrm{max} = \lim_{\eta \to \infty} G(\eta)$ in the absence of dissipation, collapses  when plotted as a function of the scaled single-spin cooperativity $\sqrt{N} \eta$. 
		(c) Fits of the threshold values $\eta_\mathrm{thres}(f)$, defined by $G(\eta_\mathrm{thres}) / G_\mathrm{max} = f$, to extract the exponent of $N$ leading to optimal collapse of the curves in (b).
		The fits have been performed over the range $100 \leq N \leq 10^4$ and are based on the MFT results. 
	}
	\label{fig:Amplification}
\end{figure*}

Comparing the dissipation rates of the TC scheme with those of the SCF scheme (see Table~\ref{tbl:Rates}), we find that the collective decay of the TC scheme has been replaced by collective dephasing, and that single-spin relaxation is counteracted by an equally strong single-spin excitation process. 
Therefore, if no signal is applied, $\cerw{\hat{S}_z} = 0$ remains zero and there is no growth of $\cerw{\hat{S}_y}$. 
Consequently, no background subtraction is needed in the SCF scheme and we can define the gain according to Eq~\eqref{eqn:DefinitionGain}.
However, the dynamics is still different from the purely unitary case given by Eq.~\eqref{eqn:HeisenbergOATSy} since $\cerw{\hat{S}_z}$ is no longer a constant of motion of $\mathcal{L}_\mathrm{spin}^\mathrm{SCF}$ and dephasing will reduce the $\cerw{\hat{S}_x}$ and $\cerw{\hat{S}_y}$ polarizations.
Therefore, even though the amplification process for SCF-based OAT dynamics appears to be more robust than the one of the TC scheme, one still expects it to be compromised in the presence of collective and single-spin dissipation. 
To quantify this effect, we numerically calculate the gain $G(\eta)$ by integrating the QME~\eqref{eqn:QME:EffectiveGeneral} for fixed $\chimw$ and different values of $\eta$ as well as $\delta/\kappa$, starting from the initial state $\ket{\psi(\phi)}$ defined in Eq.~\eqref{eqn:InitialStatePhi}.
We maximize the gain $G(\eta)$ over $\tamp$ and $\delta/\kappa$ (see Appendix~\ref{sec:App:AdditionalDetails:SCF} for details).
The results are shown by the markers in Fig.~\ref{fig:Amplification}.

As shown in Fig.~\ref{fig:Amplification}(a), the \emph{onset} of gain dynamics is controlled by the collective cooperativity:
When plotted as a function of $N \eta$, all curves of the bare gain $G(\eta)$ obtained for different system sizes collapse for $G(\eta) \ll 1$ and $N \eta \lesssim 10$. 
In this regime, however, collective dephasing and single-spin dissipation dominate and reduce the $\hat{S}_y$ component below the value of the initial signal encoded in $\cerw{\hat{S}_z}$, such that the ``amplification'' protocol actually attenuates the original signal and is of very limited use.

For larger collective cooperativities, the curves quickly fan out and ultimately converge to the respective ideal values $G_\mathrm{max} = \lim_{\eta \to \infty} G(\eta) \propto \sqrt{N}$ in the absence of dissipation. 
The relevant figure of merit to quantify the amplification performance is the fraction $G(\eta)/G_\mathrm{max}$ of gain that can be achieved for a given single-spin cooperativity $\eta$ compared to the maximally possible gain $G_\mathrm{max}$. 
In Fig.~\ref{fig:Amplification}(b), we show that the rescaled gain curves collapse quite well for a broad range of ensemble sizes $N$ when plotted as a function of the scaled cooperativity $\sqrt{N} \eta$ (which is smaller than the collective cooperativity by a large factor $\sqrt{N}$).   
This suggests that the condition to observe a significant fraction of the ideal gain $G_\mathrm{max}$ (e.g., $G(\eta)/G_\mathrm{max} \approx 1/2$) is 
\begin{align}
	\sqrt{N} \eta \gtrsim 1 \fullstop
	\label{eqn:ConditionCavityFeedbackAmplification}
\end{align}
Thus, even if the collective cooperativity is large, $N \eta \gg 1$, the actual gain $G(\eta)$ can still be significantly smaller than $G_\mathrm{max}$.

The exponent $\alpha = 1/2$ of the prefactor $N^\alpha$ in Eq.~\eqref{eqn:ConditionCavityFeedbackAmplification} can be rigorously extracted from the numerical data by plotting $G(\eta)/G_\mathrm{max}$ as a function of $\eta$, identifying the threshold values $\eta_\mathrm{thres}(f)$ where $G(\eta)/G_\mathrm{max} = f$ equals a certain fraction $0 < f < 1$, and fitting the corresponding values of $\eta_\mathrm{thres}(f)$ as a function of $N$.
As shown in Fig.~\ref{fig:Amplification}(c), the exponent for all considered ratios $f < 0.8$ is consistent with $\alpha = 1/2$.

\subsection{ACF implementation of OAT dynamics}

One may wonder if the ACF scheme performs better than the SCF scheme analyzed so far because it does not suffer from single-spin excitation and relaxation processes (see Table~\ref{tbl:Rates}). 
To check if the absence of these processes changes the amplification dynamics, we repeat the same analysis for the ACF scheme. 
We find that omitting single-spin excitation and relaxation does not change the physics significantly, and that the threshold condition to achieve $G(\eta)/G_\mathrm{max} \approx 1/2$ is still given by Eq.~\eqref{eqn:ConditionCavityFeedbackAmplification}. 
More details can be found in Appendix~\ref{sec:App:AdditonalDetails:ACF}.

\begin{table}
\caption{Overview of the conditions on the cooperativity to reach $\mathcal{O}(1/2)$ of the maximum possible gain $G_\mathrm{max} = \lim_{\eta \to \infty} G(\eta)$ in different amplification schemes.}
\label{tbl:Conditions}
\begin{tabular}{l|r}
	\toprule
	\multicolumn{1}{c|}{\textbf{Amplification scheme}} & \multicolumn{1}{c}{\textbf{Condition for large gain}} \\
	\midrule
	TC-based OAT \cite{Koppenhoefer2022} & $\etaphi \gg \sqrt{N}$ and $\etarel \gg N^{0.9}$ \\
	SCF-based OAT & $\eta \gtrsim 1/\sqrt{N}$ \\
	ACF-based OAT & $\eta \gtrsim 1/\sqrt{N}$ \\
	Superradiant decay \cite{Koppenhoefer2022,Koppenhoefer2023} & $\etaphi \gtrsim 1/N$ and $\etarel \gtrsim 1/N$ \\
	\bottomrule
\end{tabular}
\end{table}

\subsection{Comparison of the schemes}

A summary of the conditions that are required to reach $\mathcal{O}(1/2)$ of the maximum gain in the different amplification schemes is given in Table~\ref{tbl:Conditions}. 
Compared to a TC-based implementation of OAT, both the SCF and the ACF schemes feature an experimentally much more favorable condition on the required cooperativity, since it becomes \emph{easier} to satisfy  Eq.~\eqref{eqn:ConditionCavityFeedbackAmplification} with increasing ensemble size $N$. 
However, Eq.~\eqref{eqn:ConditionCavityFeedbackAmplification} is still more challenging than a condition in terms of the collective cooperativity, $N \eta_{\phi,\mathrm{rel}} \gtrsim 1$, which governs dissipative amplification protocols based on superradiant decay \cite{Koppenhoefer2022,Koppenhoefer2023}. 
In atomic platforms, the difference between a condition in terms of $\sqrt{N} \eta$ and one in terms of the collective cooperativity $N \eta$ may in practice not be very relevant, since large single-spin cooperativities $\eta \approx 10$ \cite{Colombo2022} can readily be achieved and ensemble sizes can be quite large, $N \gtrsim 10^5$ \cite{Hosten2016}.
In solid-state platforms, however, the single-spin cooperativities are typically much lower \cite{Eisenach2021}, such that reaching a collective cooperativity exceeding unity can already be a challenge.

Importantly, undesired dissipative processes can significantly reduce the gain $G(\eta)$ in all of the three implementations of OAT dynamics. 
This raises the question if non-perturbative effects will modify the analysis of the metrological gain performed by Davis \emph{et al.} \cite{Davis2016}, since it is based on the assumption of a constant gain $G(\eta) \approx G_\mathrm{max}$. 
We will address this question in Sec.~\ref{sec:MetrologicalGain} below.

\section{Analysis and interpretation using mean-field theory}
\label{sec:MFT}

To get a better analytical understanding for the unusual scaling relation of the gain, Eq.~\eqref{eqn:ConditionCavityFeedbackAmplification}, which is in between a collective-cooperativity and a single-spin-cooperativity criterion, and to analyze larger ensemble sizes $N$ that are out of reach for a numerical integration of the QME~\eqref{eqn:QME:EffectiveGeneral}, we now analyze the amplification dynamics  using second-order MFT \cite{Kubo1962,Zens2019,Groszkowski2022}.
This approach allows us to approximate the quantum dynamics by a closed set of differential equations of motion for the first moments $S_\alpha = \cerw{\hat{S}_\alpha}$  and the (co)variances $C_{\alpha\beta} = \cerw{\cakomm{\hat{S}_\alpha}{\hat{S}_\beta}}/2 - \cerw{\hat{S}_\alpha}\cerw{\hat{S}_\beta}$ with $\alpha,\beta \in \{x,y,z\}$.
The complete set of MFT equations of motion corresponding to the QME~\eqref{eqn:QME:EffectiveGeneral} is given in Appendix~\ref{sec:App:MFT}. 
Here, we only list the results for the first moments for $\Gammarel = 0$, 
\begin{subequations}%
\begin{align}%
	\frac{\d}{\d t} S_x 
		&=  - 2 \chi (C_{yz} + S_y S_z) 
		- \frac{\Gammaphi + \gammalocm + \gammalocp + 4 \gammalocz}{2} S_x \comma 
		\label{eqn:MFT:Sx} \displaybreak[1] \\
	\frac{\d}{\d t} S_y 
		&= + 2 \chi (C_{xz} + S_x S_z) 
		- \frac{\Gammaphi + \gammalocm + \gammalocp + 4 \gammalocz}{2} S_y \comma 
		\label{eqn:MFT:Sy} \displaybreak[1] \\
	\frac{\d}{\d t} S_z 
		&= - \gammalocm \left( S_z + \frac{N}{2} \right) - \gammalocp \left( S_z - \frac{N}{2} \right) \fullstop
\end{align}%
\label{eqn:MFT:Moments}%
\end{subequations}%
Note that, in the absence of dissipation, $\Gammaphi = \gammalocp = \gammalocm = \gammalocz = 0$, and ignoring the covariance $C_{xz}$, Eq.~\eqref{eqn:MFT:Sy} reproduces the Heisenberg equation of motion given in Eq.~\eqref{eqn:HeisenbergOATSy}.

The MFT results for the gain, shown by the solid curves in Fig.~\ref{fig:Amplification}, are obtained as follows. 
The system is initialized in a coherent-spin state pointing along the $x$ axis, $S_x = N/2$, $C_{yy} = C_{zz} = N/4$, and all other moments and (co)variances are set to zero. 
Starting from this initial state, the MFT equations of motion listed in Appendix~\ref{sec:App:MFT} are numerically integrated for fixed $\chimw$ and different values of $\eta$ as well as $\delta/\kappa$.
For each value of $\eta$ the gain $G(\eta)$ defined in Eq.~\eqref{eqn:DefinitionGain} is calculated and maximized over the evolution time $\tamp$ and the ratio $\delta/\kappa$ determining the relative strength between collective dephasing and single-spin dissipation. 

\newtext{%
Since our MFT is based on the assumption that the state of the system can be approximated by a Gaussian state, deviations from the exact dynamics based on the full QME will occur at long evolution times. 
However, they provide a reasonable approximation of the full dynamics until the time of maximum gain.
Therefore, by
}%
solving the MFT equations of motion numerically, one can typically obtain accurate predictions of the exponents of scaling laws in $N$, but the numerical prefactors may be off.
As shown in Figs.~\ref{fig:Amplification}(a) and (b), the MFT results agree very well with the results obtained for small $N$ by solving Eq.~\eqref{eqn:QME:EffectiveGeneral} numerically, but 
there is a small offset in the prediction of the gain in the ideal limit $\eta \to \infty$, $G_\mathrm{max}^\mathrm{ideal}\vert_\mathrm{MFT} > G_\mathrm{max}^\mathrm{ideal}$. 
Similarly, $\tamp^\mathrm{ideal}$ obtained from a MFT simulation differs from the analytical result $\tamp^\mathrm{ideal} \approx 1/\chi \sqrt{N}$.
Therefore, we extract $G_\mathrm{max}^\mathrm{ideal}$ (as well as the corresponding limits of all other quantities, e.g., $\tamp^\mathrm{ideal}$) from the numerical data at large $\eta$ rather than using the analytical results for rescaling.

Given the good agreement between MFT results and the data based on a numerically exact integration of the QME~\eqref{eqn:QME:EffectiveGeneral}, we can use the MFT equations of motion to gain analytical insight into the amplification physics by deriving an intuitive argument for the surprising condition shown in Eq.~\eqref{eqn:ConditionCavityFeedbackAmplification}. 
For $\sqrt{N} \eta \gtrsim 1$, the performance of the SCF scheme is limited by single-spin dephasing, which can be inferred from the optimal value of $\delta/\kappa \approx 1/2$ for $\eta \lesssim 1$, shown in Appendix~\ref{sec:App:AdditionalDetails:SCF}.
We will therefore ignore collective dephasing, $\Gammaphi \approx 0$, and set $\gammalocm = \gammalocp = 2 \gammalocz = \chi/\eta$. 
In order to be insensitive to the single-spin dissipation and to obtain the maximum possible gain, the OAT dynamics should amplify the signal much faster than the dissipative timescale $\eta / \chi$, 
\begin{align}
	\frac{1}{\sqrt{N} \chi} \ll \frac{\eta}{\chi} \comma
\end{align}
which immediately leads to the condition $\sqrt{N} \eta \gg 1$.

The essence of this argument also emerges if we perform a more quantitative analysis that is based on an approximate solution of the MFT equations of motion. 
At short times, we can ignore the covariances $C_{yz}$ and $C_{xz}$ in Eqs.~\eqref{eqn:MFT:Sx} and~\eqref{eqn:MFT:Sy} and integrate the resulting equations of motion:
\begin{align}
	S_x(t) 
		&= e^{- \left( \Gammaphi + \gammalocm + \gammalocp + 4 \gammalocz \right) t / 2} \nonumber \\
		&\phantom{=}\ \times \left[ S_x(0) \cos g(t) - S_y(0) \sin g(t) \right] \comma \\
	S_y(t) 
		&= e^{- \left( \Gammaphi + \gammalocm + \gammalocp + 4 \gammalocz \right) t / 2} \nonumber \\
		&\phantom{=}\ \times \left[ S_y(0) \cos g(t) + S_x(0) \sin g(t) \right] \comma \\
	S_z(t) &= e^{- (\gammalocp + \gammalocm) t} S_z(0) \comma 
\end{align}
where we introduced the auxiliary function
\begin{align}
	g(t) = \frac{\chi}{2 (\gammalocp + \gammalocm)} S_z(0) \left[ 1 - e^{- (\gammalocp + \gammalocm) t} \right] \fullstop
\end{align}
Starting from an initial state $S_x(0) = N/2$, $S_y(0) = 0$, $S_z(0) = N \phi/2$ with $\phi \ll 1$, neglecting collective dephasing, $\Gammaphi = 0$, and setting  $\gammalocm = \gammalocp = 2 \gammalocz = \chi/\eta$, we can now evaluate Eq.~\eqref{eqn:DefinitionGain} and obtain the following expression for the gain,
\begin{align}
	G(t,\eta) \approx e^{- 2 \chi t/\eta} \frac{N \eta}{4} \left( 1 - e^{- 2 \chi t/\eta} \right) \fullstop
\end{align}
Inserting the OAT amplification time of an ideal system, $\tamp^\mathrm{ideal} = 1/\chi \sqrt{N}$, and expanding the expression for $\sqrt{N} \eta \gg 1$, we find the approximate expression
\begin{align}
	G(1/\chi \sqrt{N},\eta) \approx e^{-2 /\eta \sqrt{N}} \frac{\sqrt{N}}{2} \comma
\end{align}
which approaches the correct scaling relation for the ideal gain if the condition~\eqref{eqn:ConditionCavityFeedbackAmplification} holds, $\lim_{\eta \to \infty} G(1/\chi \sqrt{N},\eta) = \sqrt{N}/2$, and clearly displays an exponential suppression if $\sqrt{N} \eta$ is small.

\section{Metrological gain in the presence of dissipation}
\label{sec:MetrologicalGain}

As discussed in Sec.~\ref{sec:Analysis:IdealProtocol}, the ultimate figure of merit for the performance of a OAT twist-untwist sensing scheme is the metrological gain $\Gmet$ defined in Eq.~\eqref{eqn:Definition:MetrologicalGain}. 
In the regime of large readout noise, $\Xidetsq \gg 1 + \sigma_\mathrm{diss}^2$, the metrological gain $\Gmet$ is proportional to the squared gain $G$ and, consequently, a large $\Gmet$ will be achieved if the conditions summarized in Table~\ref{tbl:Conditions} are satisfied.

Given that the gain physics in the presence of dissipation is already quite intricate and leads to an unusual scaling with cooperativity, cf.\ Eq.~\eqref{eqn:ConditionCavityFeedbackAmplification}, we now turn our attention to the regime of moderate readout noise, $\Xi_\mathrm{det}^2 \lesssim 1 + \sigma_\mathrm{diss}^2$.
In this regime, Davis \emph{et al.} \cite{Davis2016} argued that the metrological gain of the SCF scheme should be controlled by the collective cooperativity, $\Gmet \propto \sqrt{N \eta}$. 
Since their argument neglects the reduction of the gain $G(\eta)$ by dissipation, corrections to this result are to be expected. 
Indeed, Fig.~\ref{fig:comparison} surprisingly shows that $\Gmet \ll \sqrt{N \eta}$ is possible even though both the single-spin and the collective cooperativity are large.

\begin{figure}
	\centering
	\includegraphics[width=0.33\textwidth]{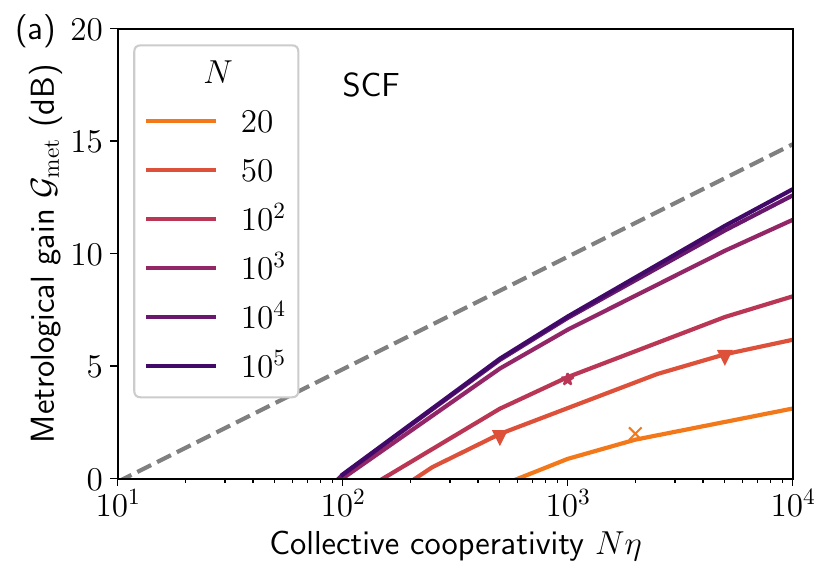}
	\includegraphics[width=0.33\textwidth]{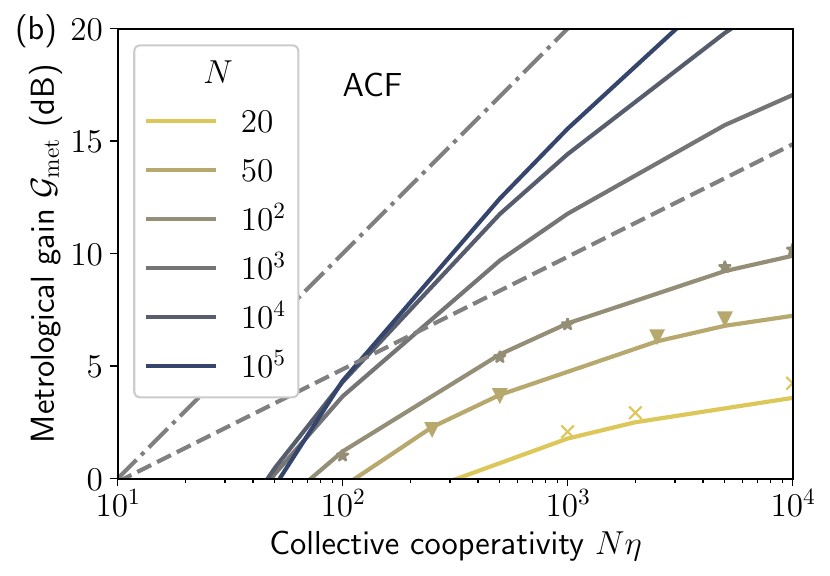}
	\caption{
		Metrological gain $\Gmet$, defined in Eq.~\eqref{eqn:Definition:Gmeteta}, as a function of the collective cooperativity $N \eta$ for (a) the SCF scheme and (b) the ACF scheme and different spin numbers $N$. 
		In both subplots, the dashed gray line indicates the estimate $\Gmetest$ for the SCF scheme, defined in Eq.~\eqref{eqn:GmetEmily}, which has been derived in Ref.~\onlinecite{Davis2016}.
		It overestimates the onset of metrological gain even though the collective cooperativities $N \eta$ are large. 
		The dash-dotted gray line in (b) indicates a $\Gmet = \FigOATACFDoubleMetrologicalGainVsCollectiveCooperativityFormulaAsymptoteDashDot$ scaling. 
		The solid lines have been generated using the MFT equations of motion and the markers have been calculated by numerically exact integration of the QME~\eqref{eqn:QME:EffectiveGeneral} for small ensemble sizes $N$. 
		The level of readout noise is $\Xidetsq = \FigOATSCFDoubleMetrologicalGainVsCollectiveCooperativityValueXidetsq$.
		\newtext{Note that the abrupt changes in the slope of $\Gmet$ for the MFT curves are artifacts of the small number of sample points.} 
	}
	\label{fig:MetrologicalGain}
\end{figure}

We analyze this behavior more closely in Fig.~\ref{fig:MetrologicalGain}(a), which shows the maximum metrological gain $\Gmet(\eta)$ for the SCF scheme as a function of the collective cooperativity $N \eta$. 
The data shown by the markers represent a simulation of the full OAT twist-untwist sensing protocol sketched in Fig.~\ref{fig:sketch_OAT_protocol} and described in Sec.~\ref{sec:Analysis:IdealProtocol}, which has been obtained as follows:
The system is initialized in a coherent-spin state $\ket{\psi_0}$ polarized in the $x$ direction and is evolved by integrating the QME~\eqref{eqn:QME:EffectiveGeneral} for a time $\tsqz$ with a OAT strength $\chi$ to obtain a spin-squeezed state $\hat{\rho}_1$. 
Subsequently, the signal $\phi$ is encoded into the $\hat{S}_z$ component by rotating this spin-squeezed state about the $\hat{S}_y$ axis, 
\begin{align}
	\hat{\rho}_2(\phi) = e^{i \phi \hat{S}_y} \hat{\rho}_1 e^{-i \phi \hat{S}_y} \fullstop
	\label{eqn:rotatedrho2}
\end{align}
Finally, the rotated state $\hat{\rho}_2(\phi)$ is evolved by integrating Eq.~\eqref{eqn:QME:EffectiveGeneral} for a time $\tunsqz = \tsqz$ using the OAT strength $-\chi$ to obtain the final state $\hat{\rho}_3(\phi)$ at time $t_\mathrm{end} = \tsqz + \tunsqz = 2 t_\mathrm{(un)sqz}$. 
This procedure is repeated for different values of $\eta$ and $\lambda = \delta/\kappa$.
As described in Sec.~\ref{sec:Analysis:Dissipation}, the gain $G$ and the additional fluctuations $\sigma_\mathrm{diss}^2$ can be obtained from the expectation value $\cerw{\hat{S}_y(t_\mathrm{end})}$ and the variance $(\bfDelta \hat{S}_y)^2$ of $\hat{\rho}_3(\phi)$ using the definition 
\begin{align}
	G = \lim_{\phi \to 0} \frac{\cerw{\hat{S}_y(t_\mathrm{end})}}{N \phi/2}
	\label{eqn:GainForGmet}
\end{align}
and Eq.~\eqref{eqn:DefinitionSigmadisssq}, respectively.
Using these results as well as Eqs.~\eqref{eqn:Definition:MetrologicalGain} and~\eqref{eqn:CompleteModelDeltaphisq}, the optimum metrological gain $\Gmet(\eta)$ for a given single-spin cooperativity $\eta$ can be calculated using 
\begin{align}
	\Gmet(\eta) = \max_{t_\mathrm{(un)sqz}} \max_{\lambda} \frac{G^2}{1 + \sigma_\mathrm{diss}^2 + \Xidetsq} \fullstop
	\label{eqn:Definition:Gmeteta}
\end{align}
The solid lines in Fig.~\ref{fig:MetrologicalGain} correspond to a MFT simulation of the dynamics generated by Eq.~\eqref{eqn:QME:EffectiveGeneral}, which allows us to investigate larger ensemble sizes $N$.
Additional details on the numerical optimization of the metrological gain can be found in Appendix~\ref{sec:App:MetrologicalGain}.

We compare our simulations with the dashed gray curve, which represents the result $\Gmetest$ for the SCF scheme [cf.\ Eq.~\eqref{eqn:GmetEmily}], which has been derived in Ref.~\onlinecite{Davis2016} using phenomenological arguments. 
While it does capture the scaling of the metrological gain correctly in the regime $\Gmetideal \gg \Gmet(\eta) \ggg 1$, it is overly optimistic in a broad regime of parameters where the collective cooperativity is large, $N \eta \gg 1$, but the metrological gain is small, $\Gmet \gtrsim 1$. 
In particular, the onset of metrological gain is highly overestimated:
$\Gmetest$ suggests that a moderate metrological gain of $5\,\mathrm{dB}$ requires a collective cooperativity of $N \eta \approx 100$ for $N=100$ spins. 
However, we find that the collective cooperativity that is actually required to reach $\Gmet = 5\,\mathrm{dB}$ is almost an order of magnitude higher, $N \eta \approx 1000$. 
Finally, for small $N \lesssim 10^4$, the metrological gain depends not only on the collective cooperativity $N \eta$, but also on the number $N$ of spins in the ensemble. 
Only for the highest considered spin numbers, the curves of $\Gmet$ for different $N$ converge to a common behavior that grows $\propto \sqrt{N \eta}$ for $\Gmet \gg 1$ before fanning out and saturating to the spin-number-dependent maximum metrological gain $\Gmetideal = N / [e (1 + \Xidetsq)]$ at large collective cooperativity.

Interestingly, the different dissipative processes of the SCF and the ACF schemes manifest themselves in the behavior of the metrologial gain for $\Xi_\mathrm{det}^2 \lesssim 1 + \sigma_\mathrm{diss}^2$, as shown in Fig.~\ref{fig:MetrologicalGain}(b):
The metrological gain of the ACF exceeds that of the SCF scheme and scales approximately linearly with the collective cooperativity $\Gmet \propto N \eta$ [given by the dash-dotted gray line in Fig.~\ref{fig:MetrologicalGain}(b)] before saturating to $\Gmetideal$. 
This corrobates the results reported in Ref.~\onlinecite{Colombo2022}. 

In Fig.~\ref{fig:comparison}, we compare the TC scheme with the ACF and SCF schemes for a fixed system size. 
We find that the TC scheme follows a similar $\Gmet \propto \sqrt{N \eta}$ scaling as the SCF scheme if $\Gmetideal \gg \Gmet(\eta) \ggg 1$, i.e., in an intermediate regime where the metrological gain is large but not yet saturating to its ideal value $\Gmetideal$.
However, as for the SCF scheme, the onset of metrological gain is highly overestimated. 
Moreover, the TC-based implementation of a OAT twist-untwist protocol requires a much higher collective cooperativity than the SCF and ACF schemes to reach similar levels of $\Gmet$. 
This is a result of the collective decay term in the TC scheme (cf.\ Table~\ref{tbl:Rates}) which highly disrupts the amplification dynamics (as described in Sec.~\ref{sec:Gain:TC}) and also led to a much more stringent requirement to achieve significant levels of gain (cf.\ Table~\ref{tbl:Conditions}).

\section{Conclusion}
\label{sec:Conclusion}

In conclusion, we analyzed three different implementations of the OAT Hamiltonian, shown in Fig.~\ref{fig:system}, and compared their ability to amplify a small signal encoded in a nonzero $\cerw{\hat{S}_z}$ polarization.
The three schemes differ in the form of undesired dissipative terms that accompany the unitary OAT interaction, which are summarized in Table~\ref{tbl:Rates}. 
Despite the fact that these differences do not lead to different conditions to generate spin squeezing, they do manifest themselves in very different conditions that need to be satisfied in order to see significant levels of gain $G$, which are summarized in Table~\ref{tbl:Conditions}. 
Interestingly, the SCF and ACF schemes provide large gain if the scaled cooperativity $\sqrt{N} \eta$ is large. 
We used second-order MFT to provide an intuitive explanation for this unusual scaling. 
While this condition is less challenging than the one required by the TC scheme, it is experimentally more challenging than reaching a large collective cooperativity, which is required by amplification schemes based on collective dissipation \cite{Koppenhoefer2022,Koppenhoefer2023}.

Motivated by these surprising results, we revisited the metrological gain $\Gmet$ of the SCF, ACF, and TC schemes. 
In the regime where readout noise dominates, $\Xidetsq \ggg 1 + \sigma_\mathrm{diss}^2$, all implementations of the OAT twist-untwist scheme have $\Gmet \propto G^2/\Xidetsq$, i.e., large metrological gain can be achieved whenever $G$ is large. 
In the opposite regime, $\Xidetsq \lesssim 1 + \sigma_\mathrm{diss}^2$, we found that the ACF scheme provides parametrically larger $\Gmet$ than the SCF and TC schemes.
Moreover, we found that the prediction $\Gmet \propto \sqrt{N \eta}$, which had been derived in Refs.~\onlinecite{Davis2016,Chu2021} 
\newtext{%
using a phenomenologically motivated model and the approximation that the action of different dissipative processes can be analyzed separately, respectively,
}%
highly overestimates the performance of the SCF and TC schemes when $\Gmet \gtrsim 1$: 
Even though the collective cooperativity may be large, MFT simulations and numerically exact integration of the quantum master equation~\eqref{eqn:QME:EffectiveGeneral} for the SCF and TC schemes show that the dissipative processes can interrupt the amplification dynamics and lead to a much smaller metrological gain than previously appreciated.

Our study highlights that the form of undesired dissipative processes can have a huge impact on the performance of a OAT twist-untwist sensing protocol and that a careful deliberation is needed when choosing the specific implementation of OAT dynamics on a given quantum sensing platform.
AMO systems provide large single-spin cooperativities and low readout noise, such that the ACF scheme can be implemented.
Solid-state sensing platforms, however, are typically plagued by much higher levels of readout noise and reaching a large single-spin cooperativity can be significantly harder than achieving a large collective cooperativity. 
Therefore, even though cavity-feedback schemes could in principle be implemented in solid-state systems \cite{Zou2014}, amplification schemes based on collective dissipation \cite{Koppenhoefer2022,Koppenhoefer2023} are expected to be more effective on these platforms.

The twist-untwist sensing protocol can also be implemented using two-axis-twist dynamics \cite{Anders2018,MunozArias2023} or twist-and-turn dynamics \cite{Mirkhalaf2018}. 
Even though the two-axis-twist Hamiltonian generates higher levels of spin squeezing than OAT \cite{Kitagawa1993}, it leads to parametrically the same metrological gain in a unitary twist-untwist sensing protocol \cite{Anders2018}.
In the future, it would be interesting to check if this remains true if undesired dissipative processes are taken into account, and to compare the  levels of metrological gain that can be achieved in different implementations of the two-axis-twist Hamiltonian in the presence of dissipation.

\begin{acknowledgments}
This work was primarily supported by the DOE Q-NEXT Center (Grant No. DOE 1F-60579). 
We also acknowledge support from the Simons Foundation (Grant No. 669487, A. C.).
We thank Peter Groszkowski and Tony Jin for fruitful discussions.
\end{acknowledgments}

\appendix

\section{Derivation of the effective quantum master equation}
\label{sec:App:DerivationEffectiveQME}

In this Appendix, we derive the effective QMEs for the TC, SCF, and ACF schemes shown in Fig.~\ref{fig:system} by adiabatically eliminating the cavity mode $\hat{a}$ and the higher levels $\{\ket{e_j}\}_{j=1,\dots,N}$. 
All three cases can be rewritten in the form of Eq.~\eqref{eqn:QME:EffectiveGeneral} with the rates shown in Table~\ref{tbl:Rates}. 

\subsection{TC coupling scheme}

For the TC coupling scheme defined by the QME~\eqref{eqn:QME:EffectiveGeneral} and Eq.~\eqref{eqn:Details:TC}, we can diagonalize $\hat{H}_0 + \hat{H}_1^\mathrm{TC}$ to leading order in $g/\Deltatc$ using a Schrieffer-Wolff transformation $\hat{H}^\mathrm{SCF} =  e^{\hat{W}} (\hat{H}_0 + \hat{H}_1^\mathrm{TC}) e^{- \hat{W}}$ with the generator
\begin{align}
	\hat{W} = \frac{g}{\Deltatc} \left( \hat{a}^\dagger \hat{S}_- - \hat{a} \hat{S}_+ \right) \fullstop
\end{align}
Up to correction terms of $\mathcal{O}(g^3/\Deltatc^2)$, this gives rise to the Hamiltonian
\begin{align}
	\hat{H}^\mathrm{TC}
		&= \omc \hat{a}^\dagger \hat{a} + \oms \hat{S}_z - \chitc \left( \vec{\hat{S}}^2 - \hat{S}_z^2 + \hat{S}_z \right) \nonumber \\
		&\phantom{=} - 2\chitc \hat{a}^\dagger \hat{a} \hat{S}_z \comma
\end{align}
where we defined the OAT strength
\begin{align}
	\chitc = \frac{g^2}{\Deltatc} \fullstop
\end{align}
The total-angular-momentum operator $\vec{\hat{S}}^2 = \hat{S}_x^2 + \hat{S}_y^2 + \hat{S}_z^2$ causes an irrelevant constant energy shift, which we will ignore in the following.
Applying the Schrieffer-Wolff transformation to the cavity-decay term $\kappa \mathcal{D}[\hat{a}] \hat{\rho}$ in Eq.~\eqref{eqn:Details:QMEStartingPoint} and performing a rotating wave approximation, we find the QME
\begin{align}
	\frac{\d}{\d t} \hat{\rho} &= - i \komm{\hat{H}^\mathrm{TC}}{\hat{\rho}} + \kappa \mathcal{D}[\hat{a}] \hat{\rho} + \chitc \frac{\kappa}{\Delta} \mathcal{D}[\hat{S}_-] \hat{\rho} \fullstop
\end{align}
In contrast to the SCF and ACF schemes, the cavity is undriven and therefore in its ground state. 
This allows us to ignore the $- 2 \chitc \hat{a}^\dagger \hat{a} \hat{S}_z$ term in $\hat{H}^\mathrm{TC}$ (which would otherwise lead to dephasing of the spins due to photon-number fluctuations in the cavity).
Moreover, we have $\abs{\chitc} \ll \oms$, such that we can ignore the small renormalization of the spin frequency. 
After an adiabatic elimination of the cavity mode, we thus find the effective QME
\begin{align}
	\frac{\d}{\d t} \hat{\rho} 
		&= - i \komm{\oms \hat{S}_z + \chitc \hat{S}_z^2}{\hat{\rho}} \nonumber \\
		&\phantom{=}\ + \chitc \frac{\kappa}{\Deltatc} \mathcal{D}[\hat{S}_-] \hat{\rho} + \mathcal{L}_\mathrm{spin}^\mathrm{TC} \hat{\rho} \fullstop
\end{align}
This result can be rewritten in the form of Eq.~\eqref{eqn:QME:EffectiveGeneral} using the frequencies and rates shown in Table~\ref{tbl:Rates}.

\subsection{SCF coupling scheme}
\label{sec:App:DerivationEffectiveQME:SCF}

For the SCF coupling scheme defined by the QME~\eqref{eqn:Details:QMEStartingPoint} and Eq.~\eqref{eqn:Details:SCF}, we can diagonalize the undriven Hamiltonian $\hat{H}_0 + \hat{H}_1^\mathrm{SCF} \vert_{\beta_\mathrm{in}(t) = 0}$ to leading order in $g/\Delta$ using a Schrieffer-Wolff transformation $\hat{H}^\mathrm{SCF} =  e^{\hat{W}} (\hat{H}_0 + \hat{H}_1^\mathrm{SCF} \vert_{\beta_\mathrm{in}(t) = 0}) e^{- \hat{W}}$ with the generator 
\begin{align}
	\hat{W} = \frac{g}{\Delta} \sum_{j=1}^N \left[ \hat{a}^\dagger \left( \ket{\uparrow_j} \bra{e_j} - \ket{\downarrow_j}\bra{e_j} \right) - \hc \right] \fullstop
	\label{eqn:DerivationEffectiveQME:SCF:SWGenerator}
\end{align}
Up to correction terms of $\mathcal{O}(g^3/\Delta^3)$, this gives rise to the Hamiltonian
\begin{align}
	\hat{H}^\mathrm{SCF} 
		&= \omc \hat{a}^\dagger \hat{a} + \oms \hat{S}_z + \Omegamw \hat{a}^\dagger \hat{a} \hat{S}_z \comma
\end{align}
where we defined the dispersive coupling strength 
\begin{align}
	\Omegamw = \frac{2 g^2}{\Delta} \comma
\end{align}
and ignored the energy term of the $\{\ket{e_j}\}_{j=1,\dots,N}$ levels. 
Applying the Schrieffer-Wolff transformation to $\mathcal{L}_\mathrm{spin}^\mathrm{SCF}$ and projecting onto the subspace containing all levels except $\{ \ket{e}_j \}_{j=1,\dots,N}$, we find the QME 
\begin{align}
	\frac{\d}{\d t} \hat{\rho} 
		&= - i \komm{\hat{H}^\mathrm{SCF} + \hat{H}_\mathrm{dr}}{\hat{\rho}} + \kappa \mathcal{D}[\hat{a}] \hat{\rho} \nonumber \\
		&\phantom{=}\ + \frac{\Gamma g^2}{\Delta^2} \sum_{j=1}^N \mathcal{D} \left[ \hat{a} \left( \ket{\uparrow_j} \bra{\uparrow_j} - \ket{\uparrow_j} \bra{\downarrow_j} \right) \right] \hat{\rho} \nonumber \\
		&\phantom{=}\ + \frac{\Gamma g^2}{\Delta^2} \sum_{j=1}^N \mathcal{D} \left[ \hat{a} \left( \ket{\downarrow_j} \bra{\downarrow_j} - \ket{\downarrow_j} \bra{\uparrow_j} \right) \right] \hat{\rho} \comma
\end{align}
where $\hat{H}_\mathrm{dr} = \sqrt{\kappa} \left[ \beta_\mathrm{in}(t) \hat{a}^\dagger + \beta_\mathrm{in}^*(t) \hat{a} \right]$.
For $\abs{\Delta} \gg \kappa, \Gamma, g$, we can expand the dissipators and perform a rotating wave approximation to neglect cross terms, such that we obtain 
\begin{align}
	\frac{\d}{\d t} \hat{\rho} 
		&= - i \komm{\hat{H}^\mathrm{SCF} + \hat{H}_\mathrm{dr}}{\hat{\rho}} + \kappa \mathcal{D}[\hat{a}] \hat{\rho} \nonumber \\
		&\phantom{=}\ + \frac{\Gamma g^2}{\Delta^2} \sum_{j=1}^N \left( \frac{1}{4} \mathcal{D} \left[ \hat{a} (\hat{\sigma}^{(j)}_z + \hat{\1}^{(j)}) \right] \hat{\rho} + \mathcal{D} \left[ \hat{a} \hat{\sigma}_+^{(j)} \right] \hat{\rho} \right) \nonumber \\
		&\phantom{=}\ + \frac{\Gamma g^2}{\Delta^2} \sum_{j=1}^N \left( \frac{1}{4} \mathcal{D} \left[ \hat{a} (\hat{\sigma}^{(j)}_z - \hat{\1}^{(j)}) \right] \hat{\rho} + \mathcal{D} \left[ \hat{a} \hat{\sigma}_-^{(j)} \right] \hat{\rho} \right) \comma
\end{align}
where $\hat{\1}^{(j)} = \ket{\uparrow_j}\bra{\uparrow_j} + \ket{\downarrow_j}\bra{\uparrow_j}$ is the identity operator in the subspace spanned by the states $\ket{\uparrow_j}$ and $\ket{\downarrow_j}$.

We now consider a cavity drive at frequency $\omega_\mathrm{dr} = \omc + \delta$, as given by Eq.~\eqref{eqn:DefinitionDrive}, switch to a frame rotating at the drive frequency, and decompose the cavity field into a semiclassical amplitude $a = \sqrt{\kappa} \overline{\beta}_\mathrm{in} / (\delta + i \kappa/2)$ and quantum fluctuations $\hat{d}$, 
\begin{align}
	\hat{a} = e^{-i (\omc + \delta) t} \left( a + \hat{d} \right) \fullstop
	\label{eqn:DisplacedFrame}
\end{align}
Keeping only leading-order terms in the dissipators, we find the new QME
\begin{align}
	\frac{\d}{\d t} \hat{\rho}
		&= - i \komm{- \delta \hat{d}^\dagger \hat{d} + (\oms + \Omegamw \ncav) \hat{S}_z}{\hat{\rho}} \nonumber \\
		&\phantom{=}\ - i \komm{\Omegamw \sqrt{\ncav} (\hat{d} + \hat{d}^\dagger) \hat{S}_z + \Omegamw \hat{d}^\dagger \hat{d} \hat{S}_z}{\hat{\rho}} \nonumber \\
		&\phantom{=}\ + \frac{\Gamma g^2}{\Delta^2} \ncav \sum_{j=1}^N \left( \frac{1}{2} \mathcal{D}[\hat{\sigma}_z^{(j)}] + \mathcal{D}[\hat{\sigma}_+^{(j)}] + \mathcal{D}[\hat{\sigma}_-^{(j)}] \right) \hat{\rho} \nonumber \\
		&\phantom{=}\ + \kappa \mathcal{D}[\hat{d}] \hat{\rho} 	\comma 
\end{align}
where we defined the intracavity photon number $\ncav = \abs{a}^2$ due to the drive, cf.\ Eq.~\eqref{eqn:Definitionncav}. 
To ignore higher-order correction terms to the Schrieffer-Wolff transformation defined in Eq.~\eqref{eqn:DerivationEffectiveQME:SCF:SWGenerator}, the condition $\abs{a}^2 \ll (\Delta/g)^2$ must hold. 
This implies $\Omegamw \ncav \ll 2 \Delta$ and thus allows us to ignore the renormalization of the spin transition frequency. 
Finally, we adiabatically eliminate the cavity fluctuations $\hat{d}$ by assuming that the cavity relaxes to its steady state fast compared to the dissipative dynamics of the spins, which yields the effective QME
\begin{align}
	\frac{\d}{\d t} \hat{\rho} 
		&= - i \komm{\oms \hat{S}_z + \chimw \hat{S}_z^2}{\hat{\rho}} + \chimw \frac{\kappa}{\delta} \mathcal{D}[\hat{S}_z] \hat{\rho} \nonumber \\
		&\phantom{=}\ + \gammalocmw \sum_{j=1}^N \left( \frac{1}{2} \mathcal{D}[\hat{\sigma}_z^{(j)}] + \mathcal{D}[\hat{\sigma}_+^{(j)}] + \mathcal{D}[\hat{\sigma}_-^{(j)}] \right) \hat{\rho} \comma
		\label{eqn:QME_final:MW}
\end{align}
where we defined the OAT strength 
\begin{align}
	\chimw = \Omegamw^2 \ncav \frac{\delta}{\delta^2 + \kappa^2/4} 
\end{align}
and the local dissipation rate
\begin{align}
	\gammalocmw =  \ncav \frac{ \Gamma g^2}{\Delta^2} = \chimw \frac{\delta^2 + \kappa^2/4}{\kappa \delta \eta} \fullstop
\end{align}
The single-spin cooperativity $\eta$ has been defined in Eq.~\eqref{eqn:Definitioneta}.
This result can be rewritten in the form of Eq.~\eqref{eqn:QME:EffectiveGeneral} using the frequencies and rates shown in Table~\ref{tbl:Rates}.

In contrast to the previous work by Zhang \emph{et al.} \cite{Zhang2015}, we apply the adiabatic elimination of the $\{\ket{e}_j\}_{j=1,\dots,N}$ levels to the dissipative terms, too, such that we obtain an effective QME that acts only on the subspace of the two-level systems $\{ \ket{\uparrow}_j, \ket{\downarrow}_j\}_{j=1,\dots,N}$ of interest and reveals the dependence of the local dissipation rate $\gammalocmw$ on $\delta$.
In the limit $\Gamma \to \infty$, we find that $\gammalocmw$ is minimized at the detuning $\delta = 1/2$.
For finite $\Gamma$, the optimal detuning grows, $\delta > 1/2$, until it converges to the optimal value $\delta \to \infty$ for $\Gamma \to 0$ (assuming one can increase $\ncav$ accordingly to keep $\chi$ constant).
The last result, i.e., the limit $\Gamma \to 0$, has already been pointed out by Zhang \emph{et al.} \cite{Zhang2015}.

\subsection{ACF coupling scheme}

\begin{figure*}
	\centering
	\includegraphics[width=0.245\textwidth]{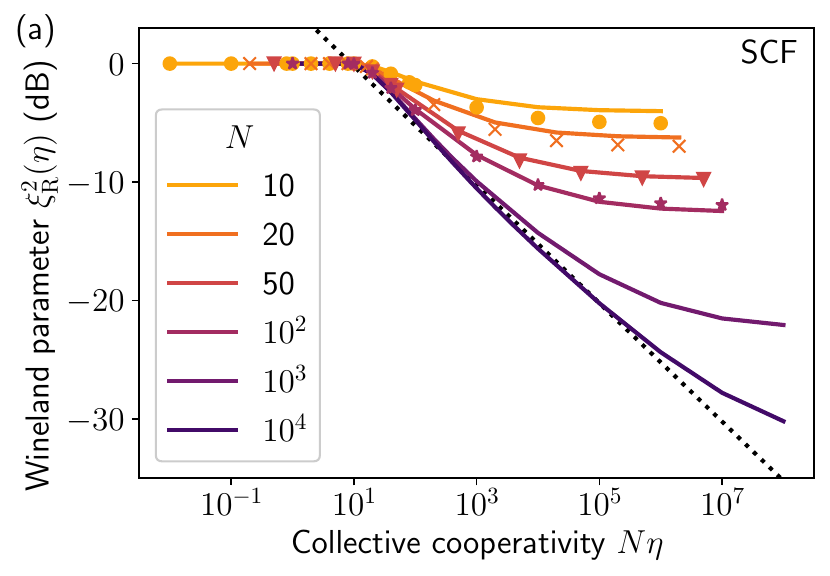}
	\includegraphics[width=0.245\textwidth]{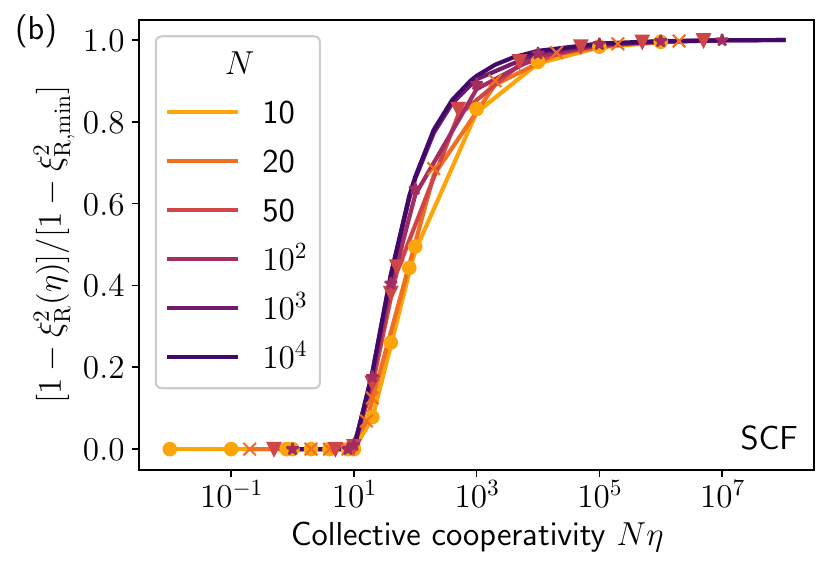}
	\includegraphics[width=0.245\textwidth]{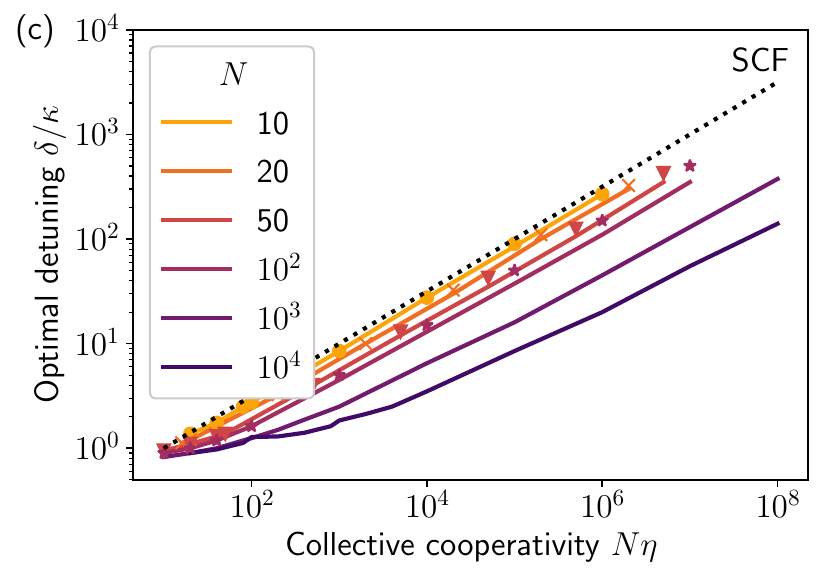}
	\includegraphics[width=0.245\textwidth]{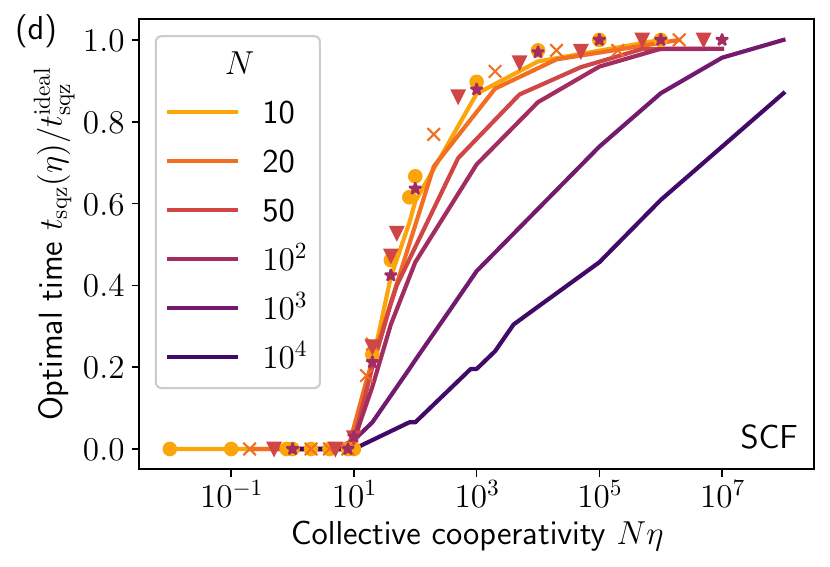}
	\caption{
		(a) Minimum Wineland parameter $\Wineland(\eta)$ and 
		(b) squeezing improvement $1 - \Wineland(\eta)$, compared to its ideal value $1 - \xi_\mathrm{R,min}^2$ [where $\xi_\mathrm{R,min}^2 = \lim_{\eta \to \infty} \Wineland(\eta)$ is the Wineland parameter in the absence of any single-spin dissipation] for the SCF scheme shown in Fig.~\ref{fig:system}(b), 
		as a function of the collective cooperativity $N \eta$ for different ensemble sizes $N$. 
		The solid lines represent data obtained by solving the mean-field-theory equations of motion, whereas the big markers of the respective color correspond to optimizations based on integration of the quantum master equation~\eqref{eqn:QME:EffectiveGeneral} for small ensemble sizes. 
		(c) Corresponding optimal values of the detuning $\delta/\kappa$ and (d) the evolution time $\tsqz(\eta)$ [relative to the time $\tsqz^\mathrm{ideal} = \lim_{\eta \to \infty} t(\eta) \propto 1/\chi N^{2/3}$ in the absence of dissipation] that minimize $\Wineland$.
		The dotted black lines in (a) and (c) indicate $\FigOATSCFSqueezingWinelandVsCFormulaAsymptote$ and $\delta/\kappa = \FigOATSCFOptimalDetuningVsCFormulaAsymptote$, respectively. 
	}
	\label{fig:SpinSqueezing}
\end{figure*}

The adiabatic elimination performed in Sec.~\ref{sec:App:DerivationEffectiveQME:SCF} can also be applied to the ACF coupling scheme defined by the QME~\eqref{eqn:Details:QMEStartingPoint} and Eq.~\eqref{eqn:Details:ACF}. 
In this case, the generator of the Schrieffer-Wolff transformation diagonalizing $\hat{H}_0 + \hat{H}_1^\mathrm{ACF}\vert_{\beta_\mathrm{in}(t) = 0}$ to leading order in $g/\Delta$ is
\begin{align}
	\hat{W} = \frac{g}{\Delta} \sum_{j=1}^N \Big( \hat{a}^\dagger \ket{\uparrow_j}\bra{e_j} - \hat{a} \ket{e_j}\bra{\uparrow_j} \Big) \fullstop 
\end{align}
Up to correction terms of $\mathcal{O}(g^2/\Delta^3)$, this gives rise to the Hamiltonian
\begin{align}
	\hat{H}^\mathrm{ACF} 
		&= \omc \hat{a}^\dagger \hat{a} + \oms \left(\hat{S}_z - \frac{N}{2} \right) + \Omegaopt \hat{a}^\dagger \hat{a} \left( \hat{S}_z + \frac{N}{2} \right) \comma
\end{align}
where we defined the dispersive coupling strength 
\begin{align}
	\Omegaopt = \frac{g^2}{\Delta} 
\end{align} 
and ignored the energy term of the $\{\ket{e_j}\}_{j=1,\dots,N}$ levels. 
Applying the Schrieffer-Wolff transformation to $\mathcal{L}_\mathrm{spin}^\mathrm{ACF}$ and projecting onto the subspace containing all levels except $\{\ket{e_j}\}_{j=1,\dots,N}$, we find the QME
\begin{align}
	\frac{\d}{\d t} \hat{\rho} 
		&= - i \komm{\hat{H}^\mathrm{ACF} + \hat{H}_\mathrm{dr}}{\hat{\rho}} + \kappa \mathcal{D}[\hat{a}] \hat{\rho} \nonumber \\
		&\phantom{=}\ + \frac{\Gamma g^2}{\Delta^2} \sum_{j=1}^N \frac{1}{4} \mathcal{D} \left[ \hat{a} (\hat{\sigma}_z^{(j)} + \hat{\1}^{(j)}) \right] \hat{\rho} \fullstop 
\end{align}

Again, we consider the cavity drive given in Eq.~\eqref{eqn:DefinitionDrive}, switch to a rotating and displaced frame as given in Eq.~\eqref{eqn:DisplacedFrame}, and adiabatically eliminate the cavity fluctuations $\hat{d}$.
We find the effective QME
\begin{align}
	\frac{\d}{\d t} \hat{\rho} 
		&= - i \komm{\left( \oms + N \chiopt \right) \hat{S}_z + \chiopt \hat{S}_z^2}{\hat{\rho}} \nonumber \\
		&\phantom{=}\ +\chiopt \frac{\kappa}{\delta} \mathcal{D}[\hat{S}_z] \hat{\rho} + \gammalocopt \sum_{j=1}^N \mathcal{D}[\hat{\sigma}_z^j] \hat{\rho} \comma
		\label{eqn:QME_final:OPT}
\end{align}
where we ignored a small renormalization of the spin frequency, since $\Omegaopt \ncav \ll \oms$, and defined the OAT strength
\begin{align}
	\chiopt = \Omegaopt^2 \ncav \frac{\delta}{\delta^2 + \kappa^2/4} 
\end{align}
as well as the single-spin dephasing rate
\begin{align}
	\gammalocopt = \chiopt \frac{\delta^2 + \kappa^2/4}{\delta \kappa \eta} \fullstop
\end{align}
Since we are coupling only the level $\ket{\uparrow_j}$ to $\ket{e_j}$, the effective Hamiltonian contains a constant frequency shift $N \chiopt$, which can be easily measured and subsequently be subtracted by working in a rotating frame \cite{Li2022}. 
The effective QME~\eqref{eqn:QME_final:OPT} can again be rewritten in the form of Eq.~\eqref{eqn:QME:EffectiveGeneral} using the frequencies and rates shown in Table~\ref{tbl:Rates}.

Comparing Eqs.~\eqref{eqn:QME_final:OPT} and~\eqref{eqn:QME_final:MW}, we find that coupling only the level $\ket{\uparrow}_j$ to $\ket{e_j}$ leads to a reduction of the OAT strength by a factor of four, since $\Omegaopt = \Omegamw/2$ for the same detuning $\Delta$, but eliminates the possiblitiy of spin flips $\ket{\uparrow}_j \leftrightarrow \ket{e}_j \leftrightarrow \ket{\downarrow}_j$ via the $\ket{e}_j$ level, which cause the additional dissipators $\mathcal{D}[\hat{\sigma}_j^\pm]$ in Eq.~\eqref{eqn:QME_final:MW}.

\section{Spin squeezing in the SCF scheme}
\label{sec:App:SpinSqueezing}

In this Appendix, we numerically optimize the Wineland parameter \cite{Wineland1992,Pezze2018}, defined in Eq.~\eqref{eqn:Definition:Wineland}, in the presence of dissipative terms in the SCF scheme.
In the absence of any bad dissipation, $\Gammaphi = \Gammarel = \gammalocz = \gammalocp = \gammalocm = 0$ in Eq.~\eqref{eqn:QME:EffectiveGeneral}, the OAT Hamiltonian generates a spin-squeezed state with a minimum Wineland parameter of $\Wineland \propto N^{-2/3}$ at the optimal evolution time $\chi \tsqz \propto N^{-2/3}$ \cite{Kitagawa1993,Pezze2018}. 
Collective dephasing and single-spin dissipation compete with the OAT dynamics and reduce the amount of spin squeezing. 
For TC-based OAT spin-squeezing dynamics, it has been shown that $\Wineland < 1$ can be achieved if the collective cooperativity satisfies $N \etarel, N \etaphi \gtrsim \mathcal{O}(10^2)$ and that the Wineland parameter close to the threshold scales like $\Wineland \propto 1/\sqrt{N \etaphi}$ ($\Wineland \propto 1/\sqrt[3]{N \etarel}$) if local dephasing (local relaxation) is the dominant single-spin dissipation process \cite{LewisSwan2018}.

A similar perturbative analytical analysis for SCF-based OAT spin-squeezing dynamics has been performed in Refs.~\onlinecite{Davis2016,Chu2021}, which predicted a threshold condition in terms of the collective cooperativity, $N \eta \gg 1$, and a scaling $\Wineland \propto 1/\sqrt{N \eta}$ close to this threshold.
Here, we check these predictions by numerical integration of the QME~\eqref{eqn:QME:EffectiveGeneral} for a given OAT strength $\chimw$ and a given single-spin cooperativity $\eta$, starting from an initial coherent-spin state $\ket{\psi_0}$ pointing along the $x$ direction.
We then minimize $\Wineland$ over the evolution time $\tsqz$ and the ratio $\kappa/\delta$ determining the relative strength of the collective and local dissipation terms.
For larger system sizes, we repeat the same analysis using the second-order MFT equations of motion listed in Appendix~\ref{sec:App:MFT}.

The results of this optimization are shown in Fig.~\ref{fig:SpinSqueezing}.
A Wineland parameter below unity can be achieved if the collective cooperativity satisfies $N \eta \gtrsim 10$.
Close to the threshold, the Wineland parameter decreases approximately like the square-root of the collective cooperativity, $\Wineland \propto 1/(N \eta)^{0.56}$, similar to the analytical predictions \cite{Davis2016,Chu2021} and to the scaling of $\Wineland$ for TC-based OAT dynamics in the presence of dominant single-spin dissipation \cite{LewisSwan2018}.
Intuitively, this result can be understood by noting that, even though single-spin excitation and relaxation processes are present in the SCF scheme, their rates are exactly balanced and thus lead to a mere relaxation of any initial polarization to a completely mixed state, i.e., no relaxation towards the ground state.
With increasing collective cooperativity, the Wineland parameter ultimately departs from the scaling $1/(N \eta)^{0.56}$ scaling and approaches its minimum value $\Wineland \propto 1/N^{2/3}$ in the limit $\eta \to \infty$.

Thus, there appears to be no significant difference between SCF-based OAT and single-spin-dephasing-dominated TC-based OAT despite the differences in their dissipative processes shown in Table~\ref{tbl:Rates}.

\section{Additional details on the gain in the SCF scheme}
\label{sec:App:AdditionalDetails:SCF}

In this Appendix, we provide additional details on the numerical optimization that has been used to generate Fig.~\ref{fig:Amplification}.

\begin{figure}
	\centering
	\includegraphics[width=0.23\textwidth]{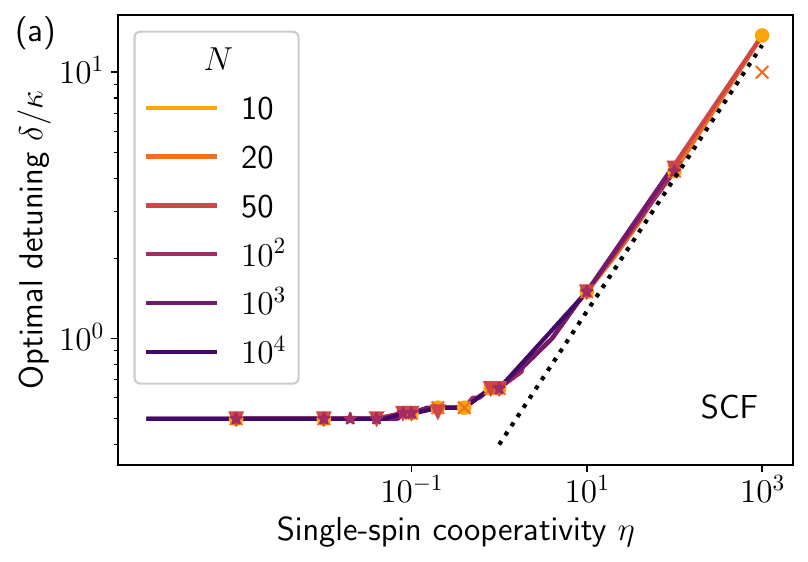}
	\includegraphics[width=0.24\textwidth]{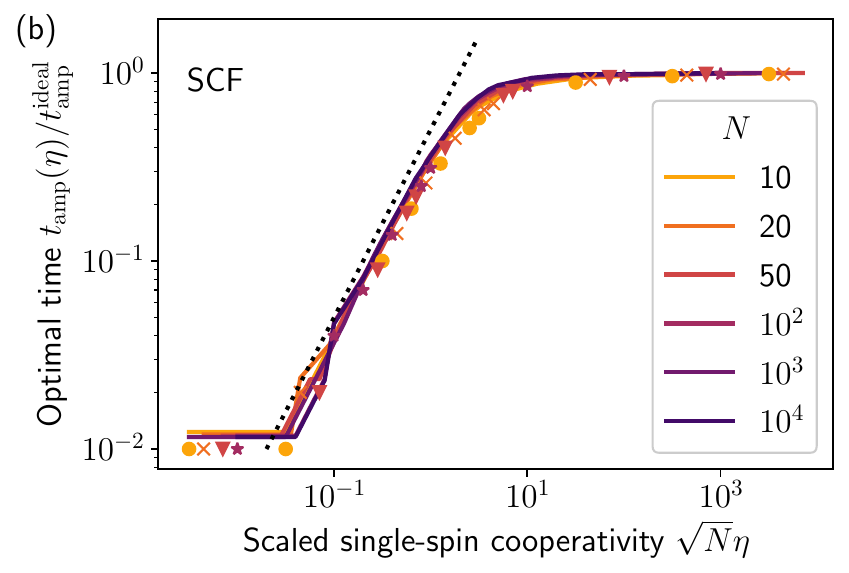}
	\caption{
		Additional details on the numerical maximization of the gain $G(\eta)$, which has been performed to generate the data for Fig.~\ref{fig:Amplification}. 
		(a) Optimal detuning $\delta/\kappa$ and (b) optimal evolution time $\tamp(\eta)$ relative to the ideal time $\tamp^\mathrm{ideal} = \lim_{\eta \to \infty} \tamp(\eta)$ in the absence of dissipation.
		The solid lines represent MFT results and the markers have been obtained by numerical integration of the QME~\eqref{eqn:QME:EffectiveGeneral}. 
		The dotted black line indicates $\delta/\kappa = \FigSCFSingleGainOptimalDetuningVsCooperativityFormulaAsymptote$ in (a) and $\tamp(\eta)/\tamp^\mathrm{ideal} = \FigSCFSingleGainOptimalTimeVsRescaledCooperativityFormulaAsymptote$ in (b). 
	}
	\label{fig:Amplification:AdditionalDetails}
\end{figure}

As discussed in Sec.~\ref{sec:Analysis:IdealProtocol}, we initialize the system in a coherent-spin state $\ket{\psi(\phi)}$ rotated away from the positive $x$-axis by a small signal $\phi \ll 1$, see Eq.~\eqref{eqn:InitialStatePhi}.
We evolve this state using the QME~\eqref{eqn:QME:EffectiveGeneral} for a time $\tamp$ using $\chimw = 1$ and different values of $\eta$ as well as $\delta/\kappa$. 
For each value of $\eta$, we maximize $G(\eta)$ over $\tamp$ and $\delta/\kappa$. 
Figure~\ref{fig:Amplification:AdditionalDetails} complements Fig.~\ref{fig:Amplification} and shows the resulting optimal values for $\delta/\kappa$ and $\tamp$.
For $\eta \ll 1$, single-spin dissipation is the dominant dissipative process, whose rate is minimized by choosing the optimal detuning $\delta/\kappa = 1/2$. 
With increasing single-spin cooperativity, the collective dephasing term becomes dominant and the optimal detuning scales as $\delta/\kappa \propto \sqrt{\eta}$. 
For $\sqrt{N} \eta \gtrsim 0.03$, the gain exceeds unity and the optimal amplification time increases, $\tamp \propto \sqrt{N} \eta$, until it converges to the ideal result $\tamp^\mathrm{ideal} = 1/\chi \sqrt{N}$ for large cooperativities.
For large $N$, the rescaled gain curves shown in Fig.~\ref{fig:Amplification} can be fitted to a hyperbolic tangent function of the form, 
\begin{align}
	f(\sqrt{N} \eta) = \frac{1}{2} \left( 1+ \tanh \left[  a \log_{10}(\sqrt{N} \eta) - b \right] \right) \comma
	\label{eqn:FittingFunction}
\end{align}
where $a \approx \FigSCFSingleGainRescaledGainVsRescaledCooperativityFitParameterA$ and $b \approx \FigSCFSingleGainRescaledGainVsRescaledCooperativityFitParameterB$.

As mentioned in Sec.~\ref{sec:Gain}, one may wonder if the gain physics described so far changes if one considers the full twist-untwist scheme proposed in Ref.~\onlinecite{Davis2016}. 
In this case, the initial state of the amplification step is a spin-squeezed state rotated about the $\hat{S}_y$ axis, and the ``untwist'' step both amplifies the signal and brings the fluctuations back to the level of a coherent-spin state. 
To check if the first ``twist'' step and the resulting squeezed quantum fluctuations affect the gain dynamics, we repeated the numerical optimization of the gain $G$ using a simulation of the full twist-untwist sensing protocol. 
We found no qualitative differences, i.e., different curves of the rescaled gain $G(\eta)/G_\mathrm{max}$ still collapse when plotted as a function of $\sqrt{N} \eta$, and they can be fitted to the hyperbolic tangent function shown in Eq.~\eqref{eqn:FittingFunction} with $a \approx \FigSCFdoubleGainRescaledGainVsRescaledCooperativityFitParameterA$ and $b \approx \FigSCFdoubleGainRescaledGainVsRescaledCooperativityFitParameterB$.
Moreover, the optimal detuning still scales asymptotially as $\delta/\kappa \propto \sqrt{\eta}$, and the optimal amplification time close to threshold scales as $\tamp(\eta)/\tamp^\mathrm{ideal} \propto \sqrt{N} \eta$.

\section{Additional details on the gain in the ACF scheme}
\label{sec:App:AdditonalDetails:ACF}

\begin{figure*}
	\centering
	\includegraphics[width=0.31\textwidth]{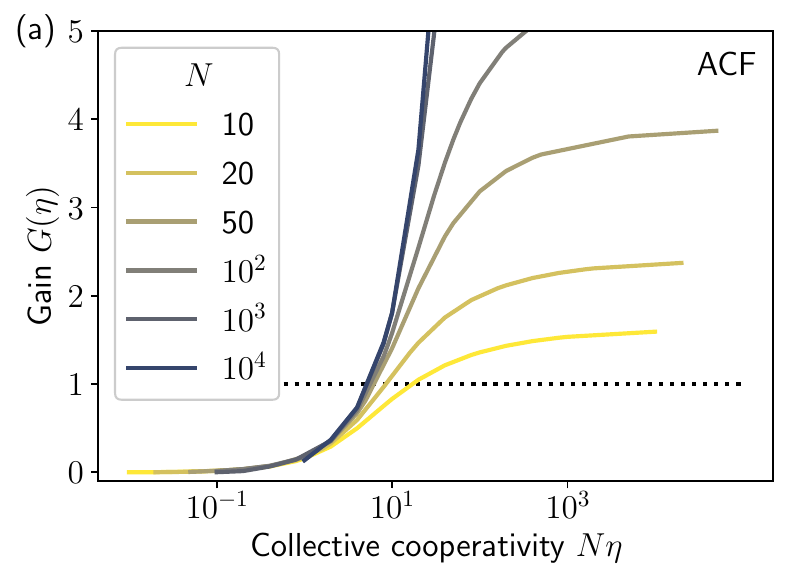}
	\includegraphics[width=0.32\textwidth]{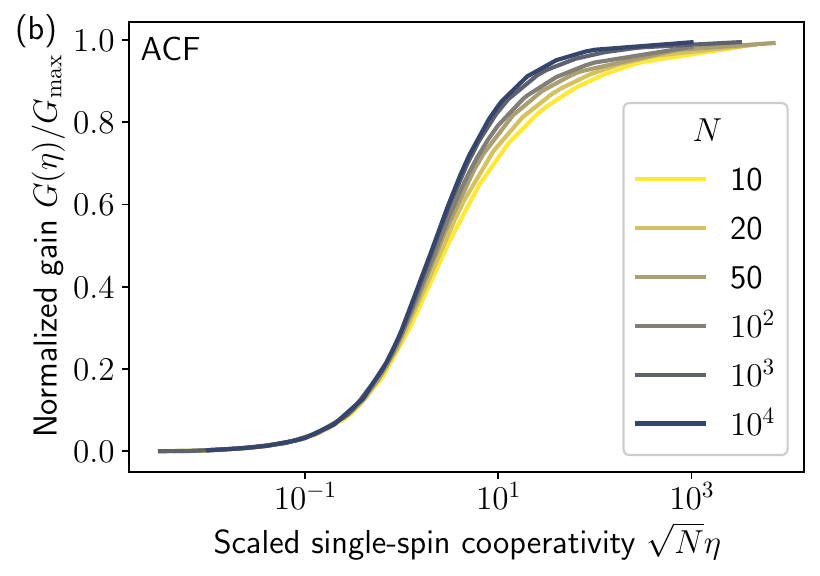}
	\includegraphics[width=0.325\textwidth]{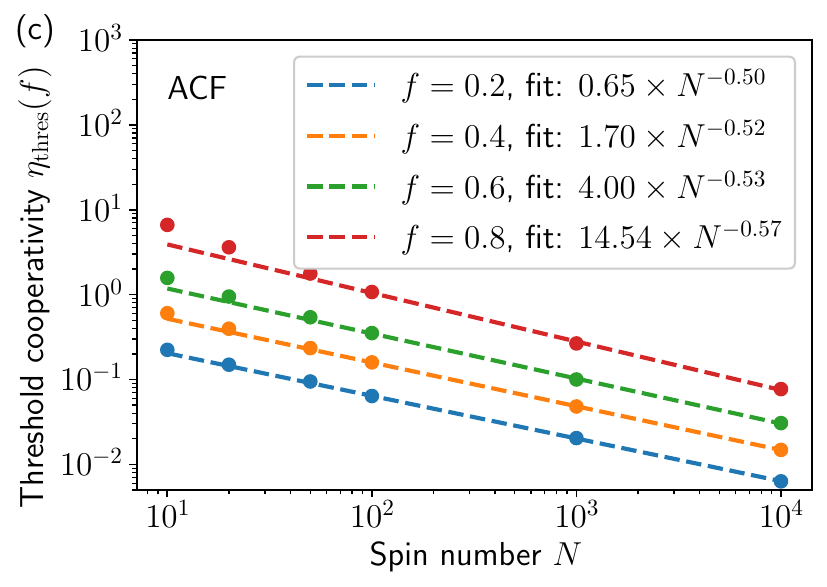}
	\includegraphics[width=0.315\textwidth]{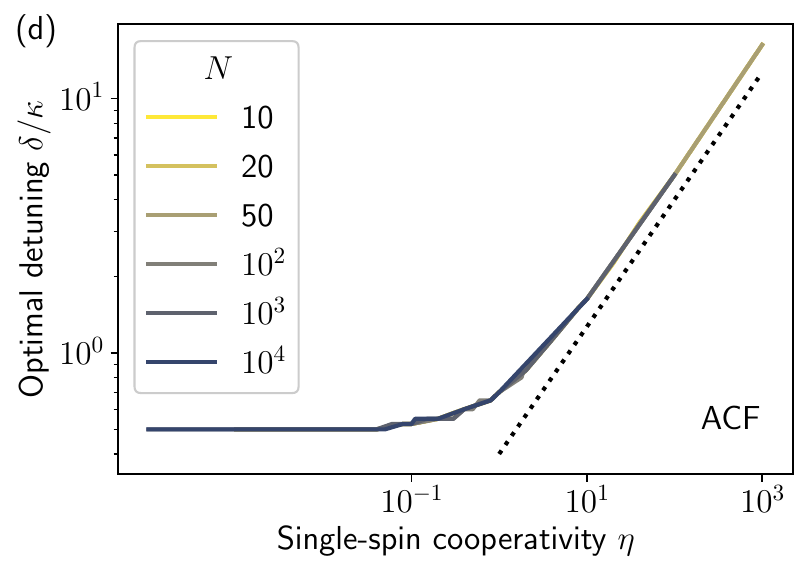}
	\includegraphics[width=0.33\textwidth]{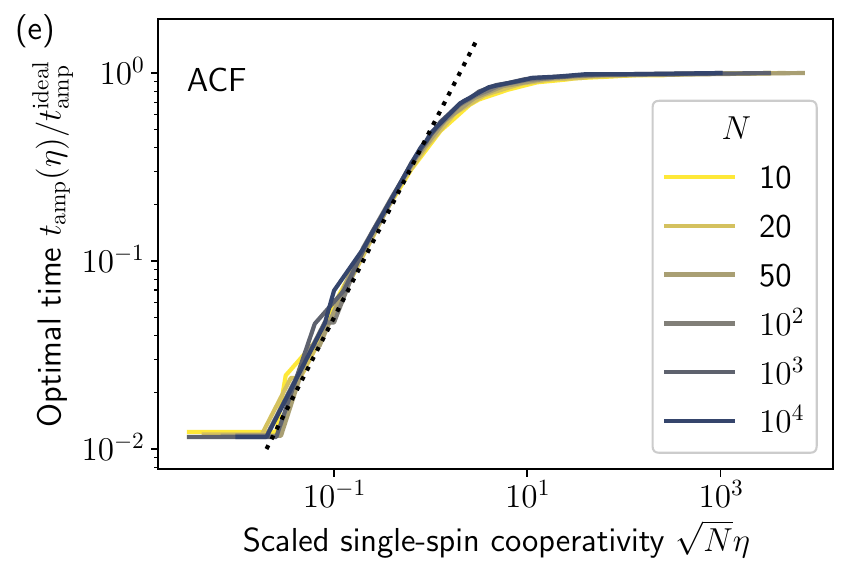}
	\caption{
		Same plots as in Figs.~\ref{fig:Amplification} 
		and~\ref{fig:Amplification:AdditionalDetails} based on MFT simulations of the ACF scheme. 
		The dotted black lines indicate a gain of unity in (a), $\delta/\kappa = \FigACFSingleGainOptimalDetuningVsCooperativityFormulaAsymptote$ in (d), and  $\tamp(\eta)/\tamp^\mathrm{ideal} = \FigACFSingleGainOptimalTimeVsRescaledCooperativityFormulaAsymptote$ in (e). 
	}
	\label{fig:Amplification:ACF}
\end{figure*}

In this Appendix, we analyze the maximum gain in the ACF scheme. 
Since only the level $\ket{\uparrow_j}$ couples to the auxiliary level $\ket{e_j}$, the decay of the auxiliary level cannot cause spin flips such that the single-spin excitation and relaxation rates are exactly zero (cf.\ Table~\ref{tbl:Rates}). 
Therefore, $\hat{S}_z$ is a constant of motion for the ACF scheme, in contrast to the SCF scheme analyzed in Sec.~\ref{sec:Gain:SCF}.
However, amplification generated by the unitary OAT dynamics still competes with collective and single-spin dephasing, which damp out any transverse polarization.

Figure~\ref{fig:Amplification:ACF} shows that the absence of single-spin excitation and relaxation processes in the ACF scheme does not change the amplification physics significantly. 
We still find that the normalized gain $G(\eta) / G_\mathrm{max}$ for different ensemble sizes $N$ collapses onto a single curve when plotted as a function of $\sqrt{N} \eta$.
Therefore, the ACF scheme has the same condition~\eqref{eqn:ConditionCavityFeedbackAmplification} to achieve significant levels of gain as the SCF scheme. 
Moreover, comparing Fig.~\ref{fig:Amplification:AdditionalDetails} with Figs.~\ref{fig:Amplification:ACF}(d) and~(e), we find that the scaling of the optimal detuning and the optimal amplification time is the same for the ACF and SCF schemes up to numerical prefactors.

\section{Mean-field theory equations of motion}
\label{sec:App:MFT}

In this Appendix, we list the complete set of MFT equations of motion that correspond to the QME~\eqref{eqn:QME:EffectiveGeneral}. 
As described in Sec.~\ref{sec:MFT}, they have been used to calculate all MFT curves in the different figures of this paper.

\begin{widetext}
\begin{subequations}%
\begin{align}%
	\frac{\d}{\d t} S_x 
		&=  - 2 \chi (C_{yz} + S_y S_z) 
		- \frac{\Gammaphi + \gammalocm + \gammalocp + 4 \gammalocz}{2} S_x 
		+  \Gammarel \left[ C_{xz} + \left(S_z - \frac{1}{2} \right) S_x \right] \comma \\
	\frac{\d}{\d t} S_y 
		&= + 2 \chi (C_{xz} + S_x S_z) 
		- \frac{\Gammaphi + \gammalocm + \gammalocp + 4 \gammalocz}{2} S_y 
		+ \Gammarel \left[ C_{yz} + \left(S_z - \frac{1}{2} \right) S_y \right] \comma \\
	\frac{\d}{\d t} S_z 
		&= - \gammalocm \left( S_z + \frac{N}{2} \right) 
		- \gammalocp \left( S_z - \frac{N}{2} \right) 
		- \Gammarel \left( C_{xx} + C_{yy} + S_x^2 + S_y^2 + S_z \right) \comma \displaybreak[1]\\
	\frac{\d}{\d t} C_{xx} 
		&= - 4 \chi \left( C_{xz} S_y + C_{xy} S_z \right) 
		- \left( \gammalocp + \gammalocm + 4 \gammalocz \right) \left( C_{xx} - \frac{N}{4} \right) \nonumber \\
		&\phantom{=}\ + \Gammaphi \left( C_{yy} - C_{xx} +  S_y^2 \right) 
		+ \Gammarel \left[ C_{zz} - C_{xx} + S_z^2 - \frac{S_z}{2} + 2 \left( C_{xz} S_x + C_{xx} S_z \right) \right] \comma \displaybreak[1] \\
	\frac{\d}{\d t} C_{xy} 
		&= + 2 \chi \left( C_{xz} S_x - C_{yz} S_y + C_{xx} S_z - C_{yy} S_z \right) 
		- \left( \gammalocm + \gammalocp + 4 \gammalocz \right) C_{xy} \nonumber \\
		&\phantom{=}\ - \Gammaphi \left( 2 C_{xy} + S_x S_y \right) 
		+ \Gammarel \left( C_{yz} S_x + C_{xz} S_y + C_{xy} S_z - C_{xy} \right)  \comma \displaybreak[1]\\
	\frac{\d}{\d t} C_{xz} 
		&= - 2 \chi \left( C_{zz} S_y + C_{yz} S_z - \frac{S_y}{4} \right) 
		- \gammalocp \left( \frac{3}{2} C_{xz} + \frac{S_x}{2} \right) 
		- \gammalocm \left( \frac{3}{2} C_{xz} - \frac{S_x}{2} \right) 
		- 2 \gammalocz C_{xz} \nonumber \\
		&\phantom{=}\ - \frac{\Gammaphi}{2} C_{xz} 
		- \Gammarel \left( 2 C_{xx} S_x  + 2 C_{xy} S_y - C_{zz} S_x - C_{xz} S_z + \frac{5}{2} C_{xz} - \frac{S_x}{4}  + S_x S_z \right) \comma \displaybreak[1]\\
	\frac{\d}{\d t} C_{yy} 
		&= + 4 \chi \left( C_{yz} S_x + C_{xy} S_z \right)
		- \left( \gammalocm + \gammalocp + 4 \gammalocz \right) \left( C_{yy} - \frac{N}{4} \right) \nonumber \\
		&\phantom{=}\ + \Gammaphi \left( C_{xx} - C_{yy} + S_x^2 \right) 
		+ \Gammarel \left[ C_{zz} - C_{yy} + S_z^2 - \frac{S_z}{2} + 2 \left( C_{yz} S_y + C_{yy} S_z \right) \right] \comma \displaybreak[1]\\
	\frac{\d}{\d t} C_{yz}
		&= + 2 \chi \left( C_{zz} S_x + C_{xz} S_z - \frac{S_x}{4} \right) 
		- \gammalocp \left( \frac{3}{2} C_{yz} + \frac{S_y}{2} \right) 
		- \gammalocm \left( \frac{3}{2} C_{yz} - \frac{S_y}{2} \right) 
		- 2 \gammalocz C_{yz} \nonumber \\
		&\phantom{=}\  - \frac{\Gammaphi}{2} C_{yz}  
		- \Gammarel \left( 2 C_{yy} S_y + 2 C_{xy} S_x - C_{zz} S_y - C_{yz} S_z + \frac{5}{2} C_{yz} - \frac{S_y}{4} + S_y S_z \right) \comma \displaybreak[1]\\
	\frac{\d}{\d t} C_{zz} 
		&= - \gammalocp \left( 2 C_{zz} + S_z - \frac{N}{2} \right)
		- \gammalocm \left( 2 C_{zz} - S_z - \frac{N}{2} \right) \nonumber \\
		&\phantom{=}\ - \Gammarel \left[ 2 C_{zz} - C_{xx} - C_{yy} - S_x^2 - S_y^2 - S_z - 4 \left( C_{xz} S_x + C_{yz} S_y \right) \right] \fullstop
\end{align}%
\label{eqn:MFT:Full}%
\end{subequations}%
\end{widetext}

\section{Additional details on the optimization of the metrological gain}
\label{sec:App:MetrologicalGain}

\begin{figure}
	\centering
	\includegraphics[width=0.23\textwidth]{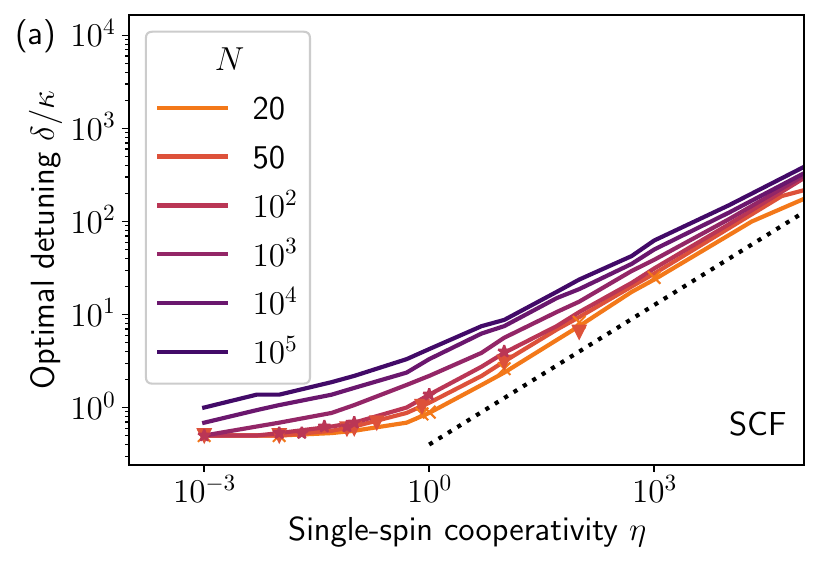}
	\includegraphics[width=0.24\textwidth]{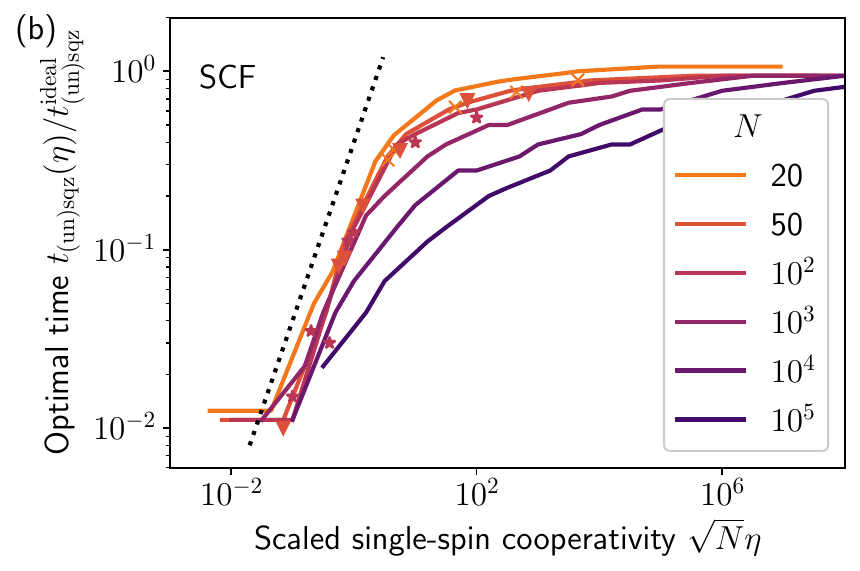}
	\includegraphics[width=0.23\textwidth]{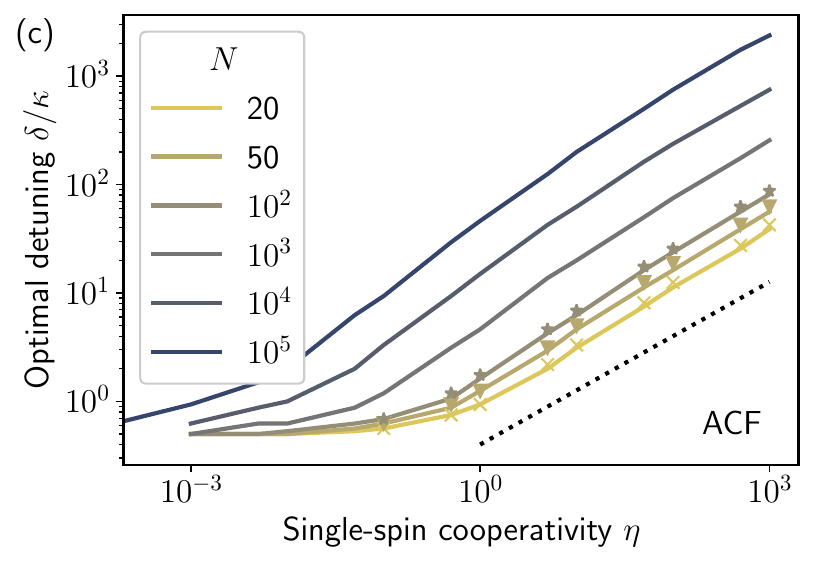}
	\includegraphics[width=0.24\textwidth]{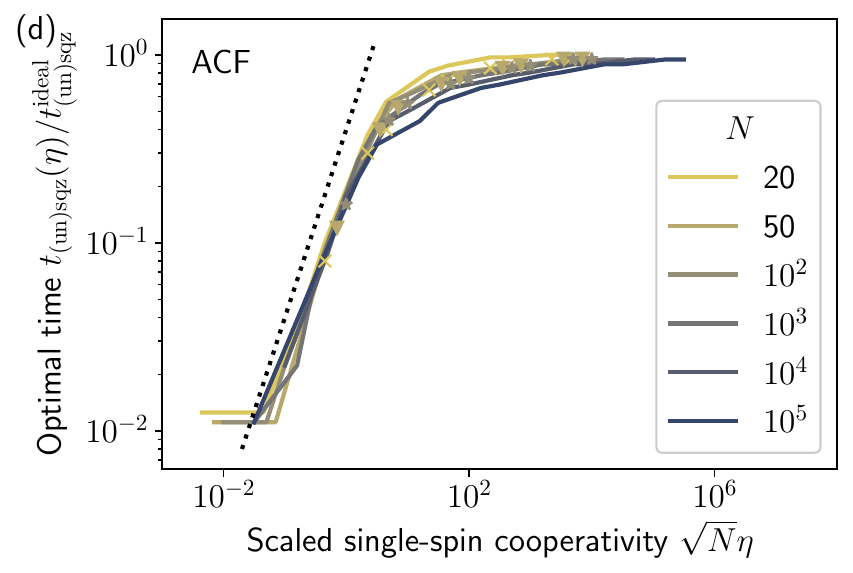}
	\caption{
		Additional details on the numerical maximization of the metrological gain $\Gmet$, which has been performed to generate the data for Fig.~\ref{fig:MetrologicalGain}.
		(a) Optimal detuning $\delta/\kappa$ and 
		(b) optimal evolution time $t_\mathrm{(un)sqz}(\eta)$ relative to the ideal time $t_\mathrm{(un)sqz}^\mathrm{ideal} = \lim_{\eta \to \infty} t_\mathrm{(un)sqz}(\eta)$ in the absence of dissipation for the SCF scheme.
		(c) Optimal detuning $\delta/\kappa$ and 
		(d) optimal evolution time $t_\mathrm{(un)sqz}(\eta)$ for the ACF scheme. 
		The dotted black line indicates $\delta/\kappa = 0.4 \sqrt{\eta}$ in (a,c) and $t_\mathrm{amp}(\eta) / t_\mathrm{amp}^\mathrm{ideal} = 0.4 \sqrt{N} \eta$ in (b,d). 
	}
	\label{fig:MetrologicalGain:OptimalParameters}
\end{figure}

In this Appendix, we provide additional information on the numerical optimization of the metrological gain $\Gmet(\eta)$ shown in Fig.~\ref{fig:MetrologicalGain}. 
The MFT results, represented by the solid lines in Fig.~\ref{fig:MetrologicalGain}, have been obtained as follows.
We initialize the system in a coherent-spin state pointing along the $x$ axis, i.e., $S_x = N/2$, $C_{yy} = C_{zz} = N/4$, and all other moments and (co)variances are set to zero.
We use the MFT equations of motion listed in Appendix~\ref{sec:App:MFT} to evolve this state for a time $\tsqz$ with a positive OAT interaction strength, $\chi > 0$. 
We then apply the signal $\phi$ by rotating the resulting spin-squeezed state about the $\hat{S}_y$ axis, as defined in Eq.~\eqref{eqn:rotatedrho2}. 
Subsequently, we reverse the sign of the OAT interaction, $\chi \to -\chi$ and untwist the state by integrating the MFT equations of motion for a time $\tunsqz = \tsqz$.
Based on the final values of $S_y(t_\mathrm{end})$ and $C_{yy}(t_\mathrm{end}) = (\bfDelta S_y)^2(t_\mathrm{end})$ at time  $t_\mathrm{end} = \tsqz + \tunsqz = 2 t_\mathrm{(un)sqz}$, the gain $G$ and the enhanced fluctuations $\sigma_\mathrm{diss}^2$ can be calculated using Eqs.~\eqref{eqn:GainForGmet} and~\eqref{eqn:DefinitionSigmadisssq}, respectively. 
This procedure is repeated for different values of $\eta$ and $\delta/\kappa$ such that we can maximize $\Gmet(\eta)$, defined in Eq.~\eqref{eqn:Definition:Gmeteta}, over $t_\mathrm{(un)sqz}$ and $\lambda = \delta/\kappa$. 
The corresponding optimal values are shown in Fig.~\ref{fig:MetrologicalGain:OptimalParameters}.

For the MFT result of the TC scheme, shown in Fig.~\ref{fig:comparison}, we need to work with a background-subtracted gain analogous to Eq.~\eqref{eqn:Gsub}, $G^\mathrm{sub} = [S_y(t_\mathrm{end},\phi) - S_y(t_\mathrm{end},0)] / (N \phi/2)$ for $\phi \ll 1$. 
In analogy to Eq.~\eqref{eqn:Definition:Gmeteta}, the corresponding metrological gain is given by
\begin{align}
	\Gmet^\mathrm{sub}(\eta) = \max_{t_\mathrm{(un)sqz}} \max_{\lambda} \frac{(G^\mathrm{sub})^2}{1 + \sigma_\mathrm{diss}^2 + \Xidetsq} \fullstop
\end{align}

\bibliography{citations}

\begin{thebibliography}{40}%
\makeatletter
\providecommand \@ifxundefined [1]{%
 \@ifx{#1\undefined}
}%
\providecommand \@ifnum [1]{%
 \ifnum #1\expandafter \@firstoftwo
 \else \expandafter \@secondoftwo
 \fi
}%
\providecommand \@ifx [1]{%
 \ifx #1\expandafter \@firstoftwo
 \else \expandafter \@secondoftwo
 \fi
}%
\providecommand \natexlab [1]{#1}%
\providecommand \enquote  [1]{``#1''}%
\providecommand \bibnamefont  [1]{#1}%
\providecommand \bibfnamefont [1]{#1}%
\providecommand \citenamefont [1]{#1}%
\providecommand \href@noop [0]{\@secondoftwo}%
\providecommand \href [0]{\begingroup \@sanitize@url \@href}%
\providecommand \@href[1]{\@@startlink{#1}\@@href}%
\providecommand \@@href[1]{\endgroup#1\@@endlink}%
\providecommand \@sanitize@url [0]{\catcode `\\12\catcode `\$12\catcode
  `\&12\catcode `\#12\catcode `\^12\catcode `\_12\catcode `\%12\relax}%
\providecommand \@@startlink[1]{}%
\providecommand \@@endlink[0]{}%
\providecommand \url  [0]{\begingroup\@sanitize@url \@url }%
\providecommand \@url [1]{\endgroup\@href {#1}{\urlprefix }}%
\providecommand \urlprefix  [0]{URL }%
\providecommand \Eprint [0]{\href }%
\providecommand \doibase [0]{https://doi.org/}%
\providecommand \selectlanguage [0]{\@gobble}%
\providecommand \bibinfo  [0]{\@secondoftwo}%
\providecommand \bibfield  [0]{\@secondoftwo}%
\providecommand \translation [1]{[#1]}%
\providecommand \BibitemOpen [0]{}%
\providecommand \bibitemStop [0]{}%
\providecommand \bibitemNoStop [0]{.\EOS\space}%
\providecommand \EOS [0]{\spacefactor3000\relax}%
\providecommand \BibitemShut  [1]{\csname bibitem#1\endcsname}%
\let\auto@bib@innerbib\@empty
\bibitem [{\citenamefont {Giovannetti}\ \emph {et~al.}(2006)\citenamefont
  {Giovannetti}, \citenamefont {Lloyd},\ and\ \citenamefont
  {Maccone}}]{Giovannetti2006}%
  \BibitemOpen
  \bibfield  {author} {\bibinfo {author} {\bibfnamefont {V.}~\bibnamefont
  {Giovannetti}}, \bibinfo {author} {\bibfnamefont {S.}~\bibnamefont {Lloyd}},\
  and\ \bibinfo {author} {\bibfnamefont {L.}~\bibnamefont {Maccone}},\
  }\bibfield  {title} {\bibinfo {title} {Quantum metrology},\ }\href
  {https://doi.org/10.1103/PhysRevLett.96.010401} {\bibfield  {journal}
  {\bibinfo  {journal} {Phys. Rev. Lett.}\ }\textbf {\bibinfo {volume} {96}},\
  \bibinfo {pages} {010401} (\bibinfo {year} {2006})}\BibitemShut {NoStop}%
\bibitem [{\citenamefont {Degen}\ \emph {et~al.}(2017)\citenamefont {Degen},
  \citenamefont {Reinhard},\ and\ \citenamefont {Cappellaro}}]{Degen2017}%
  \BibitemOpen
  \bibfield  {author} {\bibinfo {author} {\bibfnamefont {C.~L.}\ \bibnamefont
  {Degen}}, \bibinfo {author} {\bibfnamefont {F.}~\bibnamefont {Reinhard}},\
  and\ \bibinfo {author} {\bibfnamefont {P.}~\bibnamefont {Cappellaro}},\
  }\bibfield  {title} {\bibinfo {title} {Quantum sensing},\ }\href
  {https://doi.org/10.1103/RevModPhys.89.035002} {\bibfield  {journal}
  {\bibinfo  {journal} {Rev. Mod. Phys.}\ }\textbf {\bibinfo {volume} {89}},\
  \bibinfo {pages} {035002} (\bibinfo {year} {2017})}\BibitemShut {NoStop}%
\bibitem [{\citenamefont {Pezz\`e}\ \emph {et~al.}(2018)\citenamefont
  {Pezz\`e}, \citenamefont {Smerzi}, \citenamefont {Oberthaler}, \citenamefont
  {Schmied},\ and\ \citenamefont {Treutlein}}]{Pezze2018}%
  \BibitemOpen
  \bibfield  {author} {\bibinfo {author} {\bibfnamefont {L.}~\bibnamefont
  {Pezz\`e}}, \bibinfo {author} {\bibfnamefont {A.}~\bibnamefont {Smerzi}},
  \bibinfo {author} {\bibfnamefont {M.~K.}\ \bibnamefont {Oberthaler}},
  \bibinfo {author} {\bibfnamefont {R.}~\bibnamefont {Schmied}},\ and\ \bibinfo
  {author} {\bibfnamefont {P.}~\bibnamefont {Treutlein}},\ }\bibfield  {title}
  {\bibinfo {title} {Quantum metrology with nonclassical states of atomic
  ensembles},\ }\href {https://doi.org/10.1103/RevModPhys.90.035005} {\bibfield
   {journal} {\bibinfo  {journal} {Rev. Mod. Phys.}\ }\textbf {\bibinfo
  {volume} {90}},\ \bibinfo {pages} {035005} (\bibinfo {year}
  {2018})}\BibitemShut {NoStop}%
\bibitem [{\citenamefont {Kitagawa}\ and\ \citenamefont
  {Ueda}(1993)}]{Kitagawa1993}%
  \BibitemOpen
  \bibfield  {author} {\bibinfo {author} {\bibfnamefont {M.}~\bibnamefont
  {Kitagawa}}\ and\ \bibinfo {author} {\bibfnamefont {M.}~\bibnamefont
  {Ueda}},\ }\bibfield  {title} {\bibinfo {title} {Squeezed spin states},\
  }\href {https://doi.org/10.1103/PhysRevA.47.5138} {\bibfield  {journal}
  {\bibinfo  {journal} {Phys. Rev. A}\ }\textbf {\bibinfo {volume} {47}},\
  \bibinfo {pages} {5138} (\bibinfo {year} {1993})}\BibitemShut {NoStop}%
\bibitem [{\citenamefont {Bollinger}\ \emph {et~al.}(1996)\citenamefont
  {Bollinger}, \citenamefont {Itano}, \citenamefont {Wineland},\ and\
  \citenamefont {Heinzen}}]{Bollinger1996}%
  \BibitemOpen
  \bibfield  {author} {\bibinfo {author} {\bibfnamefont {J.~J.}\ \bibnamefont
  {Bollinger}}, \bibinfo {author} {\bibfnamefont {W.~M.}\ \bibnamefont
  {Itano}}, \bibinfo {author} {\bibfnamefont {D.~J.}\ \bibnamefont
  {Wineland}},\ and\ \bibinfo {author} {\bibfnamefont {D.~J.}\ \bibnamefont
  {Heinzen}},\ }\bibfield  {title} {\bibinfo {title} {Optimal frequency
  measurements with maximally correlated states},\ }\href
  {https://doi.org/10.1103/PhysRevA.54.R4649} {\bibfield  {journal} {\bibinfo
  {journal} {Phys. Rev. A}\ }\textbf {\bibinfo {volume} {54}},\ \bibinfo
  {pages} {R4649} (\bibinfo {year} {1996})}\BibitemShut {NoStop}%
\bibitem [{\citenamefont {Leroux}\ \emph {et~al.}(2010)\citenamefont {Leroux},
  \citenamefont {Schleier-Smith},\ and\ \citenamefont
  {Vuleti\'{c}}}]{Leroux2010}%
  \BibitemOpen
  \bibfield  {author} {\bibinfo {author} {\bibfnamefont {I.~D.}\ \bibnamefont
  {Leroux}}, \bibinfo {author} {\bibfnamefont {M.~H.}\ \bibnamefont
  {Schleier-Smith}},\ and\ \bibinfo {author} {\bibfnamefont {V.}~\bibnamefont
  {Vuleti\'{c}}},\ }\bibfield  {title} {\bibinfo {title} {Implementation of
  cavity squeezing of a collective atomic spin},\ }\href
  {https://doi.org/10.1103/PhysRevLett.104.073602} {\bibfield  {journal}
  {\bibinfo  {journal} {Phys. Rev. Lett.}\ }\textbf {\bibinfo {volume} {104}},\
  \bibinfo {pages} {073602} (\bibinfo {year} {2010})}\BibitemShut {NoStop}%
\bibitem [{\citenamefont {Gross}\ \emph {et~al.}(2010)\citenamefont {Gross},
  \citenamefont {Zibold}, \citenamefont {Nicklas}, \citenamefont {Est{\`e}ve},\
  and\ \citenamefont {Oberthaler}}]{Gross2010}%
  \BibitemOpen
  \bibfield  {author} {\bibinfo {author} {\bibfnamefont {C.}~\bibnamefont
  {Gross}}, \bibinfo {author} {\bibfnamefont {T.}~\bibnamefont {Zibold}},
  \bibinfo {author} {\bibfnamefont {E.}~\bibnamefont {Nicklas}}, \bibinfo
  {author} {\bibfnamefont {J.}~\bibnamefont {Est{\`e}ve}},\ and\ \bibinfo
  {author} {\bibfnamefont {M.~K.}\ \bibnamefont {Oberthaler}},\ }\bibfield
  {title} {\bibinfo {title} {Nonlinear atom interferometer surpasses classical
  precision limit},\ }\href {https://doi.org/10.1038/nature08919} {\bibfield
  {journal} {\bibinfo  {journal} {Nature}\ }\textbf {\bibinfo {volume} {464}},\
  \bibinfo {pages} {1165} (\bibinfo {year} {2010})}\BibitemShut {NoStop}%
\bibitem [{\citenamefont {Riedel}\ \emph {et~al.}(2010)\citenamefont {Riedel},
  \citenamefont {B{\"o}hi}, \citenamefont {Li}, \citenamefont {H{\"a}nsch},
  \citenamefont {Sinatra},\ and\ \citenamefont {Treutlein}}]{Riedel2010}%
  \BibitemOpen
  \bibfield  {author} {\bibinfo {author} {\bibfnamefont {M.~F.}\ \bibnamefont
  {Riedel}}, \bibinfo {author} {\bibfnamefont {P.}~\bibnamefont {B{\"o}hi}},
  \bibinfo {author} {\bibfnamefont {Y.}~\bibnamefont {Li}}, \bibinfo {author}
  {\bibfnamefont {T.~W.}\ \bibnamefont {H{\"a}nsch}}, \bibinfo {author}
  {\bibfnamefont {A.}~\bibnamefont {Sinatra}},\ and\ \bibinfo {author}
  {\bibfnamefont {P.}~\bibnamefont {Treutlein}},\ }\bibfield  {title} {\bibinfo
  {title} {Atom-chip-based generation of entanglement for quantum metrology},\
  }\href {https://doi.org/10.1038/nature08988} {\bibfield  {journal} {\bibinfo
  {journal} {Nature}\ }\textbf {\bibinfo {volume} {464}},\ \bibinfo {pages}
  {1170} (\bibinfo {year} {2010})}\BibitemShut {NoStop}%
\bibitem [{\citenamefont {Strobel}\ \emph {et~al.}(2014)\citenamefont
  {Strobel}, \citenamefont {Muessel}, \citenamefont {Linnemann}, \citenamefont
  {Zibold}, \citenamefont {Hume}, \citenamefont {Pezz{\`e}}, \citenamefont
  {Smerzi},\ and\ \citenamefont {Oberthaler}}]{Strobel2014}%
  \BibitemOpen
  \bibfield  {author} {\bibinfo {author} {\bibfnamefont {H.}~\bibnamefont
  {Strobel}}, \bibinfo {author} {\bibfnamefont {W.}~\bibnamefont {Muessel}},
  \bibinfo {author} {\bibfnamefont {D.}~\bibnamefont {Linnemann}}, \bibinfo
  {author} {\bibfnamefont {T.}~\bibnamefont {Zibold}}, \bibinfo {author}
  {\bibfnamefont {D.~B.}\ \bibnamefont {Hume}}, \bibinfo {author}
  {\bibfnamefont {L.}~\bibnamefont {Pezz{\`e}}}, \bibinfo {author}
  {\bibfnamefont {A.}~\bibnamefont {Smerzi}},\ and\ \bibinfo {author}
  {\bibfnamefont {M.~K.}\ \bibnamefont {Oberthaler}},\ }\bibfield  {title}
  {\bibinfo {title} {Fisher information and entanglement of non-gaussian spin
  states},\ }\href {https://doi.org/10.1126/science.1250147} {\bibfield
  {journal} {\bibinfo  {journal} {Science}\ }\textbf {\bibinfo {volume}
  {345}},\ \bibinfo {pages} {424} (\bibinfo {year} {2014})}\BibitemShut
  {NoStop}%
\bibitem [{\citenamefont {Muessel}\ \emph {et~al.}(2015)\citenamefont
  {Muessel}, \citenamefont {Strobel}, \citenamefont {Linnemann}, \citenamefont
  {Zibold}, \citenamefont {Juli\'a-D\'{\i}az},\ and\ \citenamefont
  {Oberthaler}}]{Muessel2015}%
  \BibitemOpen
  \bibfield  {author} {\bibinfo {author} {\bibfnamefont {W.}~\bibnamefont
  {Muessel}}, \bibinfo {author} {\bibfnamefont {H.}~\bibnamefont {Strobel}},
  \bibinfo {author} {\bibfnamefont {D.}~\bibnamefont {Linnemann}}, \bibinfo
  {author} {\bibfnamefont {T.}~\bibnamefont {Zibold}}, \bibinfo {author}
  {\bibfnamefont {B.}~\bibnamefont {Juli\'a-D\'{\i}az}},\ and\ \bibinfo
  {author} {\bibfnamefont {M.~K.}\ \bibnamefont {Oberthaler}},\ }\bibfield
  {title} {\bibinfo {title} {Twist-and-turn spin squeezing in bose-einstein
  condensates},\ }\href {https://doi.org/10.1103/PhysRevA.92.023603} {\bibfield
   {journal} {\bibinfo  {journal} {Phys. Rev. A}\ }\textbf {\bibinfo {volume}
  {92}},\ \bibinfo {pages} {023603} (\bibinfo {year} {2015})}\BibitemShut
  {NoStop}%
\bibitem [{\citenamefont {Hosten}\ \emph
  {et~al.}(2016{\natexlab{a}})\citenamefont {Hosten}, \citenamefont
  {Krishnakumar}, \citenamefont {Engelsen},\ and\ \citenamefont
  {Kasevich}}]{Hosten2016}%
  \BibitemOpen
  \bibfield  {author} {\bibinfo {author} {\bibfnamefont {O.}~\bibnamefont
  {Hosten}}, \bibinfo {author} {\bibfnamefont {R.}~\bibnamefont
  {Krishnakumar}}, \bibinfo {author} {\bibfnamefont {N.~J.}\ \bibnamefont
  {Engelsen}},\ and\ \bibinfo {author} {\bibfnamefont {M.~A.}\ \bibnamefont
  {Kasevich}},\ }\bibfield  {title} {\bibinfo {title} {Quantum phase
  magnification},\ }\href {https://doi.org/10.1126/science.aaf3397} {\bibfield
  {journal} {\bibinfo  {journal} {Science}\ }\textbf {\bibinfo {volume}
  {352}},\ \bibinfo {pages} {1552} (\bibinfo {year}
  {2016}{\natexlab{a}})}\BibitemShut {NoStop}%
\bibitem [{\citenamefont {Hosten}\ \emph
  {et~al.}(2016{\natexlab{b}})\citenamefont {Hosten}, \citenamefont {Engelsen},
  \citenamefont {Krishnakumar},\ and\ \citenamefont {Kasevich}}]{Hosten2016b}%
  \BibitemOpen
  \bibfield  {author} {\bibinfo {author} {\bibfnamefont {O.}~\bibnamefont
  {Hosten}}, \bibinfo {author} {\bibfnamefont {N.~J.}\ \bibnamefont
  {Engelsen}}, \bibinfo {author} {\bibfnamefont {R.}~\bibnamefont
  {Krishnakumar}},\ and\ \bibinfo {author} {\bibfnamefont {M.~A.}\ \bibnamefont
  {Kasevich}},\ }\bibfield  {title} {\bibinfo {title} {Measurement noise 100
  times lower than the quantum-projection limit using entangled atoms},\ }\href
  {https://doi.org/10.1038/nature16176} {\bibfield  {journal} {\bibinfo
  {journal} {Nature}\ }\textbf {\bibinfo {volume} {529}},\ \bibinfo {pages}
  {505} (\bibinfo {year} {2016}{\natexlab{b}})}\BibitemShut {NoStop}%
\bibitem [{\citenamefont {Bohnet}\ \emph {et~al.}(2016)\citenamefont {Bohnet},
  \citenamefont {Sawyer}, \citenamefont {Britton}, \citenamefont {Wall},
  \citenamefont {Rey}, \citenamefont {Foss-Feig},\ and\ \citenamefont
  {Bollinger}}]{Bohnet2016}%
  \BibitemOpen
  \bibfield  {author} {\bibinfo {author} {\bibfnamefont {J.~G.}\ \bibnamefont
  {Bohnet}}, \bibinfo {author} {\bibfnamefont {B.~C.}\ \bibnamefont {Sawyer}},
  \bibinfo {author} {\bibfnamefont {J.~W.}\ \bibnamefont {Britton}}, \bibinfo
  {author} {\bibfnamefont {M.~L.}\ \bibnamefont {Wall}}, \bibinfo {author}
  {\bibfnamefont {A.~M.}\ \bibnamefont {Rey}}, \bibinfo {author} {\bibfnamefont
  {M.}~\bibnamefont {Foss-Feig}},\ and\ \bibinfo {author} {\bibfnamefont
  {J.~J.}\ \bibnamefont {Bollinger}},\ }\bibfield  {title} {\bibinfo {title}
  {Quantum spin dynamics and entanglement generation with hundreds of trapped
  ions},\ }\href {https://doi.org/10.1126/science.aad9958} {\bibfield
  {journal} {\bibinfo  {journal} {Science}\ }\textbf {\bibinfo {volume}
  {352}},\ \bibinfo {pages} {1297} (\bibinfo {year} {2016})}\BibitemShut
  {NoStop}%
\bibitem [{\citenamefont {Braverman}\ \emph {et~al.}(2019)\citenamefont
  {Braverman}, \citenamefont {Kawasaki}, \citenamefont {Pedrozo-Pe\~nafiel},
  \citenamefont {Colombo}, \citenamefont {Shu}, \citenamefont {Li},
  \citenamefont {Mendez}, \citenamefont {Yamoah}, \citenamefont {Salvi},
  \citenamefont {Akamatsu}, \citenamefont {Xiao},\ and\ \citenamefont
  {Vuleti\'{c}}}]{Braverman2019}%
  \BibitemOpen
  \bibfield  {author} {\bibinfo {author} {\bibfnamefont {B.}~\bibnamefont
  {Braverman}}, \bibinfo {author} {\bibfnamefont {A.}~\bibnamefont {Kawasaki}},
  \bibinfo {author} {\bibfnamefont {E.}~\bibnamefont {Pedrozo-Pe\~nafiel}},
  \bibinfo {author} {\bibfnamefont {S.}~\bibnamefont {Colombo}}, \bibinfo
  {author} {\bibfnamefont {C.}~\bibnamefont {Shu}}, \bibinfo {author}
  {\bibfnamefont {Z.}~\bibnamefont {Li}}, \bibinfo {author} {\bibfnamefont
  {E.}~\bibnamefont {Mendez}}, \bibinfo {author} {\bibfnamefont
  {M.}~\bibnamefont {Yamoah}}, \bibinfo {author} {\bibfnamefont
  {L.}~\bibnamefont {Salvi}}, \bibinfo {author} {\bibfnamefont
  {D.}~\bibnamefont {Akamatsu}}, \bibinfo {author} {\bibfnamefont
  {Y.}~\bibnamefont {Xiao}},\ and\ \bibinfo {author} {\bibfnamefont
  {V.}~\bibnamefont {Vuleti\'{c}}},\ }\bibfield  {title} {\bibinfo {title}
  {Near-unitary spin squeezing in $^{171}\mathrm{Yb}$},\ }\href
  {https://doi.org/10.1103/PhysRevLett.122.223203} {\bibfield  {journal}
  {\bibinfo  {journal} {Phys. Rev. Lett.}\ }\textbf {\bibinfo {volume} {122}},\
  \bibinfo {pages} {223203} (\bibinfo {year} {2019})}\BibitemShut {NoStop}%
\bibitem [{\citenamefont {Pedrozo-Penafiel}\ \emph {et~al.}(2020)\citenamefont
  {Pedrozo-Penafiel}, \citenamefont {Colombo}, \citenamefont {Shu},
  \citenamefont {Adiyatullin}, \citenamefont {Li}, \citenamefont {Mendez},
  \citenamefont {Braverman}, \citenamefont {Kawasaki}, \citenamefont
  {Akamatsu}, \citenamefont {Xiao},\ and\ \citenamefont
  {Vuleti\'{c}}}]{PedrozoPenafiel2020}%
  \BibitemOpen
  \bibfield  {author} {\bibinfo {author} {\bibfnamefont {E.}~\bibnamefont
  {Pedrozo-Penafiel}}, \bibinfo {author} {\bibfnamefont {S.}~\bibnamefont
  {Colombo}}, \bibinfo {author} {\bibfnamefont {C.}~\bibnamefont {Shu}},
  \bibinfo {author} {\bibfnamefont {A.~F.}\ \bibnamefont {Adiyatullin}},
  \bibinfo {author} {\bibfnamefont {Z.}~\bibnamefont {Li}}, \bibinfo {author}
  {\bibfnamefont {E.}~\bibnamefont {Mendez}}, \bibinfo {author} {\bibfnamefont
  {B.}~\bibnamefont {Braverman}}, \bibinfo {author} {\bibfnamefont
  {A.}~\bibnamefont {Kawasaki}}, \bibinfo {author} {\bibfnamefont
  {D.}~\bibnamefont {Akamatsu}}, \bibinfo {author} {\bibfnamefont
  {Y.}~\bibnamefont {Xiao}},\ and\ \bibinfo {author} {\bibfnamefont
  {V.}~\bibnamefont {Vuleti\'{c}}},\ }\bibfield  {title} {\bibinfo {title}
  {Entanglement on an optical atomic-clock transition},\ }\href
  {https://doi.org/10.1038/s41586-020-3006-1} {\bibfield  {journal} {\bibinfo
  {journal} {Nature}\ }\textbf {\bibinfo {volume} {588}},\ \bibinfo {pages}
  {414} (\bibinfo {year} {2020})}\BibitemShut {NoStop}%
\bibitem [{\citenamefont {Colombo}\ \emph {et~al.}(2022)\citenamefont
  {Colombo}, \citenamefont {Pedrozo-Pe{\~{n}}afiel}, \citenamefont
  {Adiyatullin}, \citenamefont {Li}, \citenamefont {Mendez}, \citenamefont
  {Shu},\ and\ \citenamefont {Vuleti{\'{c}}}}]{Colombo2022}%
  \BibitemOpen
  \bibfield  {author} {\bibinfo {author} {\bibfnamefont {S.}~\bibnamefont
  {Colombo}}, \bibinfo {author} {\bibfnamefont {E.}~\bibnamefont
  {Pedrozo-Pe{\~{n}}afiel}}, \bibinfo {author} {\bibfnamefont {A.~F.}\
  \bibnamefont {Adiyatullin}}, \bibinfo {author} {\bibfnamefont
  {Z.}~\bibnamefont {Li}}, \bibinfo {author} {\bibfnamefont {E.}~\bibnamefont
  {Mendez}}, \bibinfo {author} {\bibfnamefont {C.}~\bibnamefont {Shu}},\ and\
  \bibinfo {author} {\bibfnamefont {V.}~\bibnamefont {Vuleti{\'{c}}}},\
  }\bibfield  {title} {\bibinfo {title} {Time-reversal-based quantum metrology
  with many-body entangled states},\ }\href
  {https://doi.org/10.1038/s41567-022-01653-5} {\bibfield  {journal} {\bibinfo
  {journal} {Nature Physics}\ }\textbf {\bibinfo {volume} {18}},\ \bibinfo
  {pages} {925} (\bibinfo {year} {2022})}\BibitemShut {NoStop}%
\bibitem [{\citenamefont {Hines}\ \emph {et~al.}(2023)\citenamefont {Hines},
  \citenamefont {Rajagopal}, \citenamefont {Moreau}, \citenamefont {Wahrman},
  \citenamefont {Lewis}, \citenamefont {Marković},\ and\ \citenamefont
  {Schleier-Smith}}]{Hines2023}%
  \BibitemOpen
  \bibfield  {author} {\bibinfo {author} {\bibfnamefont {J.~A.}\ \bibnamefont
  {Hines}}, \bibinfo {author} {\bibfnamefont {S.~V.}\ \bibnamefont
  {Rajagopal}}, \bibinfo {author} {\bibfnamefont {G.~L.}\ \bibnamefont
  {Moreau}}, \bibinfo {author} {\bibfnamefont {M.~D.}\ \bibnamefont {Wahrman}},
  \bibinfo {author} {\bibfnamefont {N.~A.}\ \bibnamefont {Lewis}}, \bibinfo
  {author} {\bibfnamefont {O.}~\bibnamefont {Marković}},\ and\ \bibinfo
  {author} {\bibfnamefont {M.}~\bibnamefont {Schleier-Smith}},\ }\bibfield
  {title} {\bibinfo {title} {Spin squeezing by rydberg dressing in an array of
  atomic ensembles},\ }\href {https://doi.org/10.1103/PhysRevLett.131.063401}
  {\bibfield  {journal} {\bibinfo  {journal} {Phys. Rev. Lett.}\ }\textbf
  {\bibinfo {volume} {131}},\ \bibinfo {pages} {063401} (\bibinfo {year}
  {2023})}\BibitemShut {NoStop}%
\bibitem [{\citenamefont {Eckner}\ \emph {et~al.}(2023)\citenamefont {Eckner},
  \citenamefont {Oppong}, \citenamefont {Cao}, \citenamefont {Young},
  \citenamefont {Milner}, \citenamefont {Robinson}, \citenamefont {Ye},\ and\
  \citenamefont {Kaufman}}]{Eckner2023}%
  \BibitemOpen
  \bibfield  {author} {\bibinfo {author} {\bibfnamefont {W.~J.}\ \bibnamefont
  {Eckner}}, \bibinfo {author} {\bibfnamefont {N.~D.}\ \bibnamefont {Oppong}},
  \bibinfo {author} {\bibfnamefont {A.}~\bibnamefont {Cao}}, \bibinfo {author}
  {\bibfnamefont {A.~W.}\ \bibnamefont {Young}}, \bibinfo {author}
  {\bibfnamefont {W.~R.}\ \bibnamefont {Milner}}, \bibinfo {author}
  {\bibfnamefont {J.~M.}\ \bibnamefont {Robinson}}, \bibinfo {author}
  {\bibfnamefont {J.}~\bibnamefont {Ye}},\ and\ \bibinfo {author}
  {\bibfnamefont {A.~M.}\ \bibnamefont {Kaufman}},\ }\bibfield  {title}
  {\bibinfo {title} {Realizing spin squeezing with rydberg interactions in a
  programmable optical clock},\ }\href
  {https://doi.org/10.1038/s41586-023-06360-6} {\bibfield  {journal} {\bibinfo
  {journal} {Nature}\ }\textbf {\bibinfo {volume} {621}},\ \bibinfo {pages}
  {734} (\bibinfo {year} {2023})}\BibitemShut {NoStop}%
\bibitem [{\citenamefont {Franke}\ \emph {et~al.}(2023)\citenamefont {Franke},
  \citenamefont {Muleady}, \citenamefont {Kaubruegger}, \citenamefont {Kranzl},
  \citenamefont {Blatt}, \citenamefont {Rey}, \citenamefont {Joshi},\ and\
  \citenamefont {Roos}}]{Franke2023}%
  \BibitemOpen
  \bibfield  {author} {\bibinfo {author} {\bibfnamefont {J.}~\bibnamefont
  {Franke}}, \bibinfo {author} {\bibfnamefont {S.~R.}\ \bibnamefont {Muleady}},
  \bibinfo {author} {\bibfnamefont {R.}~\bibnamefont {Kaubruegger}}, \bibinfo
  {author} {\bibfnamefont {F.}~\bibnamefont {Kranzl}}, \bibinfo {author}
  {\bibfnamefont {R.}~\bibnamefont {Blatt}}, \bibinfo {author} {\bibfnamefont
  {A.~M.}\ \bibnamefont {Rey}}, \bibinfo {author} {\bibfnamefont {M.~K.}\
  \bibnamefont {Joshi}},\ and\ \bibinfo {author} {\bibfnamefont {C.~F.}\
  \bibnamefont {Roos}},\ }\bibfield  {title} {\bibinfo {title}
  {Quantum-enhanced sensing on an optical transition via emergent collective
  quantum correlations},\ }\href {https://doi.org/10.1038/s41586-023-06472-z}
  {\bibfield  {journal} {\bibinfo  {journal} {Nature}\ }\textbf {\bibinfo
  {volume} {621}},\ \bibinfo {pages} {740} (\bibinfo {year}
  {2023})}\BibitemShut {NoStop}%
\bibitem [{\citenamefont {Bennett}\ \emph {et~al.}(2013)\citenamefont
  {Bennett}, \citenamefont {Yao}, \citenamefont {Otterbach}, \citenamefont
  {Zoller}, \citenamefont {Rabl},\ and\ \citenamefont {Lukin}}]{Bennett2013}%
  \BibitemOpen
  \bibfield  {author} {\bibinfo {author} {\bibfnamefont {S.~D.}\ \bibnamefont
  {Bennett}}, \bibinfo {author} {\bibfnamefont {N.~Y.}\ \bibnamefont {Yao}},
  \bibinfo {author} {\bibfnamefont {J.}~\bibnamefont {Otterbach}}, \bibinfo
  {author} {\bibfnamefont {P.}~\bibnamefont {Zoller}}, \bibinfo {author}
  {\bibfnamefont {P.}~\bibnamefont {Rabl}},\ and\ \bibinfo {author}
  {\bibfnamefont {M.~D.}\ \bibnamefont {Lukin}},\ }\bibfield  {title} {\bibinfo
  {title} {Phonon-induced spin-spin interactions in diamond nanostructures:
  Application to spin squeezing},\ }\href
  {https://doi.org/10.1103/PhysRevLett.110.156402} {\bibfield  {journal}
  {\bibinfo  {journal} {Phys. Rev. Lett.}\ }\textbf {\bibinfo {volume} {110}},\
  \bibinfo {pages} {156402} (\bibinfo {year} {2013})}\BibitemShut {NoStop}%
\bibitem [{\citenamefont {Dooley}\ \emph {et~al.}(2016)\citenamefont {Dooley},
  \citenamefont {Yukawa}, \citenamefont {Matsuzaki}, \citenamefont {Knee},
  \citenamefont {Munro},\ and\ \citenamefont {Nemoto}}]{Dooley2016}%
  \BibitemOpen
  \bibfield  {author} {\bibinfo {author} {\bibfnamefont {S.}~\bibnamefont
  {Dooley}}, \bibinfo {author} {\bibfnamefont {E.}~\bibnamefont {Yukawa}},
  \bibinfo {author} {\bibfnamefont {Y.}~\bibnamefont {Matsuzaki}}, \bibinfo
  {author} {\bibfnamefont {G.~C.}\ \bibnamefont {Knee}}, \bibinfo {author}
  {\bibfnamefont {W.~J.}\ \bibnamefont {Munro}},\ and\ \bibinfo {author}
  {\bibfnamefont {K.}~\bibnamefont {Nemoto}},\ }\bibfield  {title} {\bibinfo
  {title} {A hybrid-systems approach to spin squeezing using a highly
  dissipative ancillary system},\ }\href
  {https://doi.org/10.1088/1367-2630/18/5/053011} {\bibfield  {journal}
  {\bibinfo  {journal} {New Journal of Physics}\ }\textbf {\bibinfo {volume}
  {18}},\ \bibinfo {pages} {053011} (\bibinfo {year} {2016})}\BibitemShut
  {NoStop}%
\bibitem [{\citenamefont {Lewis-Swan}\ \emph {et~al.}(2018)\citenamefont
  {Lewis-Swan}, \citenamefont {Norcia}, \citenamefont {Cline}, \citenamefont
  {Thompson},\ and\ \citenamefont {Rey}}]{LewisSwan2018}%
  \BibitemOpen
  \bibfield  {author} {\bibinfo {author} {\bibfnamefont {R.~J.}\ \bibnamefont
  {Lewis-Swan}}, \bibinfo {author} {\bibfnamefont {M.~A.}\ \bibnamefont
  {Norcia}}, \bibinfo {author} {\bibfnamefont {J.~R.~K.}\ \bibnamefont
  {Cline}}, \bibinfo {author} {\bibfnamefont {J.~K.}\ \bibnamefont
  {Thompson}},\ and\ \bibinfo {author} {\bibfnamefont {A.~M.}\ \bibnamefont
  {Rey}},\ }\bibfield  {title} {\bibinfo {title} {Robust spin squeezing via
  photon-mediated interactions on an optical clock transition},\ }\href
  {https://doi.org/10.1103/PhysRevLett.121.070403} {\bibfield  {journal}
  {\bibinfo  {journal} {Phys. Rev. Lett.}\ }\textbf {\bibinfo {volume} {121}},\
  \bibinfo {pages} {070403} (\bibinfo {year} {2018})}\BibitemShut {NoStop}%
\bibitem [{\citenamefont {Groszkowski}\ \emph {et~al.}(2020)\citenamefont
  {Groszkowski}, \citenamefont {Lau}, \citenamefont {Leroux}, \citenamefont
  {Govia},\ and\ \citenamefont {Clerk}}]{Groszkowski2020}%
  \BibitemOpen
  \bibfield  {author} {\bibinfo {author} {\bibfnamefont {P.}~\bibnamefont
  {Groszkowski}}, \bibinfo {author} {\bibfnamefont {H.-K.}\ \bibnamefont
  {Lau}}, \bibinfo {author} {\bibfnamefont {C.}~\bibnamefont {Leroux}},
  \bibinfo {author} {\bibfnamefont {L.~C.~G.}\ \bibnamefont {Govia}},\ and\
  \bibinfo {author} {\bibfnamefont {A.~A.}\ \bibnamefont {Clerk}},\ }\bibfield
  {title} {\bibinfo {title} {Heisenberg-limited spin squeezing via bosonic
  parametric driving},\ }\href {https://doi.org/10.1103/PhysRevLett.125.203601}
  {\bibfield  {journal} {\bibinfo  {journal} {Phys. Rev. Lett.}\ }\textbf
  {\bibinfo {volume} {125}},\ \bibinfo {pages} {203601} (\bibinfo {year}
  {2020})}\BibitemShut {NoStop}%
\bibitem [{\citenamefont {Schleier-Smith}\ \emph {et~al.}(2010)\citenamefont
  {Schleier-Smith}, \citenamefont {Leroux},\ and\ \citenamefont
  {Vuleti\'{c}}}]{SchleierSmith2010}%
  \BibitemOpen
  \bibfield  {author} {\bibinfo {author} {\bibfnamefont {M.~H.}\ \bibnamefont
  {Schleier-Smith}}, \bibinfo {author} {\bibfnamefont {I.~D.}\ \bibnamefont
  {Leroux}},\ and\ \bibinfo {author} {\bibfnamefont {V.}~\bibnamefont
  {Vuleti\'{c}}},\ }\bibfield  {title} {\bibinfo {title} {Squeezing the
  collective spin of a dilute atomic ensemble by cavity feedback},\ }\href
  {https://doi.org/10.1103/PhysRevA.81.021804} {\bibfield  {journal} {\bibinfo
  {journal} {Phys. Rev. A}\ }\textbf {\bibinfo {volume} {81}},\ \bibinfo
  {pages} {021804(R)} (\bibinfo {year} {2010})}\BibitemShut {NoStop}%
\bibitem [{\citenamefont {Zhang}\ \emph {et~al.}(2015)\citenamefont {Zhang},
  \citenamefont {Zou}, \citenamefont {Zou}, \citenamefont {Jiang},\ and\
  \citenamefont {Guo}}]{Zhang2015}%
  \BibitemOpen
  \bibfield  {author} {\bibinfo {author} {\bibfnamefont {Y.-L.}\ \bibnamefont
  {Zhang}}, \bibinfo {author} {\bibfnamefont {C.-L.}\ \bibnamefont {Zou}},
  \bibinfo {author} {\bibfnamefont {X.-B.}\ \bibnamefont {Zou}}, \bibinfo
  {author} {\bibfnamefont {L.}~\bibnamefont {Jiang}},\ and\ \bibinfo {author}
  {\bibfnamefont {G.-C.}\ \bibnamefont {Guo}},\ }\bibfield  {title} {\bibinfo
  {title} {Detuning-enhanced cavity spin squeezing},\ }\href
  {https://doi.org/10.1103/PhysRevA.91.033625} {\bibfield  {journal} {\bibinfo
  {journal} {Phys. Rev. A}\ }\textbf {\bibinfo {volume} {91}},\ \bibinfo
  {pages} {033625} (\bibinfo {year} {2015})}\BibitemShut {NoStop}%
\bibitem [{\citenamefont {Davis}\ \emph {et~al.}(2016)\citenamefont {Davis},
  \citenamefont {Bentsen},\ and\ \citenamefont {Schleier-Smith}}]{Davis2016}%
  \BibitemOpen
  \bibfield  {author} {\bibinfo {author} {\bibfnamefont {E.}~\bibnamefont
  {Davis}}, \bibinfo {author} {\bibfnamefont {G.}~\bibnamefont {Bentsen}},\
  and\ \bibinfo {author} {\bibfnamefont {M.}~\bibnamefont {Schleier-Smith}},\
  }\bibfield  {title} {\bibinfo {title} {Approaching the heisenberg limit
  without single-particle detection},\ }\href
  {https://doi.org/10.1103/PhysRevLett.116.053601} {\bibfield  {journal}
  {\bibinfo  {journal} {Phys. Rev. Lett.}\ }\textbf {\bibinfo {volume} {116}},\
  \bibinfo {pages} {053601} (\bibinfo {year} {2016})}\BibitemShut {NoStop}%
\bibitem [{\citenamefont {Chu}\ \emph {et~al.}(2021)\citenamefont {Chu},
  \citenamefont {He}, \citenamefont {Thompson},\ and\ \citenamefont
  {Rey}}]{Chu2021}%
  \BibitemOpen
  \bibfield  {author} {\bibinfo {author} {\bibfnamefont {A.}~\bibnamefont
  {Chu}}, \bibinfo {author} {\bibfnamefont {P.}~\bibnamefont {He}}, \bibinfo
  {author} {\bibfnamefont {J.~K.}\ \bibnamefont {Thompson}},\ and\ \bibinfo
  {author} {\bibfnamefont {A.~M.}\ \bibnamefont {Rey}},\ }\bibfield  {title}
  {\bibinfo {title} {Quantum enhanced cavity qed interferometer with partially
  delocalized atoms in lattices},\ }\href
  {https://doi.org/10.1103/PhysRevLett.127.210401} {\bibfield  {journal}
  {\bibinfo  {journal} {Phys. Rev. Lett.}\ }\textbf {\bibinfo {volume} {127}},\
  \bibinfo {pages} {210401} (\bibinfo {year} {2021})}\BibitemShut {NoStop}%
\bibitem [{\citenamefont {Li}\ \emph {et~al.}(2022)\citenamefont {Li},
  \citenamefont {Braverman}, \citenamefont {Colombo}, \citenamefont {Shu},
  \citenamefont {Kawasaki}, \citenamefont {Adiyatullin}, \citenamefont
  {Pedrozo-Pe\~nafiel}, \citenamefont {Mendez},\ and\ \citenamefont
  {Vuleti\'{c}}}]{Li2022}%
  \BibitemOpen
  \bibfield  {author} {\bibinfo {author} {\bibfnamefont {Z.}~\bibnamefont
  {Li}}, \bibinfo {author} {\bibfnamefont {B.}~\bibnamefont {Braverman}},
  \bibinfo {author} {\bibfnamefont {S.}~\bibnamefont {Colombo}}, \bibinfo
  {author} {\bibfnamefont {C.}~\bibnamefont {Shu}}, \bibinfo {author}
  {\bibfnamefont {A.}~\bibnamefont {Kawasaki}}, \bibinfo {author}
  {\bibfnamefont {A.~F.}\ \bibnamefont {Adiyatullin}}, \bibinfo {author}
  {\bibfnamefont {E.}~\bibnamefont {Pedrozo-Pe\~nafiel}}, \bibinfo {author}
  {\bibfnamefont {E.}~\bibnamefont {Mendez}},\ and\ \bibinfo {author}
  {\bibfnamefont {V.}~\bibnamefont {Vuleti\'{c}}},\ }\bibfield  {title}
  {\bibinfo {title} {Collective spin-light and light-mediated spin-spin
  interactions in an optical cavity},\ }\href
  {https://doi.org/10.1103/PRXQuantum.3.020308} {\bibfield  {journal} {\bibinfo
   {journal} {PRX Quantum}\ }\textbf {\bibinfo {volume} {3}},\ \bibinfo {pages}
  {020308} (\bibinfo {year} {2022})}\BibitemShut {NoStop}%
\bibitem [{\citenamefont {Zou}\ \emph {et~al.}(2014)\citenamefont {Zou},
  \citenamefont {Marcos}, \citenamefont {Diehl}, \citenamefont {Putz},
  \citenamefont {Schmiedmayer}, \citenamefont {Majer},\ and\ \citenamefont
  {Rabl}}]{Zou2014}%
  \BibitemOpen
  \bibfield  {author} {\bibinfo {author} {\bibfnamefont {L.~J.}\ \bibnamefont
  {Zou}}, \bibinfo {author} {\bibfnamefont {D.}~\bibnamefont {Marcos}},
  \bibinfo {author} {\bibfnamefont {S.}~\bibnamefont {Diehl}}, \bibinfo
  {author} {\bibfnamefont {S.}~\bibnamefont {Putz}}, \bibinfo {author}
  {\bibfnamefont {J.}~\bibnamefont {Schmiedmayer}}, \bibinfo {author}
  {\bibfnamefont {J.}~\bibnamefont {Majer}},\ and\ \bibinfo {author}
  {\bibfnamefont {P.}~\bibnamefont {Rabl}},\ }\bibfield  {title} {\bibinfo
  {title} {Implementation of the dicke lattice model in hybrid quantum system
  arrays},\ }\href {https://doi.org/10.1103/PhysRevLett.113.023603} {\bibfield
  {journal} {\bibinfo  {journal} {Phys. Rev. Lett.}\ }\textbf {\bibinfo
  {volume} {113}},\ \bibinfo {pages} {023603} (\bibinfo {year}
  {2014})}\BibitemShut {NoStop}%
\bibitem [{\citenamefont {Koppenh\"ofer}\ \emph {et~al.}(2022)\citenamefont
  {Koppenh\"ofer}, \citenamefont {Groszkowski}, \citenamefont {Lau},\ and\
  \citenamefont {Clerk}}]{Koppenhoefer2022}%
  \BibitemOpen
  \bibfield  {author} {\bibinfo {author} {\bibfnamefont {M.}~\bibnamefont
  {Koppenh\"ofer}}, \bibinfo {author} {\bibfnamefont {P.}~\bibnamefont
  {Groszkowski}}, \bibinfo {author} {\bibfnamefont {H.-K.}\ \bibnamefont
  {Lau}},\ and\ \bibinfo {author} {\bibfnamefont {A.~A.}\ \bibnamefont
  {Clerk}},\ }\bibfield  {title} {\bibinfo {title} {Dissipative superradiant
  spin amplifier for enhanced quantum sensing},\ }\href
  {https://doi.org/10.1103/PRXQuantum.3.030330} {\bibfield  {journal} {\bibinfo
   {journal} {PRX Quantum}\ }\textbf {\bibinfo {volume} {3}},\ \bibinfo {pages}
  {030330} (\bibinfo {year} {2022})}\BibitemShut {NoStop}%
\bibitem [{\citenamefont {Koppenhöfer}\ \emph {et~al.}(2023)\citenamefont
  {Koppenhöfer}, \citenamefont {Groszkowski},\ and\ \citenamefont
  {Clerk}}]{Koppenhoefer2023}%
  \BibitemOpen
  \bibfield  {author} {\bibinfo {author} {\bibfnamefont {M.}~\bibnamefont
  {Koppenhöfer}}, \bibinfo {author} {\bibfnamefont {P.}~\bibnamefont
  {Groszkowski}},\ and\ \bibinfo {author} {\bibfnamefont {A.~A.}\ \bibnamefont
  {Clerk}},\ }\bibfield  {title} {\bibinfo {title} {Squeezed superradiance
  enables robust entanglement-enhanced metrology even with highly imperfect
  readout},\ }\href
  {https://doi.org/https://doi.org/10.1103/PhysRevLett.131.060802} {\bibfield
  {journal} {\bibinfo  {journal} {Phys. Rev. Lett.}\ }\textbf {\bibinfo
  {volume} {131}},\ \bibinfo {pages} {060802} (\bibinfo {year}
  {2023})}\BibitemShut {NoStop}%
\bibitem [{\citenamefont {Wineland}\ \emph {et~al.}(1992)\citenamefont
  {Wineland}, \citenamefont {Bollinger}, \citenamefont {Itano}, \citenamefont
  {Moore},\ and\ \citenamefont {Heinzen}}]{Wineland1992}%
  \BibitemOpen
  \bibfield  {author} {\bibinfo {author} {\bibfnamefont {D.~J.}\ \bibnamefont
  {Wineland}}, \bibinfo {author} {\bibfnamefont {J.~J.}\ \bibnamefont
  {Bollinger}}, \bibinfo {author} {\bibfnamefont {W.~M.}\ \bibnamefont
  {Itano}}, \bibinfo {author} {\bibfnamefont {F.~L.}\ \bibnamefont {Moore}},\
  and\ \bibinfo {author} {\bibfnamefont {D.~J.}\ \bibnamefont {Heinzen}},\
  }\bibfield  {title} {\bibinfo {title} {Spin squeezing and reduced quantum
  noise in spectroscopy},\ }\href {https://doi.org/10.1103/PhysRevA.46.R6797}
  {\bibfield  {journal} {\bibinfo  {journal} {Phys. Rev. A}\ }\textbf {\bibinfo
  {volume} {46}},\ \bibinfo {pages} {R6797} (\bibinfo {year}
  {1992})}\BibitemShut {NoStop}%
\bibitem [{\citenamefont {Barry}\ \emph {et~al.}(2020)\citenamefont {Barry},
  \citenamefont {Schloss}, \citenamefont {Bauch}, \citenamefont {Turner},
  \citenamefont {Hart}, \citenamefont {Pham},\ and\ \citenamefont
  {Walsworth}}]{Barry2020}%
  \BibitemOpen
  \bibfield  {author} {\bibinfo {author} {\bibfnamefont {J.~F.}\ \bibnamefont
  {Barry}}, \bibinfo {author} {\bibfnamefont {J.~M.}\ \bibnamefont {Schloss}},
  \bibinfo {author} {\bibfnamefont {E.}~\bibnamefont {Bauch}}, \bibinfo
  {author} {\bibfnamefont {M.~J.}\ \bibnamefont {Turner}}, \bibinfo {author}
  {\bibfnamefont {C.~A.}\ \bibnamefont {Hart}}, \bibinfo {author}
  {\bibfnamefont {L.~M.}\ \bibnamefont {Pham}},\ and\ \bibinfo {author}
  {\bibfnamefont {R.~L.}\ \bibnamefont {Walsworth}},\ }\bibfield  {title}
  {\bibinfo {title} {Sensitivity optimization for nv-diamond magnetometry},\
  }\href {https://doi.org/10.1103/RevModPhys.92.015004} {\bibfield  {journal}
  {\bibinfo  {journal} {Rev. Mod. Phys.}\ }\textbf {\bibinfo {volume} {92}},\
  \bibinfo {pages} {015004} (\bibinfo {year} {2020})}\BibitemShut {NoStop}%
\bibitem [{\citenamefont {Eisenach}\ \emph {et~al.}(2021)\citenamefont
  {Eisenach}, \citenamefont {Barry}, \citenamefont {O'Keeffe}, \citenamefont
  {Schloss}, \citenamefont {Steinecker}, \citenamefont {Englund},\ and\
  \citenamefont {Braje}}]{Eisenach2021}%
  \BibitemOpen
  \bibfield  {author} {\bibinfo {author} {\bibfnamefont {E.~R.}\ \bibnamefont
  {Eisenach}}, \bibinfo {author} {\bibfnamefont {J.~F.}\ \bibnamefont {Barry}},
  \bibinfo {author} {\bibfnamefont {M.~F.}\ \bibnamefont {O'Keeffe}}, \bibinfo
  {author} {\bibfnamefont {J.~M.}\ \bibnamefont {Schloss}}, \bibinfo {author}
  {\bibfnamefont {M.~H.}\ \bibnamefont {Steinecker}}, \bibinfo {author}
  {\bibfnamefont {D.~R.}\ \bibnamefont {Englund}},\ and\ \bibinfo {author}
  {\bibfnamefont {D.~A.}\ \bibnamefont {Braje}},\ }\bibfield  {title} {\bibinfo
  {title} {Cavity-enhanced microwave readout of a solid-state spin sensor},\
  }\href {https://doi.org/10.1038/s41467-021-21256-7} {\bibfield  {journal}
  {\bibinfo  {journal} {Nature Communications}\ }\textbf {\bibinfo {volume}
  {12}},\ \bibinfo {pages} {1357} (\bibinfo {year} {2021})}\BibitemShut
  {NoStop}%
\bibitem [{\citenamefont {Kubo}(1962)}]{Kubo1962}%
  \BibitemOpen
  \bibfield  {author} {\bibinfo {author} {\bibfnamefont {R.}~\bibnamefont
  {Kubo}},\ }\bibfield  {title} {\bibinfo {title} {Generalized cumulant
  expansion method},\ }\href {https://doi.org/10.1143/JPSJ.17.1100} {\bibfield
  {journal} {\bibinfo  {journal} {Journal of the Physical Society of Japan}\
  }\textbf {\bibinfo {volume} {17}},\ \bibinfo {pages} {1100} (\bibinfo {year}
  {1962})}\BibitemShut {NoStop}%
\bibitem [{\citenamefont {Zens}\ \emph {et~al.}(2019)\citenamefont {Zens},
  \citenamefont {Krimer},\ and\ \citenamefont {Rotter}}]{Zens2019}%
  \BibitemOpen
  \bibfield  {author} {\bibinfo {author} {\bibfnamefont {M.}~\bibnamefont
  {Zens}}, \bibinfo {author} {\bibfnamefont {D.~O.}\ \bibnamefont {Krimer}},\
  and\ \bibinfo {author} {\bibfnamefont {S.}~\bibnamefont {Rotter}},\
  }\bibfield  {title} {\bibinfo {title} {Critical phenomena and nonlinear
  dynamics in a spin ensemble strongly coupled to a cavity. ii.
  semiclassical-to-quantum boundary},\ }\href
  {https://doi.org/10.1103/PhysRevA.100.013856} {\bibfield  {journal} {\bibinfo
   {journal} {Phys. Rev. A}\ }\textbf {\bibinfo {volume} {100}},\ \bibinfo
  {pages} {013856} (\bibinfo {year} {2019})}\BibitemShut {NoStop}%
\bibitem [{\citenamefont {Groszkowski}\ \emph {et~al.}(2022)\citenamefont
  {Groszkowski}, \citenamefont {Koppenh\"ofer}, \citenamefont {Lau},\ and\
  \citenamefont {Clerk}}]{Groszkowski2022}%
  \BibitemOpen
  \bibfield  {author} {\bibinfo {author} {\bibfnamefont {P.}~\bibnamefont
  {Groszkowski}}, \bibinfo {author} {\bibfnamefont {M.}~\bibnamefont
  {Koppenh\"ofer}}, \bibinfo {author} {\bibfnamefont {H.-K.}\ \bibnamefont
  {Lau}},\ and\ \bibinfo {author} {\bibfnamefont {A.~A.}\ \bibnamefont
  {Clerk}},\ }\bibfield  {title} {\bibinfo {title} {Reservoir-engineered spin
  squeezing: Macroscopic even-odd effects and hybrid-systems implementations},\
  }\href {https://doi.org/10.1103/PhysRevX.12.011015} {\bibfield  {journal}
  {\bibinfo  {journal} {Phys. Rev. X}\ }\textbf {\bibinfo {volume} {12}},\
  \bibinfo {pages} {011015} (\bibinfo {year} {2022})}\BibitemShut {NoStop}%
\bibitem [{\citenamefont {Anders}\ \emph {et~al.}(2018)\citenamefont {Anders},
  \citenamefont {Pezz\`e}, \citenamefont {Smerzi},\ and\ \citenamefont
  {Klempt}}]{Anders2018}%
  \BibitemOpen
  \bibfield  {author} {\bibinfo {author} {\bibfnamefont {F.}~\bibnamefont
  {Anders}}, \bibinfo {author} {\bibfnamefont {L.}~\bibnamefont {Pezz\`e}},
  \bibinfo {author} {\bibfnamefont {A.}~\bibnamefont {Smerzi}},\ and\ \bibinfo
  {author} {\bibfnamefont {C.}~\bibnamefont {Klempt}},\ }\bibfield  {title}
  {\bibinfo {title} {Phase magnification by two-axis countertwisting for
  detection-noise robust interferometry},\ }\href
  {https://doi.org/10.1103/PhysRevA.97.043813} {\bibfield  {journal} {\bibinfo
  {journal} {Phys. Rev. A}\ }\textbf {\bibinfo {volume} {97}},\ \bibinfo
  {pages} {043813} (\bibinfo {year} {2018})}\BibitemShut {NoStop}%
\bibitem [{\citenamefont {Munoz-Arias}\ \emph {et~al.}(2023)\citenamefont
  {Munoz-Arias}, \citenamefont {Deutsch},\ and\ \citenamefont
  {Poggi}}]{MunozArias2023}%
  \BibitemOpen
  \bibfield  {author} {\bibinfo {author} {\bibfnamefont {M.~H.}\ \bibnamefont
  {Munoz-Arias}}, \bibinfo {author} {\bibfnamefont {I.~H.}\ \bibnamefont
  {Deutsch}},\ and\ \bibinfo {author} {\bibfnamefont {P.~M.}\ \bibnamefont
  {Poggi}},\ }\bibfield  {title} {\bibinfo {title} {Phase-space geometry and
  optimal state preparation in quantum metrology with collective spins},\
  }\href {https://doi.org/10.1103/PRXQuantum.4.020314} {\bibfield  {journal}
  {\bibinfo  {journal} {PRX Quantum}\ }\textbf {\bibinfo {volume} {4}},\
  \bibinfo {pages} {020314} (\bibinfo {year} {2023})}\BibitemShut {NoStop}%
\bibitem [{\citenamefont {Mirkhalaf}\ \emph {et~al.}(2018)\citenamefont
  {Mirkhalaf}, \citenamefont {Nolan},\ and\ \citenamefont
  {Haine}}]{Mirkhalaf2018}%
  \BibitemOpen
  \bibfield  {author} {\bibinfo {author} {\bibfnamefont {S.~S.}\ \bibnamefont
  {Mirkhalaf}}, \bibinfo {author} {\bibfnamefont {S.~P.}\ \bibnamefont
  {Nolan}},\ and\ \bibinfo {author} {\bibfnamefont {S.~A.}\ \bibnamefont
  {Haine}},\ }\bibfield  {title} {\bibinfo {title} {Robustifying twist-and-turn
  entanglement with interaction-based readout},\ }\href
  {https://doi.org/10.1103/PhysRevA.97.053618} {\bibfield  {journal} {\bibinfo
  {journal} {Phys. Rev. A}\ }\textbf {\bibinfo {volume} {97}},\ \bibinfo
  {pages} {053618} (\bibinfo {year} {2018})}\BibitemShut {NoStop}%
\end{thebibliography}%

\end{document}